\documentclass[%
 reprint,
 superscriptaddress,
 amsmath,amssymb,
 aps,
]{revtex4-2}

\usepackage{amsmath,amsthm}
\usepackage{graphicx}
\usepackage{dcolumn}
\usepackage{bm}
\usepackage{braket}
\usepackage{todonotes}
\usepackage[compat=0.4]{yquant}
\usepackage{textgreek}

\usepackage{xcolor}
\definecolor{riverlane_green}{RGB}{0, 111, 98}
\definecolor{riverlane_light_green}{RGB}{0, 150, 143}
\definecolor{riverlane_orange}{RGB}{255, 117, 0}
\definecolor{riverlane_red}{RGB}{220, 68, 5}
\definecolor{riverlane_pink}{RGB}{207, 111, 127}
\usepackage{hyperref}
\hypersetup{
  colorlinks   = true, 
  urlcolor     = riverlane_green, 
  linkcolor    = riverlane_orange, 
  citecolor   = riverlane_green  
}

\usepackage{tikz}
\usetikzlibrary{quantikz}

\usepackage{bm}
\usepackage{bbm}
\usepackage{bbold}

\newtheorem{numresult}{Numerical Result}[section]

\newtheorem{theorem}{Theorem}[section]

\newtheorem{lemma}[theorem]{Lemma}

\begin{document}

\preprint{APS/123-QED}

\title{Fault-tolerant quantum simulation of generalized Hubbard models}

\author{Andreas Juul Bay-Smidt}
\email{andreas.bay-smidt@nbi.ku.dk}
\affiliation{NNF Quantum Computing Programme, Niels Bohr Institute, University of Copenhagen, Denmark}
\affiliation{Nano-Science Center and Department of Chemistry, University of Copenhagen, Denmark}
\author{Frederik Ravn Klausen}
\affiliation{NNF Quantum Computing Programme, Niels Bohr Institute, University of Copenhagen, Denmark}
\affiliation{Department of Mathematics, Princeton University, USA}
\author{Christoph Sünderhauf}
\affiliation{Riverlane, Cambridge, CB2 3BZ, UK}
\author{Róbert Izsák}
\affiliation{Riverlane, Cambridge, CB2 3BZ, UK}
\author{Gemma C. Solomon}
\email{gsolomon@chem.ku.dk}
\affiliation{NNF Quantum Computing Programme, Niels Bohr Institute, University of Copenhagen, Denmark}
\affiliation{Nano-Science Center and Department of Chemistry, University of Copenhagen, Denmark}
\author{Nick S. Blunt}
\email{nick.blunt@riverlane.com}
\affiliation{Riverlane, Cambridge, CB2 3BZ, UK}

\date{\today}

\begin{abstract}
Quantum simulations of strongly interacting fermionic systems, such as those described by the Hubbard model, are promising candidates for useful early fault-tolerant quantum computing applications. This paper presents Tile Trotterization, a generalization of plaquette Trotterization (PLAQ), which uses a set of tiles to construct Trotter decompositions of arbitrary lattice Hubbard models. The Tile Trotterization scheme also enables the simulation of more complex models, including the extended Hubbard model. We improve previous Hubbard model commutator bounds, further provide tight commutator bounds for periodic extended Hubbard models, and demonstrate the use of tensor network methods for this task. We consider applications of Tile Trotterization to simulate hexagonal lattice Hubbard models and compare the resource requirements of Tile Trotterization for performing quantum phase estimation to a qubitization-based approach, demonstrating that Tile Trotterization scales more efficiently with system size. These advancements significantly broaden the potential applications of early fault-tolerant quantum computers to models of practical interest in materials research and organic chemistry.
\end{abstract}

\maketitle

\section{\label{sec:Introduction} Introduction}
Simulating quantum systems of interacting electrons is an important and complex challenge in the study of molecules and materials. Quantum simulation of fermionic systems is considered one of the most promising applications of quantum computers \cite{Feynman1982SimulatingComputers, Lloyd1996UniversalSimulators, Kassal2008Polynomial-timeDynamics}, which has motivated resource estimates for quantum simulation of complex systems such as the FeMoco-complex and the cytochrome P450 enzyme \cite{Reiher2017ElucidatingComputers, Goings2022ReliablyComputers, Lee2021EvenHypercontraction}. Such studies have investigated the electronic structure problem in both first and second quantization, using Gaussian, plane wave or Bloch basis sets \cite{Berry2019QubitizationFactorization,Lee2021EvenHypercontraction, Chan2022Grid-basedComputer, Kivlichan2017BoundingSpace, Ivanov2023QuantumQuantization, Babbush2018Low-DepthMaterials, Rubin2023Fault-TolerantOrbitals}. Despite significant advances in recent years, general electronic structure simulations of practically relevant and classically intractable systems are estimated to require millions of physical qubits and at least billions of T gates \cite{Goings2022ReliablyComputers, Lee2021EvenHypercontraction}, which is far beyond the capabilities of today's quantum hardware.

Recently, quantum simulations of the Hubbard model \cite{Hubbard1963ElectronBands} have received increased attention due to the model's relatively low resource requirements, making it a promising candidate for early demonstrations of practical quantum advantage \cite{Yoshioka2024HuntingProblems, Toshio2024PracticalComputer, Akahoshi2024CompilationArchitecture}. The Hubbard model is a model of interacting electrons which, despite its simplicity compared with the full electronic structure Hamiltonian, is able to describe important aspects of the physics of real materials, and has long been considered a potential model for high-temperature superconductivity \cite{Dong2022MechanismStrength,Kaczmarczyk2013SuperconductivitySolution,Lee2006DopingSuperconductivity}.

Two recent papers by Kivlichan \emph{et al.} \cite{Kivlichan2020ImprovedTrotterization} and Campbell \cite{Campbell2022EarlyModel} introduce efficient Trotterization schemes \cite{Trotter1959OnOperators, Suzuki1991GeneralPhysics, Whitfield2011SimulationComputers} to implement time evolution for the Hubbard model. Hamiltonian time evolution allows for the calculation of dynamical observables such as Green's functions, which are used to access important quantities including the many-body density of states and spectral functions. Time evolution is also an important subroutine in quantum phase estimation (QPE) \cite{Kitaev1995QuantumProblem, Abrams1998AEigenvectors, Cleve1998QuantumRevisited, Aspuru-Guzik2005Chemistry:Energies}, statistical phase estimation (SPE) \cite{Somma2019QuantumAnalysis,Lin2022Heisenberg-LimitedComputers,Wan2022RandomizedEstimation,Wang2023QuantumPrecision,Blunt2023StatisticalProcessor}, and in other ground state \cite{Ding2024Single-ancillaLindbladians, Li2024DissipativeTheory} and Gibbs state \cite{Chen2023QuantumPreparation} sampling methods. QPE can also be performed using the more modern qubitization framework \cite{Low2019HamiltonianQubitization, berry_2018, poulin_2018}, which has also been developed for the Hubbard model \cite{Babbush2018EncodingComplexity}. These prior studies focus on the on-site square lattice Hubbard model \cite{Campbell2022EarlyModel, Kivlichan2020ImprovedTrotterization, Babbush2018EncodingComplexity} and exclude a broader range of lattices and more complicated electronic interaction models. Very recently, Ref.~\cite{kan_2024} introduced an extension to the square lattice Trotter scheme in Ref.~\cite{Campbell2022EarlyModel}, allowing for beyond-nearest-neighbor hopping terms and multi-orbital interactions.

Extending these simulation methods to other lattices enables the simulation of a broader range of real-life materials with complex electronic behavior. This is especially relevant for lattices that introduce frustration, such as the Kagome and triangular lattices, which are challenging to simulate classically and may provide valuable insights into exotic phases of matter \cite{Yu2023MagneticModel, Wen2022SuperconductingModel}. Introducing more complicated electron interaction models allows for studying the effects of non-local electron-electron interactions on charge order \cite{Kennedy2025, Paki2019ChargeModel}, charge density waves \cite{Ferrari2022ChargeFilling}, pair density waves \cite{Schwemmer2024SublatticeModel} and alternative types of superconductivity \cite{Schwemmer2024SublatticeModel, Profe2024KagomePerspective, Zhou2024UnconventionalLattice}. The simplest model employed to study the physics of certain Kagome metals is the Kagome-Hubbard model which includes nearest-neighbor electron interactions \cite{Ferrari2022ChargeFilling, Schwemmer2024SublatticeModel, Profe2024KagomePerspective, Kiesel2012, Kiesel2013}. Conjugated hydrocarbon molecules can also be described by the Hubbard model \cite{Schuler2013OptimalBenzene, Dallaire-Demers2019Low-depthComputer} or more complex extensions such as the extended Hubbard model \cite{Schmalz2011AEthene, Yoshida2024AbComputing, Szabo2021ExtendedOrders} or the PPP model \cite{Pople1953ElectronHydrocarbons, Pariser1953AI, Schmalz2013FromModel, Bostrom2018ChargeCorrelations}. The electronic structure of nanographene is especially interesting because it exhibits topological frustration and strong correlations leading to unconventional magnetic properties \cite{Mishra2019TopologicalNanographene, Song2024HighlyFrustration}, and can be well described by the PPP model \cite{Chiappe_2015}.

In this study, we introduce Tile Trotterization to extend the applicability of PLAQ (developed by Campbell in \cite{Campbell2022EarlyModel}) to arbitrary lattice Hubbard models, and introduce strategies for including longer range interaction terms, used for example in the extended Hubbard model. Tile Trotterization generalizes using plaquettes as the components spanning the square lattice to components of other shapes (called tiles), that can be used to cover any lattice with nearest-neighbor hopping terms to construct efficient Trotter decompositions. We define a set of tiles based on complete bipartite graphs which we show can be implemented efficiently, contrary to tiles of arbitrary shape that generally have a high implementation cost. We also provide examples demonstrating how Tile Trotterization can be used to simulate various hexagonal lattice Hubbard models, including per-Trotter-step gate costs and Trotter error norms. Tiling-based approaches have also been considered in the context of quantum simulation of spin models \cite{Burkard_2025}, and these could potentially also benefit from more advanced tiling strategies.

We also provide tight bounds on the commutator spectral norms used to estimate Trotter error. In this, we demonstrate a natural approach to divide the commutator into operators acting on a limited number of lattice sites, the spectral norm of which can be estimated by numerical techniques such as the density matrix renormalization group (DMRG) algorithm, or other methods generally. Our commutator bounds are presented in three lemmas: \textit{Lemma}~\ref{Lemma:Hubbard_Ihh} for the standard Hubbard model and \textit{Lemmas}~\ref{Lemma:Extended_Hubbard_Chh} and \ref{Lemma:extended_Hubbard_ChC} for the extended Hubbard model. We apply these lemmas to evaluate tight commutator bounds for the periodic hexagonal lattice Hubbard and extended Hubbard models. In Eqs.~(\ref{eq:W_SO2_Hubbard}) and (\ref{eq:W_SO2_extHubbard}), we provide expressions for the non-trivial parts of the Trotter errors for periodic hexagonal lattice Hubbard and extended Hubbard model Trotter steps.

We analyze the performance of Tile Trotterization by comparing it to a qubitization-based approach. Trotterization and qubitization have different dependencies on system size and simulation accuracy, leading to interesting trade-offs when comparing their performance for specific applications \cite{Aftab2024Multi-productScaling}. We construct qubitized quantum walk operators that are optimized for the Hubbard model on the periodic hexagonal lattice, building upon previous work by Babbush \emph{et al.} in \cite{Babbush2018EncodingComplexity}. We provide further optimizations that reduce the cost of the qubitized quantum walk operators and present a detailed analysis of gate and qubit costs. To compare the Trotterization and qubitization approaches, we consider the task of energy estimation by QPE for the periodic hexagonal lattice Hubbard model. We obtain $\mathcal{O}(N^{3/2} \epsilon^{-3/2})$ T-complexity for implementing QPE using Tile Trotterization, compared to $\mathcal{O}(N^2 \epsilon^{-1})$ for qubitized QPE, where $N$ is the number of lattice sites and $\epsilon$ is the target accuracy. Contrary to common belief that qubitization generally is asymptotically more efficient than Trotterization, we find that Tile Trotterization scales better when $\epsilon$ is either constant or allowed to scale with the system size.

Overall, we find that Tile Trotterization implementations of QPE for the Hubbard model and the extended Hubbard model can be performed with T gate costs in the range $10^6$--$10^7$ for classically non-trivial system sizes, making Tile Trotterization based quantum algorithms promising candidates for early fault-tolerant quantum computing applications.

The paper is structured as follows: In Section~\ref{sec:Hubbard_models} we define the generalized Hubbard model, including the standard Hubbard model and the extended Hubbard model used throughout this paper. Section~\ref{sec:tile_Trotterization} introduces Tile Trotterization while Section~\ref{sec:UT_examples} presents a concrete application of Tile Trotterization. In Section~\ref{sec:Qubitization} we present our qubitization approach, and finally, Section~\ref{sec:Phase_estimation} presents a QPE resource comparison of the simulation methods and models discussed in this paper.

\section{\label{sec:Hubbard_models} Hubbard model Hamiltonians}

We consider generalized Hubbard models of the form 
\begin{eqnarray}
    H =  H_h + H_C,
    \label{eq:Hubbard-type-model}
\end{eqnarray}
with hopping terms given by
\begin{eqnarray}
    H_h = -\tau \sum_{i,j,\sigma} R_{ij} a^{\dagger}_{i\sigma} a_{j\sigma},
    \label{eq:H_h}
\end{eqnarray}
where $\tau$ represents the hopping parameter. Here, $R_{ij}$ is the adjacency matrix of the lattice with $R_{ij}=1$ if $i$ and $j$ are neighbors and $R_{ij}=0$ otherwise. The operators $a_{i\sigma}^\dagger$ and $a_{i\sigma}$ are fermionic creation and annihilation operators acting on a spin-$\sigma$ orbital at lattice site $i$. The potential energy, or Coulomb (C), terms have the form
\begin{eqnarray}
    H_C = \sum_{i=1}^N U_i n_{i \uparrow} n_{i \downarrow} + \sum_{i\neq j} \sum_{\sigma,\sigma'}C_{i\sigma,j\sigma'} n_{i\sigma}n_{j\sigma'},
    \label{eq:H_C}
\end{eqnarray}
where $U_{i}$ is the on-site interaction strength on site $i$ and $C_{i\sigma,j\sigma'}$ is the interaction strength between electrons in spin orbitals $i\sigma$ and $j\sigma'$ on different sites. The number operator, $n_{i\sigma}$, is defined as $n_{i\sigma}=a_{i\sigma}^\dagger a_{i\sigma}$. The form of $H_C$ varies across the Hamiltonians considered in this paper.

In all systems and models discussed throughout, we represent the number of lattice sites as $N$ and assign two spin orbitals, $\sigma \in \{\uparrow,\downarrow\}$, to each lattice site, such that the total number of spin orbitals is $2N$. 

\subsection{\label{sec:Hub} The Hubbard Model}

First, we introduce the standard Hubbard model, or simply the Hubbard model, with Coulomb interactions between electrons of opposite spin on the same lattice site. The Hubbard model is defined as
\begin{eqnarray}
        H_H =  H_h + H_I,
        \label{eq:Hubbard_model}
\end{eqnarray}
with potential energy term given by
\begin{eqnarray}
    H_I = U \sum_{i=1}^N n_{i \uparrow} n_{i \downarrow}, \label{eq:on_site_Hubbard}
\end{eqnarray}
where $U$ represents the onsite interaction strength.

There are many fermion-to-qubit mappings that could be applied to this Hamiltonian, including those which exploit locality \cite{setia_2019, guaita_2025}. In this paper we work with the standard Jordan-Wigner (JW) mapping which has been considered in previous studies of the Hubbard model \cite{Kivlichan2020ImprovedTrotterization, Campbell2022EarlyModel}. Here, the number of terms in the transformed Hamiltonian can be reduced by $2N$ by applying a chemical shift to the interaction term in Eq.~(\ref{eq:on_site_Hubbard})
\begin{eqnarray}
\! \! \! \! \! \! H'_I &=& U \sum_{i=1}^N \left(n_{i\uparrow}-\frac{\mathbb{1}}{2}\right)
\left(n_{i\downarrow}-\frac{\mathbb{1}}{2}\right) =\frac{U}{4} \sum_{i=1}^N Z_{i\uparrow}Z_{i\downarrow},
\label{eq:H'_I}
\end{eqnarray}
where $Z_{i\sigma} \equiv 2n_{i\sigma}-\mathbb{1}$. The difference between the original and the shifted on-site interaction term is
\begin{equation}
    H'_I-H_I = -\frac{U}{2} \sum_{i=1}^N \Big( n_{i\uparrow}+n_{i \downarrow} - \frac{\mathbb{1}}{2} \Big).
\end{equation}
In an $\eta$-electron subspace, this difference is a constant energy shift of $\Delta E_I = \frac{U}{2} (\frac{N}{2} -\eta)$, which can be trivially corrected \cite{Campbell2022EarlyModel}. Continuing, we use the modified version \eqref{eq:H'_I} of the on-site interaction and refer to it as $H_I$.

\subsection{\label{sec:EXTHub} The Extended Hubbard Model}
We also consider the extended Hubbard model where Coulomb interactions between electrons on neighboring sites are introduced. We define this model as
\begin{equation}
    H_{EH} = H_h + H_I + H_{V} , \label{eq:extended_Hubbard_model}
\end{equation}
with $H_C=H_I+H_V$. The Coulomb interaction between neighboring sites is defined as
\begin{equation}
    H_{V} = V \sum_{\langle ij \rangle} \sum_{\sigma,\sigma'} n_{i \sigma} n_{j \sigma'},
\end{equation}
where $V$ is a parameter for the nearest-neighbor Coulomb interaction. We define the nearest-neighbor summation, $\sum_{\langle ij \rangle}$, as a sum over each each lattice pair $\{i,j\}$ once, which corresponds to summing over all bonds in the lattice. The summation over the spin indices, $\sigma$ and $\sigma'$, ensures that we include interactions between all spin orbitals in lattice pair $\{i,j\}$.

The number of terms in the extended Hubbard model after Jordan-Wigner transformation can be reduced by applying a chemical shift to $H_{V}$ to obtain
\begin{eqnarray}
        H'_{V}  &=& V \sum_{\langle ij \rangle} \sum_{\sigma,\sigma'} \left(n_{i\sigma}-\frac{\mathbb{1}}{2}\right)\left(n_{j\sigma'}-\frac{\mathbb{1}}{2}\right)  \nonumber \\
        &= & \frac{V}{4} \sum_{\langle ij \rangle} \sum_{\sigma,\sigma'} Z_{i\sigma}Z_{j\sigma'}.
    \label{eq:HV'}
\end{eqnarray}
The difference between the original and the shifted nearest neighbor interaction term is
\begin{equation}
    H'_{V}-H_{V} =  -\frac{V}{2} \sum_{\langle ij \rangle} \sum_{\sigma,\sigma'} \Big( n_{i\sigma}+n_{j\sigma'} - \frac{\mathbb{1}}{2} \Big).
    \label{eq:H_EH_differnece}
\end{equation}
Given a $k$-regular interaction graph (all sites having $k$ nearest neighbors), e.g. periodic lattice models, then $ \sum_{\langle ij \rangle} \sum_{\sigma,\sigma'}$ runs over $2kN$ nearest-neighbor interactions, and the sum over $n_{i\sigma}$ (or $n_{j\sigma'}$) can be rewritten as
\begin{equation}
    \sum_{\langle ij \rangle} \sum_{\sigma,\sigma'} n_{i\sigma} = \frac{1}{2} \sum_{i=1}^N \sum_{j:j\sim i} \sum_{\sigma,\sigma'} n_{i\sigma} = k \sum_{i=1}^N \sum_{\sigma} n_{i\sigma},
\end{equation}
where $\sum_{j:j\sim i}$ is the sum over $j$ neighbor to $i$ and $\sum_{i=1}^N \sum_{\sigma} n_{i\sigma}$ is the total electron number operator.

In an $\eta$-electron subspace, Eq.~(\ref{eq:H_EH_differnece}) can be written in terms of the total electron number, $\eta$, and the total number of lattice sites as
\begin{equation}
    \Delta E_V =  -\frac{V}{2} \left(2k\eta-\frac{2kN}{2}\right) = Vk(\frac{N}{2}-\eta), 
    \label{eq:H_EH_differnece2}
\end{equation}
which is a constant energy shift. From now on we only consider the shifted version \eqref{eq:HV'} of the nearest-neighbor Coulomb interaction term and refer to it as $H_{V}$.

\section{Tile Trotterization}
\label{sec:tile_Trotterization}

\begin{figure*}
    \centering
    \includegraphics[width=0.75\linewidth]{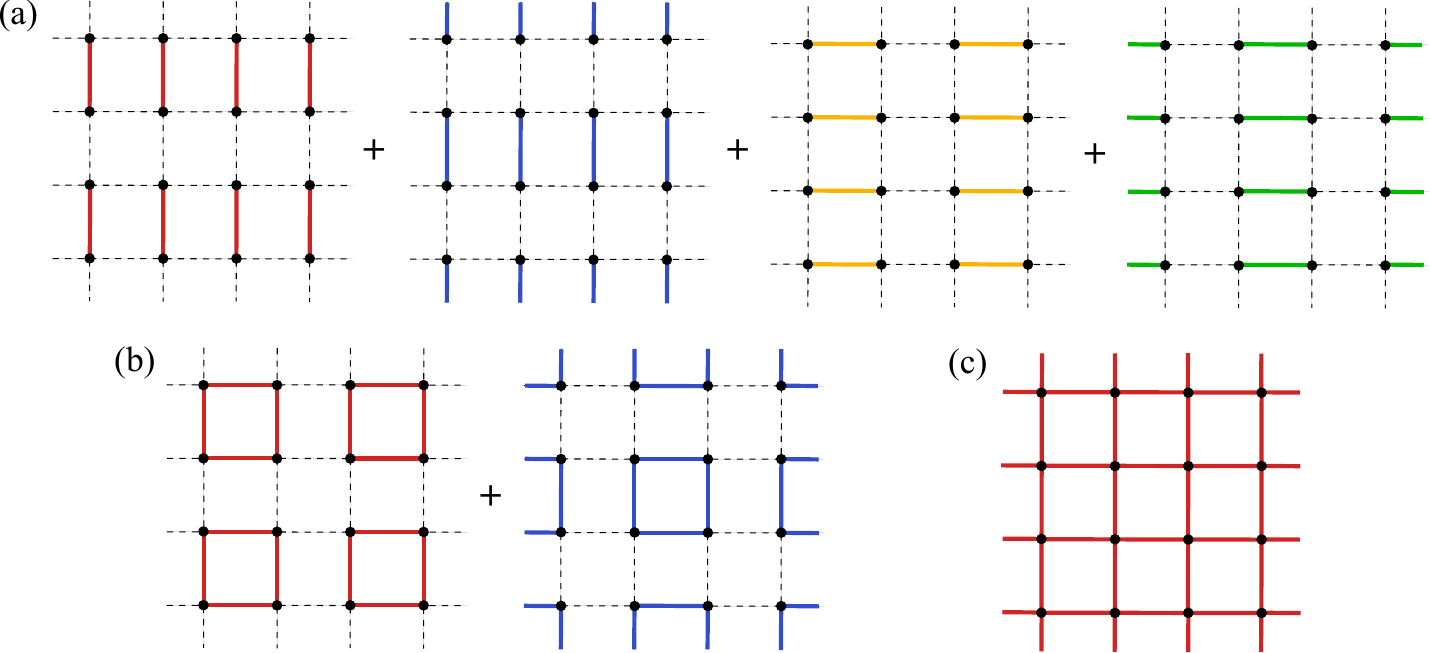}
    \caption{Visualization of three Trotterization strategies for implementing Hamiltonian simulation for the hopping Hamiltonian, $H_h$, considering a $4 \times 4$ square lattice with periodic boundaries. The hopping Hamiltonian (for each spin sector) consists of terms for each ``bond'' or ``link'' on the lattice, which do not commute if two bonds share a single lattice site. The figure presents three strategies to partition terms in $H_h$ when implementing Trotterization. In (a), $H_h$ is partitioned into four terms, each consisting of $S_1$ tiles (single bonds) for which time evolution can be implemented trivially. In contrast, (c) performs $e^{-i H_h t}$ exactly by using e.g. the fast fermionic Fourier transform (FFFT) to diagonalize $H_h$, as demonstrated in \cite{Kivlichan2020ImprovedTrotterization}. The approach of (b) shows the plaquette Trotterization method of \cite{Campbell2022EarlyModel}; here, $H_h$ is divided into just two terms, and uses $C_4$ tiles, or ``plaquettes'', which can be implemented with a particularly low non-Clifford cost. In the main text we discuss the tradeoff between these approaches.}
    \label{fig:tiling_strategies}
\end{figure*}

In this section, we show how to implement the time evolution operator of generalized Hubbard models using a special variant of second-order Trotterization. 

Trotter product formulas approximate the time evolution operator by decomposing the Hamiltonian into $m$ non-commuting terms and then applying the time evolution of these terms sequentially. Hamiltonians of the form $H=\sum^m_{j=1} H_j$ can be approximately evolved for a time $t$ using the second-order Trotter formula
\begin{equation}
   \Big\lVert e^{-iHt} - \prod_{j=1}^m e^{-iH_jt/2} \prod_{j=m}^1 e^{-iH_jt/2} \Big\rVert \leq W t^3 ,
\end{equation}
where $W$ is the Trotter error norm given by \cite{Childs2021TheoryScaling}
\begin{eqnarray}
    W &=& \frac{1}{12} \sum_{b=1}^{m-1} \Big\lVert \, \sum_{c>b,a>b} [[H_b,H_c],H_a] \, \Big\rVert \nonumber \\ &+& \frac{1}{24} \sum_{b=1}^{m-1} \Big\lVert\ \, \sum_{c>b} [[H_b,H_c],H_b] \Big\rVert ,
    \label{eq:W}
\end{eqnarray}
where $\lVert \, \cdot \, \rVert$ is the spectral norm, also known as the operator norm.

The variant of second-order Trotterization described in this paper originates from PLAQ \cite{Campbell2022EarlyModel}, which provides an efficient implementation of the time-evolution operator of square lattice Hubbard models by using a specific structure in both the Hubbard model and the square lattice. PLAQ uses ideas from the SO-FFFT method developed by Kivlichan \emph{et al.}~\cite{Kivlichan2020ImprovedTrotterization}, which reduces the Trotter error norm by diagonalizing the hopping Hamiltonian upon each application of $e^{-iH_ht}$. This allows for a simple Trotter decomposition into two non-commuting groups consisting of hopping terms and Coulomb terms. 

A similar Hamiltonian decomposition is used in PLAQ, which also separates the Coulomb and hopping terms. However, instead of simultaneously diagonalizing the hopping terms, they are further decomposed into non-commuting hopping Hamiltonian sections that each can be more easily diagonalized, at the expense of a marginally larger Trotter error norm.

To summarize the strategies available, Fig.~\ref{fig:tiling_strategies} visualizes three Trotterization approaches for the hopping Hamiltonian of a $4 \times 4$ periodic square lattice: (a) the ``naive'' approach, (b) PLAQ, and (c) SO-FFFT. PLAQ offers a compromise, increasing parallelism and reducing non-Clifford gate counts compared to (a), while also offering a more general approach than (c) (and reduced gate counts in some scenarios). As we will show, implementing each $S_1$ tile in (a) has a non-Clifford cost of $2$ arbitrary rotations, while each $C_4$ tile in (b) requires $2$ arbitrary rotations and $8$ additional T gates; since (a) uses $32$ $S_1$ tiles, while (b) uses only $8$ $C_4$ tiles, and given the large number of T gates required to synthesis an arbitrary rotation, the PLAQ method results in a significant reduction in non-Clifford gates.

Below, we introduce Tile Trotterization, generalizing the PLAQ method by allowing quantum simulation of generalized Hubbard models on arbitrary lattices, and introducing a set of tiles to minimize the non-Clifford gate count in this general setting. Although Tile Trotterization can be applied to three-dimensional lattice models, the focus in this paper is on two-dimensional models.

\subsubsection{Tile Trotterization Scheme}

\begin{figure*}
    \centering
    \includegraphics[width=0.75\linewidth]{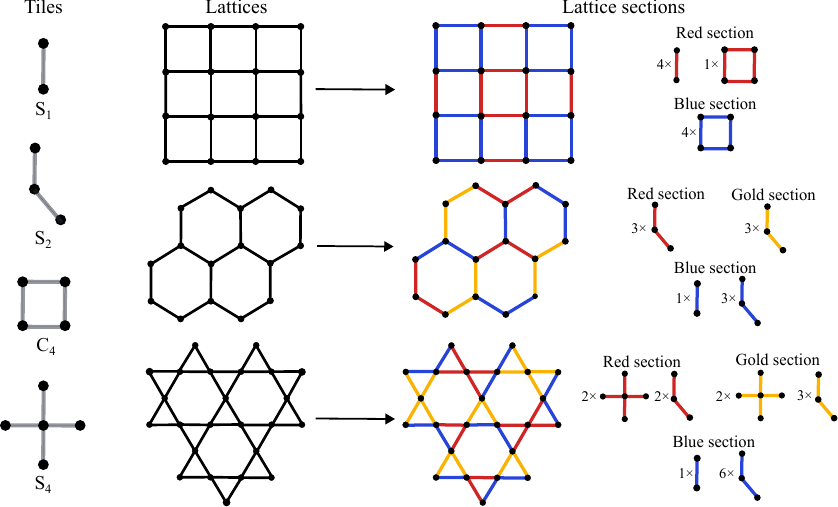}
    \caption{Illustrations of four tiles, $S_1$, $S_2$, $C_4$ and $S_4$, and how these tiles can be used to cover three examples of lattice fragments: square, hexagonal and Kagome, in order to create lattice sections. The lattice sections are indicated by different colors: red, blue and gold. On the right, we note how many tiles of the different types are used to cover each lattice section.}
    \label{fig:lattice_tiles_sections}
\end{figure*}

Given a generalized Hubbard model Hamiltonian of the form $H=H_h + H_C$, defined in Eq.~(\ref{eq:Hubbard-type-model}), the time evolution operator $e^{-iHt}$ is implemented by separating the hopping Hamiltonian into $S$ non-commuting hopping Hamiltonian sections
\begin{eqnarray}
    H_h = \sum_{s=1}^{S} H_h^s.
\end{eqnarray}
Each section, $H_h^s$, consists of $N_s$ tile Hamiltonians in each spin sector that all commute within the same section 
\begin{eqnarray}
    H_h^s = \sum_{n=1}^{N_s} \sum_\sigma H_{\sigma}^{\mathrm{tile},sn},
\end{eqnarray}
where the hopping Hamiltonian of the $n$'th tile in section $s$ is defined as
\begin{equation}
    H_{\sigma}^{\mathrm{tile},sn} = -\tau \sum_{ij} R_{ij}^{\mathrm{tile},sn} a_{i\sigma}^\dagger a_{j\sigma},
    \label{eq:H{^tile,n}}
\end{equation}
and $R^{\mathrm{tile},sn}$ is the adjacency matrix of the corresponding tile Hamiltonian. The tile Hamiltonians, Eq.~(\ref{eq:H{^tile,n}}), consist of hopping terms between neighboring lattice sites and can be represented as interaction graphs between a set of neighboring lattice points. The decomposition of the hopping Hamiltonian into sections and tiles corresponds to covering the entire lattice by tiles of different colors, where the colors are used as labels for the hopping Hamiltonian sections. These concepts are illustrated in Fig.~\ref{fig:lattice_tiles_sections}, which shows examples of tiles, lattices, and a possible way to divide these lattices into sections using tiles. The tiles have to cover the entire lattice, and are distributed such that no two tiles within a section touch the same lattice site, ensuring commutativity between all tiles within a section. Note that it is possible to mix different types of tiles to cover a lattice section.

The commutativity of the tile Hamiltonians within a section means that Hamiltonian time evolution of each section can be implemented without Trotter approximation as
\begin{eqnarray}
    e^{-iH_h^st} = \prod_n^{N_s} \prod_{\sigma} e^{-iH_{\sigma}^{\mathrm{tile},sn}t},
\end{eqnarray}
such that the cost of $e^{-iH_h^st}$ can be obtained by adding up the cost of implementing each tile within that section, e.g. the red hopping section of the square lattice in Fig.~\ref{fig:lattice_tiles_sections} requires 8 applications of $e^{-iH^{\mathrm{S_1}}_\sigma t}$ and 2 applications of $e^{-iH^{\mathrm{C_4}}_\sigma t}$, accounting for the two spin sectors.

Time evolution of a section, $e^{-iH_h^s t}$, can be implemented exactly and efficiently if time evolution of the tile Hamiltonians can also be implemented exactly and efficiently. For general tiles, a number of rotations are needed to diagonalize the tile Hamiltonian and implement time evolution, which for fault-tolerant computation must each be implemented with a high synthesis cost. We show that the implementation cost can be significantly reduced for a certain set of complete-bipartite graph tiles, which include $S_1$, $S_2$ and $S_4$ (star graphs) and $C_4$ (a circle graph) that are illustrated in Fig.~\ref{fig:lattice_tiles_sections}. The non-Clifford cost of implementing time evolution of these tiles is summarized in Table \ref{tab:Tile_non_Clifford}. Below, we describe general features of this set of tiles and in Appendix~\ref{APP:tile_trotterization} we provide a detailed costing of $e^{-iH_{\sigma}^{\mathrm{tile}}t}$. Note, we also include costing of the $S_3$ tile in Appendix~\ref{APP:tile_trotterization}, demonstrating concretely why certain other tiles have higher implementation cost.

\begin{table}
\begin{tabular}{lll}
\hline
\hline
Tile \;\;\;\;\;\; & Arbitrary rotations \;\;\; & Additional T gates \\
\hline
$S_1$ & 2 & 0 \\
$S_2$ & 2 & 4  \\
$C_4$ & 2 & 8 \\
$S_4$ & 2 & 12 \\
\hline
\hline
\end{tabular}
\caption{The non-Clifford cost of implementing $e^{-iH_{\sigma}^{\mathrm{tile}}t}$ for tiles $S_1$, $S_2$, $C_4$ and $S_4$. The number of arbitrary rotations is equal to the number of non-zero eigenvalues of the adjacency matrix of a tile. Additional gates are required to diagonalize the tile Hamiltonian; for the four tiles considered here, this can be achieved with a non-Clifford cost of just a few T gates.}
\label{tab:Tile_non_Clifford}
\end{table}
More generally, we show in Appendix~\ref{Sec:complete_bipartite_tile} that time evolution of any tile Hamiltonian with adjacency matrix corresponding to a complete bipartite graph of type $K_{2^a,2^b}$, where $a$ and $b$ are integers, can be implemented with 2 arbitrary rotations and $2^{a+2}+2^{b+2}-8$ T gates (recovering the costs given in Table \ref{tab:Tile_non_Clifford}). Notice that $K_{1,2^k}=S_{2^k}$ and $K_{2,2} = C_4$. When tiling a hopping Hamiltonian on a given lattice, we require that all hopping terms are covered by tiles. Therefore, a relevant measure when considering the efficiency of tiling is non-Clifford cost per bond for a given tile. A $K_{2^a, 2^b}$ graph covers $2^{a+b}$ bonds, resulting in an arbitrary rotation cost per bond of $2/2^{a+b}$ and a T gate cost per bond of $(2^{a+2}+2^{b+2}-8)/2^{a+b}$, meaning that larger tiles could reduce the required number of arbitrary rotations. However, these additional tiles are less relevant for two-dimensional lattices with nearest-neighbor hopping, and therefore we focus on $S_1$, $S_2$, $S_4$ and $C_4$ tiles as described above.

A Tile Trotterization step of generalized Hubbard model Hamiltonians can be implemented as
\begin{eqnarray}
        & & \! \! \! \! \! \Big\lVert e^{-iHt} \!-\! e^{-iH_C \frac{t}{2}}  \prod_{s=1}^S e^{-iH_h^s \frac{t}{2}}\prod_{s=S}^1 e^{-iH_h^s \frac{t}{2}} e^{-iH_C \frac{t}{2}} \Big\rVert \leq W_{\mathrm{tile}} t^3 . \nonumber \\ 
        & & \label{eq:Tile_Trotter_step}
\end{eqnarray}
A Trotter step includes applications of $e^{-iH_Ct}$, and since all terms of $H_C$ commute, the time evolution of each term can be implemented without Trotter error using two CNOT gates and one arbitrary $Z$-axis rotation. The Tile Trotterization error norm, $W_{\mathrm{tile}}$, can be bounded using
\begin{eqnarray}
    & & W_{\mathrm{tile}} \leq W_{\mathrm{SO2}} + W_{\mathrm{h}}, \label{eq:Tile_Trotter_error_norm}
\end{eqnarray}
as shown in Appendix~\ref{APP:tile_Trotter_decomposition}, where $W_{\mathrm{SO2}}$ arises from the decomposition into Coulomb and hopping terms
\begin{eqnarray}
    & & \! \! \! \! \! \! \! \! \! \! \! \! \! \Big\lVert e^{-i(H_h+H_C)t} - e^{-iH_C \frac{t}{2}}e^{-i H_h t}e^{-iH_C \frac{t}{2}} \Big\rVert \leq W_{\mathrm{SO2}} t^3,
    \label{eq:Trotter_decomp_WSO2}
\end{eqnarray}
which is evaluated using the double commutator formula given by Eq.~(\ref{eq:W}). The terminology ``$\mathrm{SO2}$'' is taken from Ref.~\cite{Campbell2022EarlyModel}, and indicates the split-operator decomposition of Coulomb and hopping terms with the ordering employed in Eq.~(\ref{eq:Trotter_decomp_WSO2}). We evaluate $W_{\mathrm{SO2}}$ as
\begin{eqnarray}
    & & \! \! \! \!  \! \! \! \! \! \! \! \! \! \! \! \! W_{\mathrm{SO2}} = \frac{1}{12} \Big\lVert [[H_C,H_h],H_h] \Big\rVert + \frac{1}{24} \Big\lVert [[H_C,H_h],H_C] \Big\rVert. 
    \label{eq:WSO2}
\end{eqnarray}
This expression is independent of the tiling and captures the total Trotter error norm if the entire hopping Hamiltonian is simultaneously diagonalized. The additional Trotter error, $W_h$, depends on the number of sections and the distribution of tiles within each section
\begin{eqnarray}
    \Big\lVert e^{-iH_ht} - \prod_{s=1}^S e^{-iH_h^st/2} \prod_{s=S}^1 e^{-iH_h^st/2} \Big\rVert \leq W_{h} t^3,
\end{eqnarray}
and can be obtained from Eq.~(\ref{eq:W}) by defining hopping Hamiltonian sections as $H_1 = H_h^1$, $H_2 = H_h^2$, ... , $H_S = H_h^S$, where the subscripts 1, 2 and $S$ denote the order in which the terms are applied in the Trotter decomposition. In Appendix~\ref{APP:W_h}, we explicitly write out Eq.~(\ref{eq:W}) for cases where lattices can be split into three sections and show how to compute $W_h$ efficiently. This additional hopping Hamiltonian decomposition adds low cost in the strongly interacting regimes. We highlight this point in Table~\ref{tab:W_h_W_SO2_comparison} by comparing $W_{\mathrm{SO2}}$ and $W_h$ for hexagonal lattice Hubbard and extended Hubbard models for different values of $U$ and $V$. The fact that $W_h \ll W_{\mathrm{SO2}}$ in the strongly-interacting regime is one of the motivations for Tile Trotterization over the SO-FFFT method. However, note from Table~\ref{tab:W_h_W_SO2_comparison} that in the weakly-correlated regime, it can be the case that $W_h / W_{\mathrm{SO2}}$ is close to $1$.

\section{\label{sec:UT_examples} Application of Tile Trotterization}

Here, we demonstrate how to perform costing of Tile Trotterization in practice. As an example, we consider the more challenging application of the periodic hexagonal lattice extended Hubbard model in the main text. In Appendix~\ref{APP:Tile_Trotter_applications}, we give further examples for the Hubbard model on arbitrary hexagonal lattice fragments (models without periodic boundary conditions), and the Hubbard model on a periodic hexagonal lattice. The two periodic models will then be considered in the QPE resource analysis in Section~\ref{sec:Phase_estimation_costings_examples}.

\subsection{\label{sec:UT_on_EXTHub_gates} Tile Trotterization of the periodic hexagonal lattice extended Hubbard model: gate counts}

We consider the application of Tile Trotterization for simulating the extended Hubbard model on the periodic hexagonal lattice. We use the periodic hexagonal lattice model described in Appendix~\ref{APP:hexagonallatticedivision} which is described by the parameters $L_x$ and $L_y$ and contains $N=2 L_x L_y$ lattice sites.

To apply Tile Trotterization, we cover the periodic hexagonal lattice with $S_2$ tiles to divide the lattice into three sections: blue (b), red (r) and gold (g), as shown in Fig.~\ref{Fig:hex_lattices} in Appendix~\ref{APP:hexagonallatticedivision}. This corresponds to decomposing the hopping Hamiltonian into three sections: $H_h = H_h^{\mathrm{b}} + H_h^{\mathrm{r}} + H_h^{\mathrm{g}}$. The number of $S_2$ tiles in each section is $N_{\mathrm{b}}=N_{\mathrm{r}}=N_{\mathrm{g}}=N/4$ for all periodic hexagonal lattice models considered in this paper, which have parameters $L_x=L_y=L$.

A single Trotter step of the extended Hubbard model, defined in Eq.~(\ref{eq:extended_Hubbard_model}), on the periodic hexagonal lattice is implemented as
\begin{eqnarray}
    e^{-iH_{EH}t} &\approx & e^{-i(H_I+H_V)\frac{t}{2}} e^{-iH_h^b \frac{t}{2}} e^{-iH_h^r \frac{t}{2}} e^{-iH_h^g t} \nonumber \\ & \times & e^{-iH_h^r \frac{t}{2}} e^{-iH_h^b \frac{t}{2}}e^{-i(H_I+H_V)\frac{t}{2}} ,
\end{eqnarray}
with $H_{EH} = H_h+H_I+H_V$, and where we choose to implement the sections in the order $H_1=H_h^b$, $H_2=H_h^r$ and $H_3=H_h^g$. Performing $r$ repetitions of this Trotter step leads to
\begin{eqnarray}
    \Big( e^{-iH_{EH}t} \Big)^r &\approx&  e^{-i(H_I+H_V)\frac{t}{2}} \Big( e^{-iH_h^b \frac{t}{2}}e^{-iH_h^r \frac{t}{2}} e^{-iH_h^g t}e^{-iH_h^r \frac{t}{2}} \nonumber \\ &\times& e^{-iH_h^b \frac{t}{2}}e^{-i(H_I+H_V)t} \Big)^r e^{i(H_I+H_V)\frac{t}{2}} ,
     \label{eq:Extended_hubbard_trotterstep}
\end{eqnarray}
and therefore a single Trotter step for large $r$ requires two applications of $e^{-iH_h^b t/2}$, two applications of $e^{-iH_h^r t/2}$, one application of $e^{-iH_h^g t}$ and one application of $e^{-i(H_I+H_V)t}$.

Under the Jordan-Wigner transformation, the system is represented by $2N$ qubits and $e^{-i(H_I+H_V)t}$ contains $N$ terms from $H_I$ and $6N$ terms from $H_V$, which can be implemented with $7$ layers of $N$ arbitrary $Z$-axis rotations, where the angle of all gates within a given layer are the same, and $14$ layers of $N$ CNOT gates.

The cost of implementing the time evolution of the hopping terms is obtained by counting the number of applications of $e^{-iH_\sigma^{S_2}t}$ in each section. The time evolution of the hopping Hamiltonian is decomposed into five applications of time evolution of hopping Hamiltonian sections. Each section contains $N/4$ $S_2$ tiles, so that we need $5 \times\frac{N}{4}\times 2$ applications of $e^{-iH_\sigma^{S_2}t}$, accounting for the two spin sectors. The time evolution of each $S_2$ Hamiltonian can be implemented with 2 arbitrary Z-axis rotations (of the same angle) and 4 T gates, as shown in Appendix~\ref{APP:tile_trotterization}. Therefore, the time evolution of the hopping term can be implemented with $5$ layers of $N$ arbitrary rotations of the same angle in each layer. The total non-Clifford gate cost per Trotter step is
\begin{eqnarray}
    & & N_R = 7N + 2\times\frac{10}{4}N = 12N, \label{eq:N_R_periodic_EH}\\
    & & N_T = 4\times \frac{10}{4}N = 10N, \label{eq:N_T_periodic_EH}
\end{eqnarray}
where $N_R$ is the number of arbitrary rotations and $N_T$ is the number of T gates. 

Arbitrary rotations are expensive to perform on fault-tolerant quantum computers using the surface code, most commonly being decomposed into a sequence of T and Clifford gates. Hamming weight phasing (HWP) can be used to reduce the number of arbitrary rotations by trading them for additional ancilla qubits and Toffoli gates. In particular, HWP is applicable when applying a layer of repeated rotation gates of the same angle. In this case, we can introduce ancilla qubits to calculate the Hamming weight of the logical state, and apply a smaller number of rotations to this weight \cite{Gidney2017HalvingAddition, Kivlichan2020ImprovedTrotterization}. See also Appendix~\ref{sec:PHUB} for more discussion on using HWP for Tile Trotterization.

The time evolution of the extended Hubbard model contains $12$ layers of $N$ arbitrary rotations, where the angle of all gates within a given layer are the same. Using HWP, we choose to implement $m$ arbitrary rotations simultaneously using $m-1$ ancilla qubits, with the requirement that $N$ is an integer multiple of $m$, leading to a worst case total gate cost of \cite{Campbell2022EarlyModel}
\begin{eqnarray}
    & &N_R = \frac{12N}{m} \lfloor \log_2(m)+1 \rfloor \label{eq:num_rot_hwp}, \\
    & &N_T = 10N + 4\times \frac{12N}{m}(m-1),
\end{eqnarray}
where each Toffoli gate has been converted into 4 T gates. The total number of qubits required for this Tile Trotterization implementation with HWP is $2N+(m-1)$.

\subsection{\label{sec:UT_on_EXTHub_norms} Tile Trotterization: error bounds}

\begin{table}
\begin{tabular}{llllll}
\hline
\hline
& \multicolumn{2}{c}{Model parameters} & & & \\
\cline{2-3}
Models & $U$ \; \; \;\; & $V$ \;\; & $W_{\mathrm{SO2}}$ \;\;\; & $W_h / W_{\mathrm{SO2}}$  \\
\hline
Hubbard & 1 & - & 114.0 & 0.95 \\
& 2 & - & 244.8 & 0.44 \\
& 4 & - & 556.8 & 0.20 \\
& 8 & - & 1382 & 0.08 \\
\hline
Extended Hubbard \; \;  & 1 & 0.5 & 158.9 & 0.68  \\
& 2 & 1 & 369.0 & 0.29 \\
& 4 & 2 & 942.9 & 0.12 \\
& 8 & 4 & 2705 & 0.04 \\
\hline
\hline
\end{tabular}
\caption{Comparison of upper bound estimates of $W_{\mathrm{SO2}}$ and $W_h$ for different electron-electron interaction parameters of periodic hexagonal Hubbard and extended Hubbard models ($U/V = 2$). We take $\tau = 1$ and lattice parameters $L_x = L_y = 8$ ($N = 128$) resulting in $W_h \leq 108.6$ for each of the examples considered above (using the $S_2$-tiling shown in Appendix~\ref{APP:hexagonallatticedivision}).}
\label{tab:W_h_W_SO2_comparison}
\end{table}

The Tile Trotterization error norm, $W_{\mathrm{tile}}$, is obtained by summing $W_{\mathrm{SO2}}$ and $W_h$, where $W_{\mathrm{SO2}}$ can be evaluated from the commutator bound expression in Eq.~(\ref{eq:WSO2}). In this paper we introduce new strategies for evaluating commutator bounds for the Hubbard model and the extended Hubbard model, resulting in three lemmas in Appendix~\ref{APP:COMMUTATORBOUNDS}. \textit{Lemma} \ref{Lemma:Hubbard_Ihh} can be used to evaluate $\lVert[H_I,H_h],H_h]\rVert$ for any on-site interaction Hubbard model. Our commutator bound improves upon previous state-of-the-art bounds and for the case of the periodic hexagonal lattice Hubbard model, we improve $\lVert[H_I,H_h],H_h]\rVert$ by $31.5 \%$ compared to the bound in Ref~\cite{Campbell2022EarlyModel}. Combining Lemma 2 from Ref~\cite{Campbell2022EarlyModel} with our improved commutator bound, we obtain the following final expression for $W_{\mathrm{SO2}}$ for the periodic hexagonal lattice Hubbard model,
\begin{equation}
    W_{\mathrm{SO2}} \leq \frac{U^2 \tau}{24} \lVert R \rVert_1 + \frac{9.9 U \tau^2 }{12}N ,
    \label{eq:W_SO2_Hubbard}
\end{equation}
where $\lVert \, \cdot \, \rVert_1$ is the Schatten one-norm and $R$ is the adjacency matrix of the lattice. We compute $\lVert R \rVert_1$ numerically and use these results to bound $\lVert R \rVert_1 \leq 1.59 N$ for periodic hexagonal lattices with size $4 \leq L \leq 24$.

Our second and third lemmas apply for extended Hubbard models on periodic lattices where all lattice sites have $k$ nearest neighbors (e.g. periodic hexagonal, square, Kagome and triangular lattices). \textit{Lemma} \ref{Lemma:Extended_Hubbard_Chh} and \textit{Lemma} \ref{Lemma:extended_Hubbard_ChC} give expressions that can be used to evaluate $\lVert[H_C,H_h],H_h]\rVert$ and $\lVert[H_C,H_h],H_C]\rVert$, with $H_C = H_I + H_V$. We summarize the strategies used to obtain these bounds below and give expressions for obtaining $W_{\mathrm{SO2}}$ for the periodic hexagonal extended Hubbard model. For detailed derivations, we refer to Appendix~\ref{APP:COMMUTATORBOUNDS}.

Our strategy for calculating Trotter error bounds is to partition or divide the nested commutators into a sum of terms acting on a smaller but still non-trivial number of spin orbitals, for which we can evaluate the spectral norm either exactly or with very high precision using numerical techniques. We derive expression that allow for constructing these partitions and which clearly define the resulting subspaces. This is done by evaluating the nested commutators for a subset of terms in the Hamiltonian using commutator and anti-commutator rules provided in Appendix~\ref{sec:APP_COMM_RULES}. The derived lemmas give a natural way to adjust the partitioning of commutator terms by first partitioning the lattice appropriately, for which there is significant flexibility. To obtain numerical bounds in practice, we partition the nested commutators into exactly equivalent terms (by lattice symmetry), evaluate the spectral norm of the resulting operator numerically using DMRG, and sum over partitions to obtain the total commutator bound (for details, see Appendix~\ref{APP:COMMUTATORBOUNDS}).

Combining the numerical results obtained using \textit{Lemma} \ref{Lemma:Extended_Hubbard_Chh} and \textit{Lemma} \ref{Lemma:extended_Hubbard_ChC}, we obtain the following final expression for $W_{\mathrm{SO2}}$ for the periodic hexagonal lattice extended Hubbard model with $U/V=2$ and $L \geq 4$,
\begin{equation}
    W_{\mathrm{SO2}} \leq \frac{1}{12} \Big( 2.4 U^2 \tau + 12.5 U \tau^2 \Big) N.
    \label{eq:W_SO2_extHubbard}
\end{equation}
We must also evaluate the additional hopping Hamiltonian error norm, $W_h$. We calculate this as derived in Appendix~\ref{APP:W_h}, where Eq.~(\ref{eq:W_h_written_out_for_RBG}) gives a formula for efficiently evaluating $W_h$ for general tilings of lattices with three sections. Fig.~\ref{fig:W_h_scaling} (in Appendix~\ref{APP:W_h}) plots $W_h/\tau^3$ as a function of system size using the periodic hexagonal lattice $S_2$-tiling presented in Appendix~\ref{APP:hexagonallatticedivision}. This yields a linear scaling of $W_h$ with the hexagonal lattice system size. Numerically, for even $L$ such that $4 \leq L \leq 24$, we find
\begin{equation}
    W_h \leq 0.8532 \tau^3 N.
\end{equation}
In Table~\ref{tab:W_h_W_SO2_comparison} we compare $W_h$ and $W_{\mathrm{SO2}}$ for the Hubbard and extended Hubbard model, for a periodic hexagonal lattice of size $L=8$ ($N=128$) for different values of $U$ and $V$. These results show that the Trotter decomposition of the hopping Hamiltonian contributes low additional Trotter error for systems with $U \geq 4$. We also see that $W_h$ comprises a smaller fraction of $W_{\mathrm{tile}}$ for the extended Hubbard model compared to the standard Hubbard model, due to the larger value of $W_{\mathrm{SO2}}$ obtained for the former model.

\section{\label{sec:Qubitization} Qubitization for the hexagonal Hubbard model}

The discussion so far has focused on quantum simulation using Tile Trotterization. In order to assess this approach, we will next consider qubitization, which is a state-of-the-art approach for performing quantum simulation, therefore providing an important comparison point \cite{berry_2018, poulin_2018, Babbush2018EncodingComplexity}. In Appendix~\ref{APP:qubitization}, we provide quantum circuits for the qubitized quantum walk operator for the periodic hexagonal Hubbard model, and include detailed resource estimates. Our construction of the qubitization circuit follows that developed in Ref.~\cite{Babbush2018EncodingComplexity}, but we introduce improvements that further reduce resource estimates, and perform costing of all circuit elements in detail. For brevity, in this section we briefly summarize the key costs to perform qubitized QPE using these circuits, which will be used to perform resource estimation of QPE in Section~\ref{sec:Phase_estimation}.

The quantum walk operator used in qubitization consists of SELECT, PREPARE and reflection operations. For the Hubbard model, the asymptotically dominant cost of performing qubitized QPE comes from implementing SELECT (controlled on an ancilla). For an $L_x \times L_y$ hexagonal lattice we obtain a T gate cost of $C_S = 40 L_x L_y - 4$, or, expressed in terms of the number of lattice points, $N$,
\begin{eqnarray}
    C_S = 20N - 4.
    \label{eq:CS}
\end{eqnarray}
Note that the $\mathcal{O}(N)$ term is identical to that obtained in Ref.~\cite{Babbush2018EncodingComplexity} for the square lattice, but our circuit removes an additional $\mathcal{O}(\log(N))$ contribution.

The cost of PREPARE is more involved, and we summarize the cost of each circuit element in Table~\ref{tab:walk_circuit_cost} in Appendix~\ref{APP:qubitization}. Summing the contribution from each of these elements, and taking the case where $L_x = L_y = L$, the total T gate cost of PREPARE is
\begin{equation}
    C_P = 46 \lceil \log_2 L \rceil + 4 \Theta + 4 \Gamma - 24 \eta_L - 16, \label{eq:CP}
\end{equation}
where $\eta_L$ is the largest power of $2$ that is a factor of $L$, $\Theta$ is the number of T gates per rotation in the UNIFORM state preparation gadgets, and $\Gamma$ is the number of T gates per each other rotation in PREPARE. Therefore, the cost of PREPARE is only logarithmic in the lattice dimension, $L$. For our resource estimates, we choose $\Theta = 10$ and $\Gamma = 40$. The controlled reflection operator adds an additional T gate cost of
\begin{eqnarray}
    C_R = 32 \lceil \log_2L \rceil + 77,
    \label{eq:CR}
\end{eqnarray}
using the scheme of Ref.~\cite{Khattar2024RiseCircuits} to perform the reflection with a single ancilla qubit.

The controlled walk operator also requires additional flag and ancilla qubits. Once again taking the case of $L_x = L_y = L$, the total number of qubits to implement the circuits in Appendix~\ref{APP:qubitization} is
\begin{eqnarray}
    N_{\mathrm{qubits}} = 2N + 6 \lceil \log_2 L \rceil + 15. \label{eq:qubits_walk_operator}
\end{eqnarray}

\section{\label{sec:Phase_estimation} Quantum phase estimation}

In this section, we provide resource estimates for QPE based on Tile Trotterization and qubitization. We begin by describing the QPE schemes used for each method.

\subsection{\label{sec:tile_trot_PE_costing} Trotterized QPE}

We follow the quantum phase estimation costing procedure for Trotterization methods described by Kivlichan \emph{et al.} in Ref.~\cite{Kivlichan2020ImprovedTrotterization}. This phase estimation scheme relies on adaptive phase estimation techniques \cite{Berry2009HowMeasurements, Higgins2007Entanglement-freeEstimation} which use a single control qubit. We define the maximum allowed error in the energy estimate as $\epsilon$, and following Ref.~\cite{Campbell2022EarlyModel}, assign $x \epsilon$ of the total error to rotation synthesis error and $(1-x)\epsilon$ to Trotter and phase estimation error. This allows us to determine the total number of Trotter steps required for the phase estimation procedure as \cite{Kivlichan2020ImprovedTrotterization, Campbell2022EarlyModel}
\begin{equation}
    N_{\mathrm{PE}} = 6.203 \frac{\sqrt{W}}{(1-x)^{3/2}\epsilon^{3/2}},
\end{equation}
where $W$ is the Trotter error norm.

The T gate cost per Trotter step from synthesis of rotation gates, $N_{RT}$, using repeat-until-success synthesis \cite{Bocharov2015EfficientCircuits}, can be evaluated as \cite{Campbell2022EarlyModel}
\begin{equation}
    N_{RT} = N_R \Big( 1.15 \log_2\Big(\frac{N_R\sqrt{3W}}{x \sqrt{1-x} \epsilon^{3/2}}\Big) + 9.2 \Big),
\end{equation}
where $N_R$ is the number of arbitrary rotations per Trotter step. The total T gate cost for performing Trotterized QPE is given by
\begin{equation}
    N_{\mathrm{T,Trotter}} = N_{\mathrm{PE}} (N_{RT}+N_T),
\end{equation}
where $N_T$ is the number of T gates per Trotter step that do not come from the synthesis of arbitrary rotations.

To assess the T gate complexity, note that the number of Trotter steps scales as $N_{\mathrm{PE}} = \mathcal{O}(N^{1/2}\epsilon^{-3/2})$ for the Hubbard model and extended Hubbard model, and the T gate cost per step \emph{aside from rotation synthesis} is $N_T = \mathcal{O}(N)$. If HWP is applied, then $N_R = \mathcal{O}(\log(N))$ (from Eq.~(\ref{eq:num_rot_hwp})) in which case $N_{RT}$ does not contribute asymptotically, resulting in a T gate complexity of
\begin{equation}
    \mathcal{O}(N^{3/2} \epsilon^{-3/2}). \label{eq:Tile_Trotter_scaling_HWP} 
\end{equation}
If HWP is not applied then $N_R = \mathcal{O}(N)$ (from Eq.~(\ref{eq:N_R_periodic_EH})), and the overall T complexity picks up a logarithmic factor
\begin{equation}
    \mathcal{O}(N^{3/2} \epsilon^{-3/2}\log(N\epsilon^{-1})). \label{eq:Tile_Trotter_scaling} 
\end{equation}
In practice, the precision of rotation synthesis can be improved with low additional cost, and the number of T gates per arbitrary rotation is often set as a constant, meaning the logarithmic factor can be disregarded. The scaling with and without HWP reported here applies for both the extended Hubbard model (Sec.~\ref{sec:UT_examples}) and the Hubbard model systems (\ref{sec:TILE_TROTTER_fragments} and \ref{sec:PHUB}). The Hubbard model systems however have smaller prefactors coming from the lower value of $W_{\mathrm{SO2}}$ and fewer arbitrary rotation layers. Note that the scaling of QPE with Tile Trotterization and HWP matches that for the square lattice with plaquette Trotterization, see Ref.~\cite{Campbell2022EarlyModel}, Eq.~(F10).

The phase estimation procedure used for Tile Trotterization requires two qubits in addition to the $2N$ system qubits: one for adaptive phase estimation and one for repeat-until-success synthesis \cite{Kivlichan2020ImprovedTrotterization, Campbell2022EarlyModel}. Through numerical tests of the overall T gate cost, we find that optimal values of $x$ are approximately $x=0.03$ without applying HWP and $x=0.01$ with HWP.

\subsection{\label{sec:Qubitized_QPE} Qubitized QPE}

For qubitized QPE, we follow the phase estimation scheme described in Ref.~\cite{Lee2021EvenHypercontraction}, which uses $N_{\mathcal{W}}$ repetitions of the qubitized quantum walk operator, $\mathcal{W}$, which consists of the operators controlled SELECT (S), PREPARE (P), PREPARE$^\dagger$ (P$^\dagger$) and controlled reflection (R). The number of repetitions of the qubitized quantum walk operator is given by
\begin{equation}
    N_{\mathcal{W}} = \Big\lceil \frac{\pi \lambda}{2 \epsilon} \Big\rceil,
\end{equation}
where $\lambda$ is the L1 norm of the Hamiltonian and $\epsilon$ is the allowed error in phase estimation. For the periodic hexagonal lattice Hubbard model, the L1 norm is
\begin{equation}
    \lambda = \Big( 3\tau + \frac{U}{4} \Big) N.
    \label{eq:L1norm}
\end{equation} 
The total qubitized QPE T gate cost is
\begin{equation}
    N_{\mathrm{T,N_{\mathcal{W}}}} = \Big\lceil \frac{\pi \lambda}{2 \epsilon} \Big\rceil \, (C_{S} + C_{P} + C_{P^\dagger} + C_R),
    \label{eq:qubitization_T_cost}
\end{equation}
where $C_Q$ represents the T gate cost of operation $Q$. Note that our implementation has the same cost for P and P$^\dagger$. The T gate complexity of qubitized QPE on the periodic hexagonal lattice Hubbard model scales as 
\begin{equation}
    \mathcal{O}(N^2\epsilon^{-1}). \label{eq:qubitization_scaling}
\end{equation}
This matches the asymptotic scaling for the square lattice obtained in Ref.~\cite{Babbush2018EncodingComplexity}, Eq.~61. However, our analysis also considers the non-Clifford cost of all circuit elements.

\begin{figure*}
    \centering
    \includegraphics[width=0.8\linewidth]{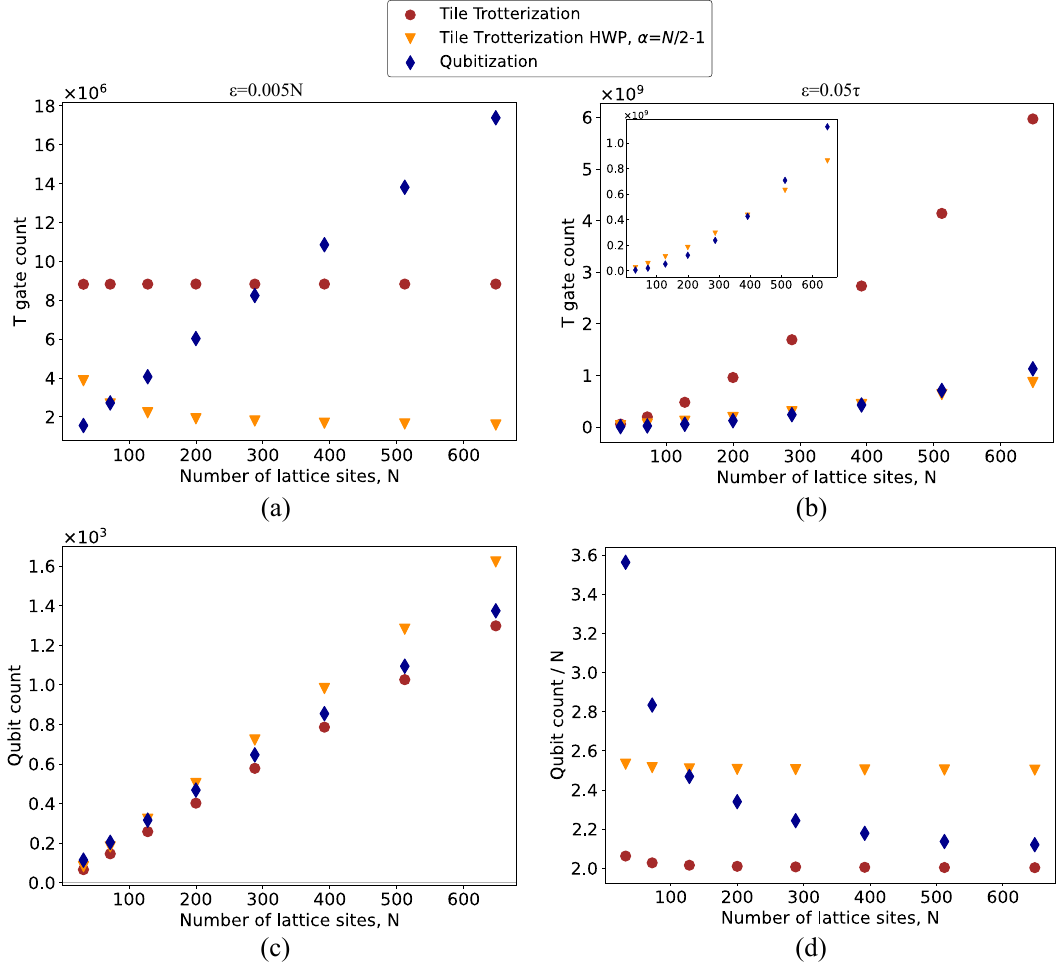}
    \caption{Resource estimates for QPE performed on a periodic hexagonal lattice Hubbard model with $L_x=L_y=L$, using Tile Trotterization (see Appendix \ref{sec:PHUB}), Tile Trotterization HWP (Hamming weight phasing; with $\alpha = N/2-1$ ancilla qubits, see Appendix \ref{sec:PHUB}), and qubitization (see Sec.~\ref{sec:Qubitization}). (a) Total T gate count (including T gates from arbitrary rotation synthesis) as a function of the number of lattice sites, $N$, with $\epsilon=0.005N$. (b)  Total T gate count (including T gates from arbitrary rotation synthesis) as a function of the number of lattice sites, $N$, with $\epsilon=0.05\tau$. The inset shows the total T gate count for Tile Trotterization HWP and qubitization. Note that the total T gate counts of Tile Trotterization based QPE are upper bounds. (c) The total number of qubits used for each method. (d) The total number of qubits used for each method, divided by $N$.
    }
    \label{fig:QUBITIZATION_VS_TROTTER_plots}
\end{figure*}

Note that this phase estimation scheme uses $\alpha_{PE}=2 \lceil \log_2(N_{\mathcal{W}}+1) \rceil-1$ ancilla qubits for the control register \cite{Lee2021EvenHypercontraction, Georges2024QuantumSet}, in addition to the qubits quoted in Eq.~(\ref{eq:qubits_walk_operator}). Therefore, the number of qubits depend on the Hubbard model parameters, $U$ and $\tau$, as well as $\epsilon$. The unary iterator over the walk operator also requires $(4 N_{\mathcal{W}} - 4)$ T gates to implement, in addition to the per-walk operator T gate counts given in Eq.~(\ref{eq:qubitization_T_cost}).

\subsection{\label{sec:Phase_estimation_costings_examples} QPE resource estimates}
In our QPE resource estimates, we disregard the cost of state preparation, although this is an important and non-trivial problem \cite{Lee2023EvaluatingChemistry}. Efforts have been made to efficiently prepare correlated fermionic states of Hubbard models on quantum computers \cite{Dallaire-Demers2019Low-depthComputer, Motta2023BridgingStructure}. The state preparation problem, and its cost for Hubbard models, is further discussed in Yoshioka \emph{et al}.~\cite{Yoshioka2024HuntingProblems}. We will continue by assuming that an initial state can be prepared with a sufficient overlap with the target state.

In Table~\ref{Table:Trotter_error_norms} of Appendix~\ref{APP:Trotter_tables} we provide Trotter error norms, qubit counts, arbitrary rotation costs and T costs per Trotter step to perform the simulations considered in this section. The qubitization resource estimates are obtained from Eq.~(\ref{eq:qubitization_T_cost}), with the cost of each quantum walk operator element given by Eqs.~(\ref{eq:CS})--(\ref{eq:CR}).

We begin our QPE resource analysis by comparing the performance of Tile Trotterization (Appendix \ref{sec:PHUB}) with our qubitization based approach (Sec.~\ref{sec:Qubitization}) for simulating the periodic hexagonal lattice Hubbard model. This is done to highlight the difference in scaling and performance of the two approaches for different system sizes $N$, and for different maximum errors, $\epsilon$. We consider two different error regimes in our QPE costing analysis: relative and additive errors. Relative errors are often appropriate for condensed phase applications, for example when calculating the energy per unit cell, while absolute errors are often considered in molecular quantum chemistry, or when studying defects in solids. For a more detailed discussion of this point, we refer to Ref.~\cite{Kivlichan2020ImprovedTrotterization}. The QPE resource comparisons are shown in Fig.~\ref{fig:QUBITIZATION_VS_TROTTER_plots} using either Tile Trotterization, Tile Trotterization with HWP ($\alpha=N/2-1$ ancillas), or qubitization for hexagonal lattice Hubbard models with parameters $L_x=L_y=L$, $U=4$ and $\tau=1$.

\begin{figure*}
    \centering
    \includegraphics[width=0.81\linewidth]{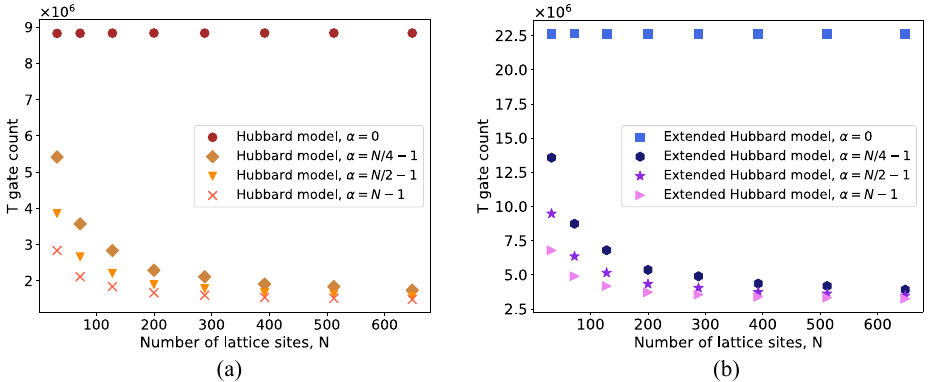}
    \caption{Upper bounds on the total T gate count (including T gates from arbitrary rotation synthesis) for performing QPE with Tile Trotterization on periodic hexagonal lattice Hubbard models with $L_x=L_y=L$ and $\epsilon=0.005N$, as a function of the number of lattice sites, $N$. We compare four different simulation approaches with $\alpha=0$, $\alpha=N/4-1$, $\alpha=N/2-1$ and $\alpha=N-1$ ancilla qubits for HWP for (a) the Hubbard model (App.~\ref{sec:PHUB}) and (b) the Extended Hubbard model (Sec.~\ref{sec:UT_examples}).}
    \label{fig:QPE_Trotter_HWP_comparison}
\end{figure*}

Fig.~\ref{fig:QUBITIZATION_VS_TROTTER_plots}(a) presents the total T gate count as a function of the number of lattice sites, $N$, and taking the relative error case with $\epsilon=0.005N$. The T gate cost of qubitized QPE is lower than for Trotterized QPE for small $N$ (and small $\epsilon$) but grows linearly with $N$ in the relative error regime, eventually becoming more expensive than both Trotterization methods. The cost of Tile Trotterization-based QPE remains constant in this error regime while the Tile Trotterization HWP simulation also asymptotically approaches a constant T gate cost. At small $N$, the total T gate cost of Tile Trotterization HWP is dominated by a $\log(N)$ term from the number of arbitrary rotations per Trotter step, $N_{RT}$, and therefore the T gate cost initially decreases as the logarithmic term is suppressed by the linearly growing $\epsilon$. Eventually, the linearly growing $N_T$ dominates the total T gate cost per Trotter step, resulting in the total T gate cost approaching a constant value.

Fig.~\ref{fig:QUBITIZATION_VS_TROTTER_plots}(b) shows the total T gate count as a function of $N$, with an absolute error of $\epsilon=0.05\tau$. The inset in the top-left corner shows the T gate cost of Tile Trotterization HWP and qubitization only, to more clearly highlight their different scaling, $\mathcal{O}(N^{3/2})$ for Tile Trotterization HWP and $\mathcal{O}(N^{2})$ for qubitization. In this error regime, we once again see that the qubitization approach performs better than Tile Trotterization HWP for the smaller systems, but Tile Trotterization HWP eventually reaches a lower T gate cost due to its better system size scaling. These results only demonstrate the total T gate cost of periodic Hubbard model simulations. We note that simulating hexagonal lattice fragments (or nanographene) with Tile Trotterization (Appendix~\ref{sec:TILE_TROTTER_fragments}) will have similar costs for models of same size.

Fig.~\ref{fig:QUBITIZATION_VS_TROTTER_plots}(c) presents the number of qubits used for the three simulation methods, and Fig.~\ref{fig:QUBITIZATION_VS_TROTTER_plots}(d) shows the qubit count divided by $N$ to highlight the differences in qubit counts. We have chosen $U=4$, $\tau=1$ and $\epsilon=0.05\tau$ for the qubitization-based qubit count; the Trotter qubit counts are independent of these parameters. These figures show that the qubit counts for the three approaches are comparable and grow similarly with the system size. However, for small lattice sizes, the qubitization-based approach requires significantly more ancilla qubits relative to the $2N$ system qubits.

Fig.~\ref{fig:QPE_Trotter_HWP_comparison} presents a comparison between Trotterized QPE resource estimates for (a) the Hubbard model and (b) the extended Hubbard model, and also the effect of varying the number of ancilla qubits, $\alpha$, used for the HWP procedure. In these plots we take a relative target accuracy of $\epsilon=0.005N$. In these QPE resource estimates, we provide four examples using $\alpha=0$, $\alpha=N/4-1$, $\alpha=N/2-1$ and $\alpha=N-1$ ancilla qubits for HWP. Note that the $\alpha=0$ and $\alpha=N/2-1$ results for the Hubbard model correspond to those also shown in Fig.~\ref{fig:QUBITIZATION_VS_TROTTER_plots}(a). The difference in T gate cost for different $\alpha$ is largest for the small system sizes because the cost in this regime is dominated by $\log(N)$ terms with different constant factors, as shown in Eq.~(\ref{eq:periodic_Hubbard_N_R}). These differences are suppressed by $\epsilon$ growing linearly with the system size. The T gate cost of the Tile Trotterization HWP simulations asymptotically approaches the same constant value dominated by $N_T$, as given by Eq.~(\ref{eq:Periodic_Hubbard_N_T}). 

In the relative error regime with $\epsilon=0.005N$, our upper bounds on the T gate cost are just below $9\times10^6$ for the Hubbard model and around $22.5\times 10^6$ for the extended Hubbard model simulations using a direct Tile Trotterization approach without HWP ($\alpha=0$).

\section{\label{sec:Discussion} Discussion and outlook}

This paper presents a Trotterization method that can be used to implement the time evolution operator for generalized Hubbard models on arbitrary lattices. Tile Trotterization allows for constructing Trotter decompositions of generalized Hubbard models by using a set of efficiently implementable tiles to decompose the hopping Hamiltonian, and provides a straightforward method to obtain per-Trotter-step gate counts. Tile Trotterization can also be combined with Hamming weight phasing to significantly reduce the number of arbitrary rotation gates required per Trotter step. In this paper, we focus on simulations of local two-dimensional Hubbard models, but the Tile Trotterization scheme allows for arbitrary density-density interactions and can be extended to include longer-range hopping terms by constructing additional hopping Hamiltonian sections. The $S_4$ tiles are a natural choice for constructing sections of the next-nearest-neighbor hopping graph of the square lattice. The Tile Trotterization scheme can also be applied to simulate three-dimensional lattice Hubbard models, however, specific examples and optimal tile coverings of three-dimensional lattices are left for future work.

We demonstrated applications of Tile Trotterization for energy estimation by QPE for both the Hubbard model and the extended Hubbard model with nearest-neighbor Coulomb interactions, showing that the method achieves a T gate complexity $\mathcal{O}(N^{3/2}\epsilon^{-3/2})$, where $N$ is the number of lattice sites and $\epsilon$ is the maximum allowed error in the energy. We demonstrated the use of numerical simulations, including tensor network methods, combined with proof techniques to partition the required nested commutators into sets of smaller but non-trivial operators, allowing us to achieve tight bounds on the Trotter error. We used our commutator bounds to evaluate the Trotter error of the periodic hexagonal lattice Hubbard model and extended Hubbard model. Our commutator bound lemmas, and the numerical techniques used to evaluate them, can also be applied to other local Hubbard models and the same complexity can be achieved, e.g. for square or Kagome lattices.

We also constructed and optimized the qubitized quantum walk operator for the hexagonal lattice Hubbard model and provided detailed Toffoli gate, T gate and qubit counts. These were used to perform resource estimation for qubitized quantum phase estimation, demonstrating a T gate complexity of $\mathcal{O}(N^2\epsilon^{-1})$. The Trotterization-based QPE approach studied in this paper has better scaling with respect to system size, which results in lower T gate counts for simulations of large systems, especially when $\epsilon$ is allowed to scale extensively with $N$.

The results presented here demonstrate that a range of classically non-trivial model Hamiltonians can be simulated with $10^6$--$10^7$ non-Clifford gates. This is several orders of magnitude less than found in many resource estimation studies of \emph{ab initio} chemical systems, suggesting that such model Hamiltonians are promising candidate applications for early fault-tolerant quantum computers. Going forward, it will be important to consider more detailed costing of this problem for early fault-tolerant architectures. Current resource estimates focus on counting non-Clifford gates, which have historically been expected to be most expensive to perform under traditional QEC schemes. However, recent proposals have questioned this understanding \cite{Gidney2024MagicGates}. Previous resource estimates have also focused on particular architectures where the cost of Clifford gates can be ignored \cite{Litinski2019ASurgery}, which may not be appropriate for early FT devices. Therefore, an interesting and important task is to consider a more detailed compilation to an early FT architecture. Here, relevant questions include: how routing can be efficiently performed under lattice surgery; how gates may be best parallelized; and efficient schemes for performing rotation gates with lower costs. An attempt at this was recently performed in Ref.~\cite{Akahoshi2024CompilationArchitecture}, using the STAR architecture \cite{Akahoshi2024PartiallyRotations} and a Trotter scheme essentially using S1 tiles.

Performing such a compilation to an early fault-tolerant architecture will introduce a number of considerations beyond those discussed in this paper. These include, for example, the cost of Clifford gates under lattice surgery and the parallelizability of a given logical circuit. We expect that the Tile Trotterization scheme here is a particularly promising candidate for such early FT applications. In addition to the low gate count, circuits are relatively simple, consisting of just a small number of rotation gate layers per Trotter step (in addition to fermionic swap layers). Such considerations will also affect the efficiency of various schemes; for example, the Hamming weight phasing method considered in this work reduces the number of rotation gates, but at the cost of additional qubits, two-qubit gates and significantly reduced parallelizability. Indeed, applying HWP to implement a single layer of $n$ rotation gates requires $n-1$ Toffoli gates in adder circuits, which cannot be readily parallelized; given these costs, it remains to be seen if applying HWP would give any advantage when implemented in a practical FT architecture. Similarly, the qubitization implementation is asymptotically dominated by unary iterators in SELECT consisting of Toffoli gates which cannot be easily parallelized. Therefore, Tile Trotterization without HWP may have greater benefits than represented in Fig.~\ref{fig:QUBITIZATION_VS_TROTTER_plots}, for example. Separately, we also note that the Tile Trotterization scheme could be performed in combination with early fault-tolerant algorithms such as statistical phase estimation to significantly reduce circuit depths compared to those presented here \cite{Ding2023SimultaneousComputers, Wang2023QuantumPrecision}. The scheme presented in this paper also allows for significantly more complicated model Hamiltonians, such as the PPP model, which provide a link to $\emph{ab initio}$ quantum chemistry. Given all of these benefits, and the significant challenge of performing dynamics of model Hamiltonians in strongly correlated regimes by conventional methods, we believe that this is a promising area for practical quantum algorithms going forward.

\section*{Data availability}

Data files for the numerical calculation of Trotter error norms are uploaded to Zenodo  with the DOI identifier https://doi.org/10.5281/zenodo.15674475.

\begin{acknowledgments}
This work is supported by the Novo Nordisk Foundation, Grant number NNF22SA0081175, NNF Quantum Computing Programme. We thank Matthew S. Teynor and Earl T. Campbell for feedback on the manuscript. We would also like to acknowledge the anonymous referees of this manuscript for their careful reviews which led to changes that improved our manuscript.
\end{acknowledgments}

\bibliography{main}

\appendix
\begin{figure*}
    \centering
    \includegraphics[width=0.95\linewidth]{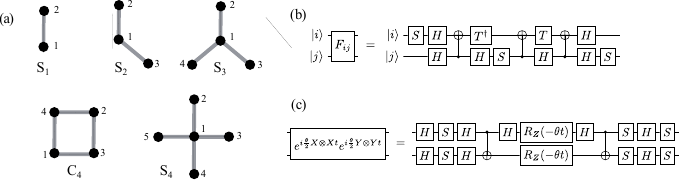}
    \caption{(a) Illustrations of the five tiles considered in this paper: $S_1$, $S_2$, $S_3$, $C_4$ and $S_4$. Each lattice site is indexed with a number that corresponds to the order of the spin orbitals in the JW string, and the edges represents nearest-neighbor hopping terms between neighboring lattice sites. The sites are ordered in this way to minimize the number of required $f_{\mathrm{swap}}$ operations when implementing $e^{-iH_{\sigma}^{\mathrm{tile}}t}$. (b) Quantum circuit diagram for implementing fermionic operator $F_{ij}$ on a two-qubit $\{ i,j\}$ subspace in the Jordan-Wigner representation. (c) Quantum circuit diagram for implementing $e^{i \frac{\theta}{2} X \otimes X t}e^{i \frac{\theta}{2} Y \otimes Y t}$, up to a global phase factor of $e^{i \frac{\pi}{2}}$. Note that when controlling this operation on an ancilla, this extra controlled phase will be a Clifford gate.}
    \label{fig:4tiles}
\end{figure*}

\section{\label{APP:tile_trotterization} Time evolution of Tile Hamiltonians}

This appendix describes how to diagonalize and perform time evolution of tile Hamiltonians for five examples of tiles: $S_1$, $S_2$, $S_3$, $S_4$ and $C_4$. We also consider a more systematic analysis using the abstract set of tiles with adjacency matrices corresponding to the complete bipartite graph $K_{n,m}$. The Hamiltonian of a tile is defined as
\begin{eqnarray}
    H_{\sigma}^{\mathrm{tile}} = - \tau \sum_{ij} R_{ij}^{\mathrm{tile}} a_{i\sigma}^\dagger a_{j\sigma},
\end{eqnarray}
where $R^{\mathrm{tile}}$ is the adjacency matrix describing the nearest-neighbor hopping terms contained in the tile Hamiltonian, meaning $R^{\mathrm{tile}}_{ij}=1$ if the tile Hamiltonian contains a hopping term between site $i$ and $j$, and otherwise $R^{\mathrm{tile}}_{ij}=0$. The time evolution operator can be implemented as
\begin{equation}
    e^{-iH_{\sigma}^{\mathrm{tile}}t} = e^{i \tau \sum_{ij} R_{ij}^{\mathrm{tile}} a_{i\sigma}^\dagger a_{j\sigma} t} = e^{i \tau M^{\mathrm{tile}}_\sigma t},
\end{equation}
where $M^{\mathrm{tile}}_\sigma$ is defined as 
\begin{equation}
    M^{\mathrm{tile}}_\sigma = \sum_{ij} R_{ij}^{\mathrm{tile}} a_{i\sigma}^\dagger a_{j\sigma}.
\end{equation}
Since the adjacency matrix, $R^{\mathrm{tile}}$, is real symmetric, it has real eigenvalues, $\lambda_e$, and eigenvectors, $v_e$, and it can be diagonalized as $R^{\mathrm{tile}} = \sum_{e} \lambda_e v_e v_e^\dagger.$
Thus, 
\begin{eqnarray}
M^{\mathrm{tile}}_\sigma &=& \sum_{ij} \sum_{e} \lambda_e  (v_e)_i (v_e)_j a_{i\sigma}^\dagger a_{j\sigma}, \\
 &=&   \sum_{e} \lambda_e \Big(\sum_{i} (v_e)_i a_{i\sigma}^\dagger \Big) \Big(\sum_{j} (v_e)_j  a_{j\sigma} \Big), \\
  &=&   \sum_{e} \lambda_e  b_{e\sigma}^\dagger b_{e\sigma},
\end{eqnarray}
for $b_{e \sigma} =  \sum_j (v_e)_j a_{j\sigma}$, where $(v_e)_j$ is the $j$'th element of eigenvector $v_e$. Whenever an eigenvalue $\lambda_e$ equals $0$, this decomposition will reduce the number of required arbitrary rotations. Throughout this section, we only focus on the eigenvectors that have non-zero eigenvalues.

Tile Hamiltonians of any type can be diagonalized and time evolution can be performed without Trotter error. However, the difference in the cost of the diagonalization procedure between different types of tiles contribute significantly to the total cost of implementing $e^{-iH_\sigma^{\mathrm{tile}}t}$. We show that tile Hamiltonians with hopping terms corresponding to star interaction graphs of type $S_1$, $S_2$ and $S_4$ and the circle interaction graph $C_{4}$, where the subscript denotes the number of edges, can be efficiently diagonalized. Fig.~\ref{fig:4tiles}(a) shows five tiles: $S_1$, $S_2$, $S_3$, $C_4$ and $S_4$. These tiles (except for $S_3$) are part of the same family of tiles: their adjacency matrices are complete bipartite graphs of type $K_{2^a,2^b}$, where $a$ and $b$ are integers (the costing of this abstract set of tiles is presented in Section~\ref{Sec:complete_bipartite_tile}). This structure has two major benefits for the cost of implementing time evolution: 1) they have a particularly cheap diagonalization procedure and 2) they all require just $2$ arbitrary rotations. We also include a costing of the $S_3$ tile Hamiltonian (which is also a complete bipartite graph but which does not have the same eigenvector structure) to show an example where the diagonalization procedure significantly increases the total cost of $e^{-iH_\sigma^{\mathrm{tile}}t}$. There might exist other tiles with an efficient implementation of $e^{-iH_\sigma^{\mathrm{tile}}t}$, but the tiles given in Fig.~\ref{fig:4tiles}(a) are sufficient for most practical purposes.

The tiles considered here, except for $S_3$, can be diagonalized using the fermionic operator $F_{ij}$ \cite{Verstraete2009QuantumSystems,Campbell2022EarlyModel}. We let $F_{ij}$ act on creation and annihilation operators defined in a two-spin-orbital subspace as
\begin{eqnarray} 
& & F_{ij} a_{i\sigma} \label{eq:Fij}
F^\dagger_{ij} = \frac{1}{\sqrt{2}} (a_{i\sigma} + a_{j\sigma}), \label{eq:F_ij_1} \\
& & F_{ij} a_{j\sigma} F^\dagger_{ij} = \frac{1}{\sqrt{2}} (a_{i\sigma} - a_{j\sigma}). \label{eq:F_ij_2}
\end{eqnarray}
These examples are given for annihilation operators and can be trivially extended to creation operators. Under the Jordan-Wigner (JW) transformation, the qubit representation of $F_{ij}$ acting on neighboring fermions $i$ and $j$ in the JW string is given by \cite{Campbell2022EarlyModel}
\begin{equation}
    F_{ij} =
        \begin{pmatrix}
    1 & 0 & 0 & 0\\
    0 & \frac{1}{\sqrt{2}} & \frac{1}{\sqrt{2}} & 0 \\
    0 & \frac{1}{\sqrt{2}} & -\frac{1}{\sqrt{2}} & 0 \\
    0 & 0 & 0 & -1
    \end{pmatrix} ,
    \label{eq:F_ij_3}
\end{equation}
which can be implemented using just $2$ T gates as shown in Fig.~\ref{fig:4tiles}(b) (also see Figure 8 of Ref~\cite{Kivlichan2020ImprovedTrotterization}). 

\begin{table*}
\begin{tabular}{l|l|l|l|l|l|l|l}
Tile & Arbitrary rotations & Other non-Clifford rotations & T gates & CNOT gates & Hadamard gates & S gates & $f_{\mathrm{swap}}$ \\
\hline
\hline
$S_1$ & 2 & 0 & 0 & 2 & 8 & 6 & 0 \\
\hline
$S_2$ & 2 & 0 & 4 & 8 & 20 & 12 & 0 \\
\hline
$C_4$ & 2 & 0 & 8 & 14 & 32 & 18 & 0 \\
\hline
$S_4$ & 2 & 0 & 12 & 20 & 44 & 24 & 2 \\
\hline
\hline
$S_3$ & 2 & 4 & 4 & 14 & 32 & 18 & 0
\end{tabular}
\caption{The cost for implementing $e^{-iH_{\sigma}^{\mathrm{tile}}t}$ for $S_1$, $S_2$, $S_3$, $C_4$ and $S_4$ tiles, assuming an initial JW-ordering of spin orbitals given in Fig.~\ref{fig:4tiles}(a). Here, ``arbitrary rotations'' refers to rotations dependent on time step, $t$, while ``other non-Clifford rotations'' are non-Clifford $Z$-axis rotations to diagonalize the tile Hamiltonian, not including T gates, which would therefore have a significant synthesis cost. Note that these tiles require two arbitrary rotations each while the other gate costs scale with the size of the tile. To obtain the cost, we assume an $F_{ij}$ implementation using the quantum circuit in Fig.~\ref{fig:4tiles}(b) and an implementation of $e^{i X\otimes X \:\theta}e^{i Y \otimes Y \:\theta}$, where $\theta$ is a parameter dependent on $\tau$ and $t$, using the quantum circuit in Fig.~\ref{fig:4tiles}(c). Further optimization of the Clifford gate costs could be performed, for example by merging gates from $F_{ij}$ with gates from $e^{i X\otimes X \:\theta}e^{i Y \otimes Y \:\theta}$.}
\label{tab:Tile_costs}
\end{table*}

The qubit implementation of fermionic operators under the JW transformation requires keeping track of the anti-symmetric fermionic properties during the computation. The qubit implementation of $F_{ij}$ is non-local in the JW string ordering, such that implementing $F_{ij}$ on non-adjacent qubits in the JW ordering would introduce additional complexity. This can be solved by using fermionic swap gates to update the JW ordering as needed, so that $F_{ij}$ is only applied on adjacent qubits. The qubit representation of the fermionic swap gate is given by \cite{Kivlichan2020ImprovedTrotterization}  
\begin{equation}
    f_{\mathrm{swap}} =
        \begin{pmatrix}
    1 & 0 & 0 & 0\\
    0 & 0 & 1 & 0 \\
    0 & 1 & 0 & 0 \\
    0 & 0 & 0 & -1
    \end{pmatrix} ,
\end{equation}
which can be implemented using a SWAP gate followed by a CZ gate, having Clifford cost only.

In our costing of $e^{-iH_{\sigma}^{\mathrm{tile}}t}$, we assume an initial JW ordering constructed to avoid using (or in the case of the $S_4$ tile, to minimize) fermionic swap gates when diagonalizing a tile Hamiltonian. The initial JW ordering within a given tile follow the labels of the sites given in Fig.~\ref{fig:4tiles}(a). In general, the Clifford costings for implementing time evolution of the tile Hamiltonian depend on the initial JW ordering of the qubits used.

It is also important to note that different hopping sections will require different JW orderings. Therefore, in a full implementation we would require layers of $f_{\mathrm{swap}}$ gates in between sections, but we do not account for this Clifford only cost in this paper.

The rest of Appendix A will show how to diagonalize the tile Hamiltonians and provide quantum circuits for implementing $e^{-iH_{\sigma}^{\mathrm{tile}}t}$ for the $S_1$, $S_2$, $S_3$, $C_4$ and $S_4$ tiles. We provide non-Clifford and Clifford gate counts for the implementation of each $e^{-iH_{\sigma}^{\mathrm{tile}}t}$, and the cost of these are summarized in Table \ref{tab:Tile_costs}.

\subsection{$S_1$ tile}
The $S_1$ tile has hopping terms between two spin orbitals $\phi_{1\sigma}$ and $\phi_{2\sigma}$, and contains hopping terms encoded by the adjacency matrix
\begin{equation}
     R^{\mathrm{S_1}} =
        \begin{pmatrix}
    0 & 1\\
    1 & 0
    \end{pmatrix},
\end{equation}
with eigenvalues $\lambda_+ = 1$ and $\lambda_- = -1$ and corresponding eigenvectors
\begin{equation}
    v_{+} =
        \begin{pmatrix}
    1/\sqrt{2} \\
    1/\sqrt{2}
    \end{pmatrix}, \;\;     v_{-} =
        \begin{pmatrix}
    1/\sqrt{2}\\
    -1/\sqrt{2}
    \end{pmatrix}.
\end{equation}
Therefore, $M^{S_1}_\sigma$ can be written as
\begin{eqnarray}
    M^{S_1}_\sigma = b^\dagger b - c^\dagger c,
\end{eqnarray}
where $b$ and $c$ are given by
\begin{eqnarray}
    & & b = \frac{1}{\sqrt{2}} (a_{1\sigma} + a_{2\sigma}), \\ & & c = \frac{1}{\sqrt{2}} (a_{1\sigma} - a_{2\sigma}).
\end{eqnarray}
We can construct $b$ and $c$ using $V=F_{12}$ as
\begin{eqnarray}
    & & b = V a_{1\sigma} V^\dagger, \\ & & c = V a_{2\sigma} V^\dagger,
\end{eqnarray}
and implement $e^{-iH_\sigma^{S_1}t}$ on $\ket{\phi_{1\sigma}\phi_{2\sigma}}$, as 
\begin{eqnarray}
    e^{-iH_\sigma^{S_1}t} = Ve^{i \tau (a_{1\sigma}^\dagger a_{1\sigma}-a_{2\sigma}^\dagger a_{2\sigma})t} V^\dagger. \label{eq:e^-iHS1}
\end{eqnarray}
Under the Jordan-Wigner transformation, we can write
\begin{eqnarray}
    e^{i \tau (a_{i\sigma}^\dagger a_{i\sigma}-a_{j\sigma}^\dagger a_{j\sigma})t} =
        \begin{pmatrix}
    1 & 0 & 0 & 0\\
    0 & e^{i \tau t} & 0 & 0 \\
    0 & 0 & e^{-i \tau t} & 0 \\
    0 & 0 & 0 & 1
    \end{pmatrix} ,
\end{eqnarray}
which means Eq.~(\ref{eq:e^-iHS1}) can be further compiled to
\begin{eqnarray}
    & & \! \! \! \! \! \! \! e^{-iH_\sigma^{S_1}t} = F_{12}e^{i \tau (a_{1\sigma}^\dagger a_{1\sigma}-a_{2\sigma}^\dagger a_{2\sigma})t}F_{12}^\dagger  =  \nonumber \\ & & \! \! \! \! \! \! \! \! \! \!  \begin{pmatrix}
    1 & 0 & 0 & 0\\
    0 & \cos(\tau t) & i \sin(\tau t) & 0 \\
    0 & i \sin(\tau t) & \cos(\tau t) & 0 \\
    0 & 0 & 0 & 1
    \end{pmatrix} = e^{i \frac{\tau}{2} X \otimes X t}e^{i\frac{\tau}{2} Y \otimes Y t}. \label{eq:S1compiling}
\end{eqnarray}
This operation can be implemented using the quantum circuit given in Fig.~\ref{fig:4tiles}(c), using $\frac{\theta}{2} = \tau$ (see Eq.~(18) in Ref.~\cite{BassmanOftelie2022Constant-depthComputers}). This results in a non-Clifford cost of 2 arbitrary rotations of the same angle and additional Clifford cost of 2 CNOT gates, 6 S gates and 8 Hadamard gates for implementing $e^{-iH_\sigma^{S_1}t}$. 

\subsection{$S_2$ tile}
The $S_2$ tile has hopping terms between three spin orbitals $\phi_{1\sigma}$,  $\phi_{2\sigma}$ and $\phi_{3\sigma}$, described by the adjacency matrix
\begin{eqnarray}
     R^{S_2} =
        \begin{pmatrix}
    0 & 1 & 1\\
    1 & 0 & 0\\
    1 & 0 & 0
    \end{pmatrix}, 
\end{eqnarray}
with eigenvalues $\lambda_0 = 0$, $ \lambda_+ = \sqrt{2}$ and $\lambda_- = -\sqrt{2}$ and corresponding eigenvectors 
\begin{equation}
     v_{+} =
        \begin{pmatrix}
    1/\sqrt{2}\\
    1/2\\
    1/2
    \end{pmatrix}, \;\;
    v_{-} =
        \begin{pmatrix}
    1/\sqrt{2}\\
    -1/2\\
    -1/2
    \end{pmatrix}. 
\end{equation}
We can write $M^{S_2}_\sigma$ as
\begin{eqnarray}
    M^{S_2}_\sigma = \sqrt{2} b^\dagger b - \sqrt{2} c^\dagger c,
\end{eqnarray}
where $b$ and $c$ are given by
\begin{eqnarray}
    & & b = \frac{1}{\sqrt{2}} a_{1\sigma} + \frac{1}{2} (a_{2\sigma}+a_{3\sigma}), \\ & &c = \frac{1}{\sqrt{2}} a_{1\sigma} - \frac{1}{2} (a_{2\sigma}+a_{3\sigma}).
\end{eqnarray}
Employing the unitary transformation $V=F_{23}F_{12}$, we can write
\begin{eqnarray}
    & & b = V a_{1\sigma} V^\dagger, \\ & & c = V a_{2\sigma} V^\dagger,
\end{eqnarray}
and so can implement the time evolution of the $S_2$ tile on $\ket{\phi_{1\sigma}\phi_{2\sigma}\phi_{3\sigma}}$ as 
\begin{eqnarray}
    e^{-iH^{S_2}_\sigma t} = Ve^{i\tau \sqrt{2} (a_{1\sigma}^\dagger a_{1\sigma}-a_{2\sigma}^\dagger a_{2\sigma})t} V^\dagger.
\end{eqnarray}
This can be further compiled using the same strategy as shown in Eq.~(\ref{eq:S1compiling}), leading to the final expression 
\begin{eqnarray}
    e^{-iH^{S_2}_\sigma t} = F_{23}e^{i \frac{\tau}{\sqrt{2}} X_1 \otimes X_2 t}e^{i\frac{\tau}{\sqrt{2}} Y_1 \otimes Y_2 t}F_{23}^\dagger.
\end{eqnarray}
The total gate cost for implementing the time evolution of the $S_2$ tile Hamiltonian is therefore two arbitrary rotations, 4 T gates, 8 CNOT gates, 12 S gates and 20 Hadamard gates.

\subsection{$C_4$ tile}
The plaquette ($C_4$) tile, which first introduced in Ref.~\cite{Campbell2022EarlyModel}, has hopping terms between four spin orbitals, $\phi_{1\sigma}$, $\phi_{2\sigma}$, $\phi_{3\sigma}$ and $\phi_{4\sigma}$. Using the ordering in  Fig.~\ref{fig:4tiles}(a), the adjacency matrix is given by
\begin{equation}
    R^{\mathrm{C_4}} =
        \begin{pmatrix}
    0 & 0 & 1 & 1 \\
    0 & 0 & 1 & 1 \\
    1 & 1 & 0 & 0 \\
    1 & 1 & 0 & 0
    \end{pmatrix} ,
\end{equation}
for the Jordan-Wigner string ordering $\ket{\phi_{1\sigma}\phi_{2\sigma}\phi_{3\sigma}\phi_{4\sigma}}$. This adjacency matrix has two zero eigenvalues, $\lambda_+ = 2$ and $\lambda_- = -2$ and corresponding eigenvectors
\begin{equation}
    v_{+} = \frac{1}{2}
        \begin{pmatrix}
    1\\
    1\\
    1\\
    1
    \end{pmatrix}, \;\;
    v_{-} =
        \frac{1}{2}
        \begin{pmatrix}
    1\\
    1\\
    -1\\
    -1
    \end{pmatrix} . 
\end{equation}
Then, $M^{\mathrm{C_4}}_\sigma$ can be written as
\begin{eqnarray}
    M^{\mathrm{C_4}}_\sigma = 2 b^\dagger b - 2 c^\dagger c,
\end{eqnarray}
with $b$ and $c$ given by
\begin{eqnarray}
    & & b = \frac{1}{2} (a_{1\sigma} + a_{2\sigma} + a_{3\sigma} + a_{4\sigma}), \\ & & c = \frac{1}{2} (a_{1\sigma} + a_{2\sigma} - a_{3\sigma} - a_{4\sigma}).
\end{eqnarray}
Employing the unitary transformation $V=F_{34}F_{21}F_{23}$, we can write
\begin{eqnarray}
    & & b = V a_{2\sigma} V^\dagger, \\ & & c = V a_{3\sigma} V^\dagger,
\end{eqnarray}
and implement the time evolution as 
\begin{eqnarray}
    e^{-iH^{C_4}_\sigma t} = Ve^{i\tau 2 (a_{2\sigma}^\dagger a_{2\sigma}-a_{3\sigma}^\dagger a_{3\sigma})t} V^\dagger.
\end{eqnarray}
This can be further compiled using the same strategy as shown in Eq.~(\ref{eq:S1compiling}). Then, the time evolution of the $C_4$ tile Hamiltonian can be implemented as
\begin{eqnarray}
    e^{-iH^{C_4}_\sigma t} = F_{34}F_{21}e^{i \tau X_2 \otimes X_3 t}e^{i \tau Y_2 \otimes Y_3 t}F_{21}^\dagger F_{34}^\dagger,
\end{eqnarray}
which requires 2 arbitrary rotations and 8 T gates. The total gate cost of $e^{-iH^{C_4}_\sigma t}$ is summarized in Table~\ref{tab:Tile_costs}. Note that the initial Jordan-Wigner ordering of the sites used here saves four $f_{\mathrm{swap}}$ operations in the implementation of the $C_4$ tile time evolution compared to the implementations in Refs. \cite{Campbell2022EarlyModel, kan_2024} (see Eq.~B21 in Ref.~\cite{kan_2024}).

\subsection{$S_4$ tile}

The $S_4$ tile has hopping terms between $5$ spin orbitals, $\phi_{1\sigma}$, $\phi_{2\sigma}$, $\phi_{3\sigma}$, $\phi_{4\sigma}$ and $\phi_{5\sigma}$, described by the adjacency matrix
\begin{equation}
    R^{S_4} =
        \begin{pmatrix}
    0 & 1 & 1 & 1 & 1\\
    1 & 0 & 0 & 0 & 0\\
    1 & 0 & 0 & 0 & 0\\
    1 & 0 & 0 & 0 & 0\\
    1 & 0 & 0 & 0 & 0\\
    \end{pmatrix}.
\end{equation}
This matrix has three zero eigenvalues, $\lambda_+ = 2$ and $\lambda_-=-2$ and corresponding eigenvectors
\begin{equation}
 v_{+} =
        \begin{pmatrix}
    1/\sqrt{2}\\
    1/(2\sqrt{2})\\
    1/(2\sqrt{2})\\
    1/(2\sqrt{2})\\
    1/(2\sqrt{2})
    \end{pmatrix}, \;\;
    v_{-} =
        \begin{pmatrix}
    1/\sqrt{2}\\
    -1/(2\sqrt{2})\\
    -1/(2\sqrt{2})\\
    -1/(2\sqrt{2})\\
    -1/(2\sqrt{2})
    \end{pmatrix} . 
\end{equation}
This means $M^{S_4}_\sigma$ can be written as
\begin{eqnarray}
    M^{S_4}_\sigma = 2 b^\dagger b - 2 c^\dagger c,
\end{eqnarray}
where $b$ and $c$ are given by
\begin{eqnarray}
    & & \! \! \! \!  b = \frac{1}{\sqrt{2}} \Big(a_{1\sigma} + \frac{1}{2}(a_{2\sigma} + a_{3\sigma} + a_{4\sigma} + a_{5\sigma}) \Big), \\ & & \! \!  \! \! c =\frac{1}{\sqrt{2}} \Big(a_{1\sigma} - \frac{1}{2}(a_{2\sigma} + a_{3\sigma} + a_{4\sigma} + a_{5\sigma}) \Big).
\end{eqnarray}
Employing the unitary transformation $V=F_{45}F_{32}F_{34}f_{\mathrm{swap}}^{23}F_{13}$, we can write
\begin{eqnarray}
    & & b = V a_{1\sigma} V^\dagger, \\ & & c = V a_{3\sigma} V^\dagger,
\end{eqnarray}
where $f_{\mathrm{swap}}^{23}$ is the fermionic swap operator that swaps the electrons in orbitals $2$ and $3$, such that the orbitals $\phi_{1\sigma}$ and $\phi_{3\sigma}$ adjacent in the Jordan-Wigner ordering after the swap performed by $V^\dagger$. The following $f_{\mathrm{swap}}^{23}$ in $V$ returns the orbitals to the initial JW-ordering.

The time evolution of the $S_4$ tile Hamiltonian can therefore be implemented on $\ket{\phi_{1\sigma}\phi_{2\sigma}\phi_{3\sigma}\phi_{4\sigma}\phi_{5\sigma}}$ as 
\begin{eqnarray}
    e^{-iH^{S_4}_\sigma t} = Ve^{i\tau 2 (a_{1\sigma}^\dagger a_{1\sigma}-a_{3\sigma}^\dagger a_{3\sigma})t} V^\dagger.
\end{eqnarray}
This can be further compiled using the same strategy as shown in Eq.~(\ref{eq:S1compiling}) (remember that $\phi_1$ and $\phi_3$ are adjacent after the swap), resulting in the following expression for the time evolution of the $S_4$ tile Hamiltonian
\begin{eqnarray}
    e^{-iH^{S_4}_\sigma t} &=& F_{45}F_{32}F_{34}f_{\mathrm{swap}}^{23} e^{i \tau X_1 \otimes X_3 t} \nonumber \\&\times& e^{i \tau Y_1 \otimes Y_3 t} f_{\mathrm{swap}}^{23}F_{34}^\dagger F_{32}^\dagger F_{45}^\dagger,
\end{eqnarray}
which requires 2 arbitrary rotations and 12 T gates. The total gate cost is summarized in Table~\ref{tab:Tile_costs}.

\subsection{$S_3$ tile}
Next, we demonstrate a method to implement the time evolution of the $S_3$ tile and highlight why this is more costly than for the other tiles described above. The $S_3$ tile has hopping terms between $4$ spin orbitals, $\phi_{1\sigma}$, $\phi_{2\sigma}$, $\phi_{3\sigma}$ and $\phi_{4\sigma}$, described by the adjacency matrix
\begin{equation}
    R^{S_3} =
        \begin{pmatrix}
    0 & 1 & 1 & 1 \\
    1 & 0 & 0 & 0 \\
    1 & 0 & 0 & 0 \\
    1 & 0 & 0 & 0 \\
    \end{pmatrix}.
\end{equation}
This matrix has two non-zero eigenvalues, $\lambda_+ = \sqrt{3}$ and $\lambda_-=-\sqrt{3}$ and corresponding eigenvectors
\begin{equation}
 v_{+} =
        \begin{pmatrix}
    1/\sqrt{2}\\
    1/\sqrt{6}\\
    1/\sqrt{6}\\
    1/\sqrt{6}
    \end{pmatrix}, \;\;
    v_{-} =
        \begin{pmatrix}
    1/\sqrt{2}\\
    -1/\sqrt{6}\\
    -1/\sqrt{6}\\
    -1/\sqrt{6}
    \end{pmatrix} . 
\end{equation}
Therefore $M^{S_3}_\sigma$ can be written as
\begin{eqnarray}
    M^{S_4}_\sigma = \sqrt{3} b^\dagger b - \sqrt{3} c^\dagger c,
\end{eqnarray}
where $b$ and $c$ are given by
\begin{eqnarray}
    & & b = \frac{1}{\sqrt{2}} a_{1\sigma} + \frac{1}{\sqrt{6}}(a_{2\sigma} + a_{3\sigma} + a_{4\sigma}) , \\ & & c = \frac{1}{\sqrt{2}} a_{1\sigma} - \frac{1}{\sqrt{6}}(a_{2\sigma} + a_{3\sigma} + a_{4\sigma}) .
\end{eqnarray}
These cannot be constructed through sequential applications of $F_{ij}$ because this operator only creates superpositions with a prefactor of $1/\sqrt{2}$. This can be solved by defining a new operator, $G_{ij}$, that acts on annihilation operators $a_i$ and $a_j$ as
\begin{eqnarray}
    & G_{ij} a_i G_{ij}^\dagger = \sqrt{\frac{1}{3}} a_i + \sqrt{\frac{2}{3}} a_j, \\
    & G_{ij} a_j G_{ij}^\dagger = \sqrt{\frac{2}{3}} a_i - \sqrt{\frac{1}{3}} a_j,
\end{eqnarray}
and has matrix representation 
\begin{equation}
     G_{ij} =
\begin{pmatrix}
    1 & 0 & 0 & 0 \\
    0 & \sqrt{\frac{1}{3}} & \sqrt{\frac{2}{3}} & 0 \\
    0 & \sqrt{\frac{2}{3}} & -\sqrt{\frac{1}{3}} & 0 \\
    0 & 0 & 0 & -1 
\end{pmatrix} .
\end{equation}
This operator can be implemented using the quantum circuit shown in Fig.~8 of Ref.~\cite{Kivlichan2020ImprovedTrotterization} when replacing T and $T^\dagger$ with $R_z(\arccos{\frac{1}{\sqrt{3}}})$ and $R_z(-\arccos{\frac{1}{\sqrt{3}}})$. In the standard fault-tolerant compilation scheme, these rotations must be synthesized in terms of Clifford + T (approximated up to a required precision), and therefore $G_{ij}$ becomes significantly more costly than $F_{ij}$.

We define $V=F_{34}G_{23}F_{12}$, which allows us to write
\begin{eqnarray}
    & & b = V a_{1\sigma} V^\dagger, \\ & & c = V a_{2\sigma} V^\dagger.
\end{eqnarray}
Then, the time evolution of the $S_3$ tile Hamiltonian can be implemented on $\ket{\phi_{1\sigma}\phi_{2\sigma}\phi_{3\sigma}\phi_{4\sigma}}$ as 
\begin{eqnarray}
    e^{-iH^{S_3}_\sigma t} = Ve^{i\tau \sqrt{3} (a_{1\sigma}^\dagger a_{1\sigma}-a_{2\sigma}^\dagger a_{2\sigma})t} V^\dagger,
\end{eqnarray}
which can be further compiled using the strategy shown in Eq.~(\ref{eq:S1compiling}), resulting in the following expression
\begin{equation}
    e^{-iH^{S_3}_\sigma t} = F_{34}G_{23} e^{i \frac{\sqrt{3}\tau}{2} X_1 \otimes X_2 t}e^{i \frac{\sqrt{3}\tau}{2} Y_1 \otimes Y_2 t}  G_{23}^{\dagger}F_{34}^{\dagger}.
\end{equation}
This implementation of the time evolution operator of the $S_3$ tile Hamiltonian requires 2 applications of $G_{ij}$ and has total non-Clifford cost of 2 arbitrary rotations, four $R_z(\theta)$ rotations of $\theta = \pm \arccos{\frac{1}{\sqrt{3}}}$ and 4 T gates. Assuming that we can implement the $\theta=\pm \arccos{\frac{1}{\sqrt{3}}}$ up to required precision $\epsilon$ with 30 T gates, the time evolution of the $S_3$ tile has a total cost of 2 arbitrary rotations and 124 T gates. This cost is far larger per hopping term implemented than especially $S_2$, $C_4$ and $S_4$. Note that synthesis of these rotations will also contribute additional errors that we would need to consider in the QPE costing.

\subsection{$K_{2^a,2^b}$ tile} \label{Sec:complete_bipartite_tile}

Lastly, we consider an abstract set of tiles which encompasses all of the above examples, but also bipartite graphs in general. The additional tiles in this set are likely not of practical use (although could be of valuable for implementing beyond-nearest-neighbor hopping), but allow for a more comprehensive analysis.

The complete bipartite graph $K_{n,m}$ has vertices $v_1 \dots v_{n+m}$ and edges between $v_{i}$ and $v_{j}$ if and only if $i\leq n$ and $j \geq n+1$ or $j\leq n$ and $i \geq n+1$. Notice that $K_{1,2^k} = S_{2^k}$, $K_{2,2} = C_4$, and that the $S_3$ tile is given by $K_{1,3}$, so this generalization covers all the cases discussed above. The adjacency matrix $R^{K_{n,m}}$ has rank 2 since it is off-diagonal with $1$ in the off-diagonal $n \times m$ blocks. It has eigenvalues $\pm\sqrt{nm},0$ with multiplicities $1,1$ and $n+m-2$, and normalized eigenvectors corresponding to the non-zero eigenvalues
\begin{equation}
    v_{+} = (\frac{1}{\sqrt{2n}}1_n, 
    \frac{1}{\sqrt{2m}}1_m), \;\;\;
    v_{-} =
       (\frac{1}{\sqrt{2n}}1_n, 
    -\frac{1}{\sqrt{2m}}1_m),
\end{equation}
where $1_{k}$ denotes the vector of 1s with length $k$.

Then, $M^{\mathrm{K_{n,m}}}_\sigma$ can be written as
\begin{eqnarray}
    M^{\mathrm{K_{n,m}}}_\sigma = \sqrt{nm} b^\dagger b - \sqrt{nm} c^\dagger c,
\end{eqnarray}
with $b$ and $c$ given by
\begin{eqnarray}
    & & b = \frac{1}{\sqrt{2n}}\sum_{i=1}^{n} a_{i \sigma}+  \frac{1}{\sqrt{2m}}\sum_{i=n+1}^{m+n} a_{i \sigma}, \\ & &c =\frac{1}{\sqrt{2n}}\sum_{i=1}^{n} a_{i \sigma}-  \frac{1}{\sqrt{2m}}\sum_{i=n+1}^{m+n} a_{i \sigma}.
\end{eqnarray}
At this point it is advantageous to consider the case where  $n=2^{a}$ and $m=2^{b}$ for non-negative integers $a$, $b$. In this case, we can define $V$ to be $\displaystyle \prod_{k=0}^{a-1} \prod_{i=1}^{2^{a-k}-1} F_{2^{k}i,2^{k}(i+1)}\prod_{l=0}^{b-1} \prod_{j=1}^{2^{b-l}-1} F_{2^{l}j+n,2^{l}(j+1)+n} F_{1,2^a+1}$, where whenever $a=0$ and $b=0$ we define the empty product to be 1. In this definition of $V$, we ignore $f_{\mathrm{swap}}$ operators that depend on the initial JW ordering of the orbitals. Then,
\begin{eqnarray}
    & & b = V a_{1\sigma} V^\dagger, \\ & & c = V a_{2^a+1,\sigma} V^\dagger,
\end{eqnarray}
and we can implement the time evolution as 
\begin{eqnarray}
   \hspace{-1cm} e^{-iH^{\mathrm{K_{n,m}}}_\sigma t} = Ve^{i\tau \lambda (a_{1\sigma}^\dagger a_{1\sigma}-a_{2^a+1,\sigma}^\dagger a_{2^a+1,\sigma})t} V^\dagger, 
\end{eqnarray}
where $\lambda = 2^{\frac{a+b}{2}}$. 
This can be further compiled using the same strategy as shown in Eq.~(\ref{eq:S1compiling}). Then, the time evolution of the $K_{2^a, 2^b}$ tile Hamiltonian can be implemented as
\begin{eqnarray}
    \hspace{-1cm} e^{-iH^{\mathrm{K_{n,m}}}_\sigma t} = \tilde V e^{i \frac{\lambda}{2} \tau X_1 \otimes X_{2^a+1} t}e^{i \tau Y_1 \otimes Y_{2^a+1} t\frac{\lambda}{2}} \tilde V^\dagger,
\end{eqnarray}
with $\tilde V =  V F_{1,2^a+1}^\dagger,$ which requires $2$ arbitrary rotations and $2^{a+2}+2^{b+2}-8$ T gates (since the total number of $F_{ij}$ operators in $\tilde V$ is $2((2^a-1) +(2^b-1))$ when canceling $F_{1,2^a+1}$ and $F_{1,2^a+1}^\dagger$). Inserting appropriate values of $a$ and $b$ one recovers the costing above. The total number of bonds covered by the tile is $2^{a+b}$. Therefore, the cost of implementing time evolution of tiles with adjacency matrices corresponding to bipartite graphs of type $K_{2^a,2^b}$ is $2/2^{a+b}$ arbitrary rotations per bond and $(2^{a+2}+2^{b+2}-8)/2^{a+b}$ T gates per bond. The T gate cost of a single arbitrary rotation is often in the range 30--50 T gates, thus, using larger tiles decreases the overall total T gate cost required for implementing time evolution of the hopping Hamiltonian. 

This section shows that tiles with adjacency matrices of type $K_{2^a,2^b}$ have a simple and, in the case of $a$ and/or $b$ $>0$, cheap implementation of time evolution of tile Hamiltonians per hopping term in a given tile. For the two-dimensional lattices considered here, we primarily find the $S_2$ ($K_{1,2}$), $S_4$ ($K_{1,4}$) and $C_4$ ($K_{2,2}$) tiles practically useful. However, in some cases larger tiles may be applied e.g. to tile three-dimensional lattices or to construct beyond nearest-neighbor hopping sections. The application of larger tiles will result in a significant saving in the number of T gates per bond.

\begin{widetext}
\section{\label{APP:tile_Trotter_decomposition} Tile Trotterization decomposition}

Here, we show that the Trotter error norm of Tile Trotterization can be obtained as $W_{\mathrm{tile}} \leq W_{\mathrm{SO2}}+W_h$, given that we can compute $W_{\mathrm{SO2}}$ and $W_h$ that come from the following Trotter partitionings:
\begin{eqnarray}
    & & \Big\lVert e^{-i(H_h+H_C)t} - e^{-iH_C t/2}e^{-i H_h t}e^{-iH_C t/2} \Big\rVert \leq W_{\mathrm{SO2}} t^3,
    \label{eq:APP_WSO2}
    \\ 
    & & \lVert e^{-iH_ht} - \prod_{s=1}^S e^{-iH_h^st/2} \prod_{s=S}^1 e^{-iH_h^st/2} \rVert \leq W_{h} t^3 . \label{eq:APP_W_h}
\end{eqnarray}
The Tile Trotterization error norm, $W_{\mathrm{tile}}$, can be calculated as
\begin{equation}
        \Big\lVert e^{-iHt} - e^{-iH_C \frac{t}{2}}  \prod_{s=1}^S e^{-iH_h^s \frac{t}{2}}\prod_{s=S}^1 e^{-iH_h^s \frac{t}{2}} e^{-iH_C \frac{t}{2}} \Big\rVert \leq W_{\mathrm{tile}} t^3 \leq (W_{\mathrm{SO2}}+W_h)t^3,
\end{equation}
which is shown by the following derivation:
\begin{eqnarray}
& &\Big\lVert e^{-i(H_h+H_C)t} - e^{-iH_Ct/2}  \prod_{s=1}^S e^{-iH_h^st/2}\prod_{s=S}^1 e^{-iH_h^st/2} e^{-iH_Ct/2}  \Big\rVert \nonumber \\
& &= \Big\lVert e^{-i(H_h+H_C)t} -  e^{-iH_Ct/2}  e^{-iH_ht}e^{-iH_Ct/2}+  e^{-iH_Ct/2}  e^{-iH_ht}e^{-iH_Ct/2} - e^{-iH_Ct/2}  \prod_{s=1}^S e^{-iH_h^st/2}\prod_{s=S}^1 e^{-iH_h^st/2} e^{-iH_Ct/2}  \Big\rVert \nonumber \\
& &\leq \Big\lVert e^{-i(H_h+H_C)t} -  e^{-iH_Ct/2}  e^{-iH_ht}e^{-iH_Ct/2} \Big\rVert +  \Big\lVert e^{-iH_Ct/2}  e^{-iH_ht}e^{-iH_Ct/2} - e^{-iH_Ct/2}  \prod_{s=1}^S e^{-iH_h^st/2}\prod_{s=S}^1 e^{-iH_h^st/2} e^{-iH_Ct/2} \Big\rVert  \nonumber \\
& &\leq W_{\mathrm{SO2}} t^3 + \Big\lVert e^{-iH_Ct/2} \left( e^{-iH_ht}  - \prod_{s=1}^S e^{-iH_h^st/2}\prod_{s=S}^1 e^{-iH_h^st/2}\right)e^{-iH_Ct/2}  \Big\rVert,   \nonumber \\
& &\leq W_{\mathrm{SO2}} t^3 + \Big\lVert e^{-iH_Ct/2} \Big\rVert  \Big\lVert ( e^{-iH_ht}  - \prod_{s=1}^S e^{-iH_h^st/2}\prod_{s=S}^1 e^{-iH_h^st/2}) \Big\rVert \Big\lVert  e^{-iH_Ct/2}  \Big\rVert,   \nonumber \\
& &\leq (W_{\mathrm{SO2}} + W_h)t^3. 
\end{eqnarray}
Therefore the Tile Trotterization error norm can be bounded by a sum of simpler error norms, $W_{\mathrm{SO2}}$ and $W_{\mathrm{h}}$.

\section{\label{APP:W_h} The hopping Hamiltonian Trotter error norm $W_h$}

Tile Trotterization splits the hopping Hamiltonian into $S$ sections and uses the following Trotter decomposition to implement the hopping Hamiltonian,
\begin{eqnarray}
    & & \lVert e^{-iH_ht} - \prod_{s=1}^S e^{-iH_h^st/2} \prod_{s=S}^1 e^{-iH_h^st/2} \rVert \leq W_{h} t^3 . \label{eq:W_h}
\end{eqnarray}

In the case where it is possible to divide a lattice into three sections of colors blue, red and gold, we have that $S=3$ and that the hopping Hamiltonian sections can be written as $H_1=H_h^b$, $H_2=H_h^r$ and $H_3=H_h^g$, where the subscripts 1, 2 and 3 imply the order in which the terms are implemented within the Trotter step.

The Trotter error norm, $W_h$, can be obtained from Eq.~(\ref{eq:W}), which can be written out using a package provided by Schubert \emph{et al.}~\cite{Schubert2023TrotterModel}, which automatically writes out Eq.~(\ref{eq:W}) for an arbitrary number of non-commuting Hamiltonian terms. We evaluate Eq.~(\ref{eq:W}) for $H_1=H_h^b$, $H_2=H_h^r$ and $H_3=H_h^g$ as
\begin{eqnarray}
    W_h &=& \frac{1}{12} \Big( \lVert [[H_h^b,H_h^r],H_h^r] \rVert + \lVert [[H_h^b,H_h^r],H_h^g] \rVert + \lVert [[H_h^b,H_h^g],H_h^r] \rVert + \lVert [[H_h^b,H_h^g],H_h^g] \rVert + \lVert [[H_h^r,H_h^g],H_h^g] \rVert \Big) \nonumber \\ &+& \frac{1}{24} \Big( \lVert [[H_h^b,H_h^r],H_h^b] \rVert + \lVert [[H_h^b,H_h^g],H_h^b] \rVert + \lVert [[H_h^r,H_h^g],H_h^r] \rVert \Big). \label{eq:W_h_h_h_h}
\end{eqnarray}
\begin{figure*}
    \centering
    \includegraphics[width=0.45\linewidth]{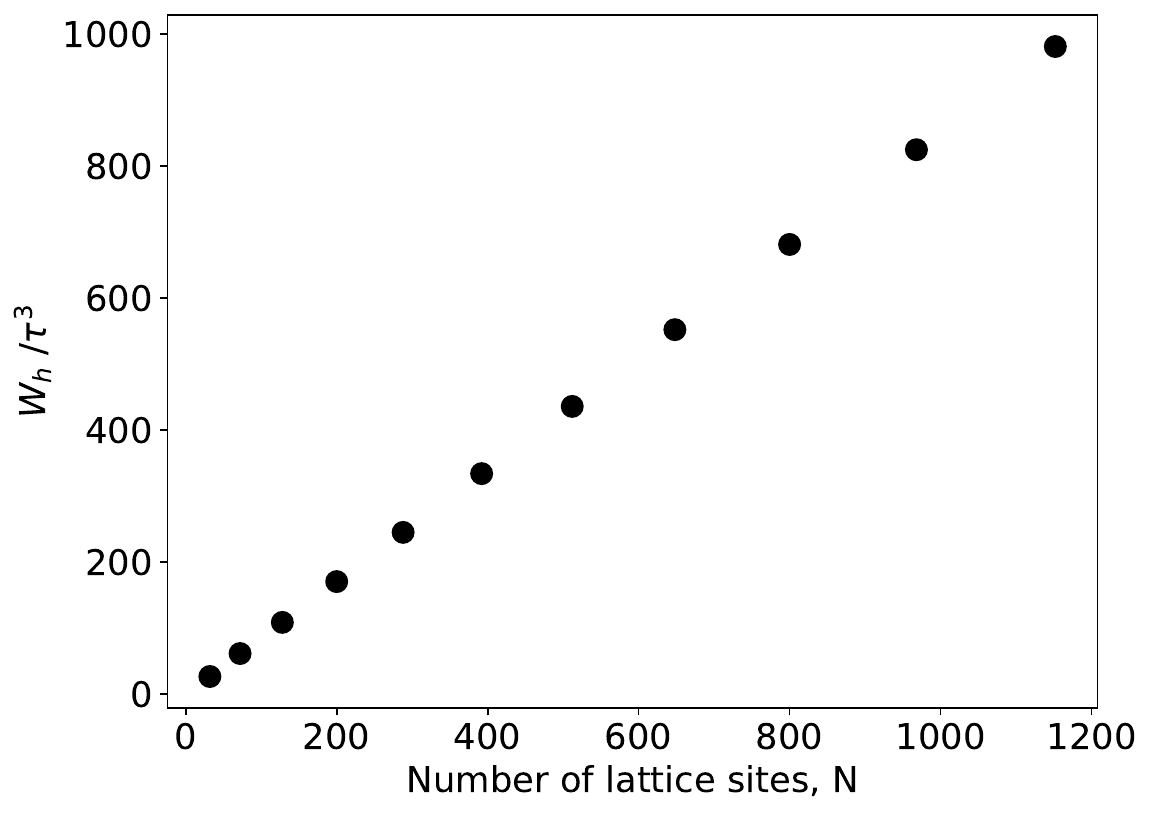}
    \caption{Plot of $W_h/\tau^3$ as a function of the number of lattice sites, $N$, for the periodic hexagonal lattice, using the division described in Appendix~\ref{APP:hexagonallatticedivision}.}
    \label{fig:W_h_scaling}
\end{figure*}
This expression can be evaluated from the adjacency matrices of the different sections. The Hamiltonian sections can be written as
\begin{eqnarray}
    & & H_h^b = -\tau \sum_{ij\sigma} R_{ij}^b a_{i\sigma}^\dagger a_{j\sigma}, \\
    & & H_h^r = -\tau \sum_{ij\sigma} R_{ij}^r a_{i\sigma}^\dagger a_{j\sigma}, \\
    & & H_h^g = -\tau \sum_{ij\sigma} R_{ij}^g a_{i\sigma}^\dagger a_{j\sigma},
\end{eqnarray}
where $R^r$, $R^b$ and $R^g$ are the adjacency matrices of the red, blue and gold lattice sections respectively. Because the Hamiltonian section operators are free fermionic, and using the argument of Ref.~\cite{Campbell2022EarlyModel}~Appendix~A, this  expression for $W_h$ can be evaluated from the adjacency matrices as
\begin{eqnarray}
    W_h &=& \frac{\tau^3}{12} \Big( \lVert [[R^b,R^r],R^r] \rVert_1 + \lVert [[R^b,R^r],R^g] \rVert_1 + \lVert [[R^b,R^g],R^r] \rVert_1 + \lVert [[R^b,R^g],R^g] \rVert_1 + \lVert [[R^r,R^g],R^g] \rVert_1 \Big) \nonumber \\ &+& \frac{\tau^3}{24} \Big( \lVert [[R^b,R^r],R^b] \rVert_1 + \lVert [[R^b,R^g],R^b] \rVert_1 + \lVert [[R^r,R^g],R^r] \rVert_1 \Big), \label{eq:W_h_written_out_for_RBG}
\end{eqnarray}
where $\lVert \, \cdot \, \rVert_1$ is the Schatten one-norm and $\tau$ is the hopping parameter. This expression can be used to evaluate $W_h$ for any model where the lattice can be divided into three lattice sections.

The hopping section error norm, $W_h$, is evaluated numerically for the periodic hexagonal lattice using the lattice division described in Appendix~\ref{APP:hexagonallatticedivision}, and scales linearly in $N$. This is shown in Fig.~\ref{fig:W_h_scaling} for a lattice with parameters $L_x=L_y=L$ in the range $4\leq L \leq 24$ for even $L$ and hopping parameter $\tau=1$. This figure shows $W_h/\tau^3$ as a function of the number of lattice sites $N$, ranging between $N=32$ and $N=1152$, demonstrating the linear dependence of $W_h$ on the number of lattice sites. We obtain a bound on $W_h$ using the largest constant factor: $W_h/\tau^3N$, of the points plotted in Fig.\ref{fig:W_h_scaling}, leading to
\begin{equation}
    W_h \leq 0.8532 \tau^3 N,
\end{equation}
for periodic hexagonal lattices with even $L$ in the range $4\leq L \leq 24$ using the $S_2$ tiling shown in Appendix~\ref{APP:hexagonallatticedivision}.
\end{widetext}

\begin{figure*}
    \centering
    \includegraphics[width=0.9\linewidth]{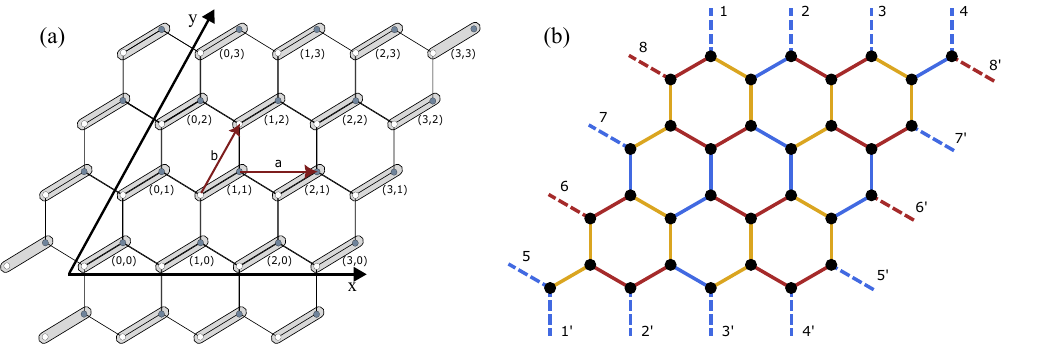}
    \caption{(a) The periodic hexagonal lattice model where each pair of lattice points are labeled with $(l_x,l_y)$ defined from the $(x,y)$ positions of the lattice points as $(x,y)=(l_x\cdot a,l_y\cdot b)$, where $a$ and $b$ are lattice vectors. The size of the lattice is given by $L_x = \max(l_x)+1$ and $L_y = \max(l_y)+1$. For this lattice, we have $L_x = L_y = 4$ and $N = 2L_xL_y = 32$. Periodic boundary conditions are applied in the $x$ and $y$ directions which is illustrated through the periodic image in the lattice sketch. (b) The division of a periodic hexagonal lattice with size $L_x=L_y=4$ into three colored sections: blue, red and gold, using $S_2$ tiles. The dashed bonds with attached number indicate the bonds that ensure periodicity. For example, the dashed bonds labeled with 1 and 1' indicate the same bond. Each section is covered by $N_{b} = N_{r} = N_{g} = N/4$ $S_2$ tiles}
    \label{Fig:hex_lattices}
\end{figure*}

\begin{figure*}
    \centering
    \includegraphics[width=1\linewidth]{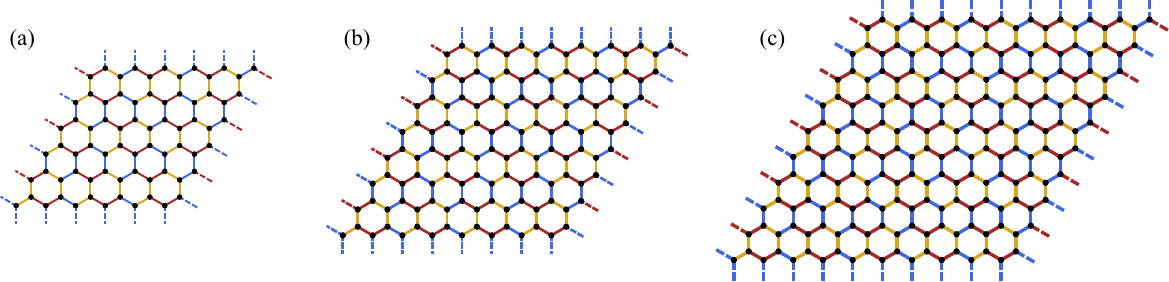}
    \caption{Lattice sections used in this paper for periodic hexagonal lattice models with lattice size $L_x = L_y = L$, where (a) $L=6$, (b) $L=8$ and (c) $L=10$, which corresponds to (a) $N=72$, (b) $N=128$, (c) $N=200$. The number of tiles in each section of the periodic hexagonal lattice models is $N_r=N_b=N_g=N/4$ $S_2$ tiles.}
    \label{fig:sections,L=6,8,10}
\end{figure*}

\section{\label{APP:hexagonallatticedivision} Periodic Hexagonal lattice model}

We consider the periodic hexagonal lattice model shown in Fig.~\ref{Fig:hex_lattices}(a). Each lattice point pair of color white and grey are labeled by $(l_x,l_y)$ which is defined from the $(x,y)$ positions of the lattice pairs as: $(x,y)=(a\cdot l_x,b \cdot l_y)$, where $a$ and $b$ are the lattice vectors. We define the lattice size using the parameters $L_x=\max(l_x)+1$ and $L_y=\max(l_y)+1$, such that the number of lattice sites is $N=2 L_x L_y$.

All lattices, both periodic and non-periodic, can be divided into sections in a number of different ways using either one type or several different types of tiles. The lattice sections influence both the Trotter error norm, $W_{\mathrm{h}}$, the parallelizability and the per-Trotter-step gate count. The effect of $W_h$ is however insignificant relative to the total Trotter error norm $W_{\mathrm{tile}}$. Therefore, to construct efficient hopping Hamiltonian decompositions we focus on two aspects: 1) minimizing the number of sections to improve parallelizability and 2) minimizing the number of tiles in each section to reduce gate counts (assuming we use $S_1$, $S_2$, $C_4$ and $S_4)$. These two points are generally not possible to optimize simultaneously. For example, the tiling of the Kagome lattice shown in Fig.~\ref{fig:lattice_tiles_sections} can be achieved using more $S_4$ tiles than the choice demonstrated. This would reduce the overall gate cost of the hopping Hamiltonian but with an additional cost of significantly more sections and reduced parallelizability. Therefore, when choosing efficient tiling of the lattice models, we aim to keep the number of sections to a minimum and try to avoid using $S_1$ tiles.

We choose to divide our periodic hexagonal lattice models into three sections of colors red, blue and gold using $S_2$ tiles as shown for $L_x = L_y=4$ in Fig.~\ref{Fig:hex_lattices}(b). The dashed bonds with labels 1 and 1' indicate a single bond that ensures periodicity of the lattice. This lattice section division ensures that we have $N_{\mathrm{b}}=N_{\mathrm{r}}=N_{\mathrm{g}}=N/4$ $S_2$ tiles in each section, which will be the case for all periodic hexagonal lattice models studied in this paper.

Figs.~\ref{fig:sections,L=6,8,10}(a--c) show the lattice divisions of the periodic hexagonal lattice model for $L=6$, $8$ and $10$, with a lattice section division following the same structure as for the $L=4$ case. 

Note that it is not a requirement that each section contains the same number of tiles, and in fact this does not necessarily give the most optimal implementation. As shown in Eq.~(\ref{eq:fragment_Trotter_step}), one Trotter step of hexagonal lattice Hubbard models with three lattice sections uses two applications of $e^{-iH_h^b t/2}$, two applications of $e^{-iH_h^r t/2}$ and one application of $e^{-iH_h^g t}$. Therefore, the gate costings of a Trotter step might be reduced by putting as many tiles as possible into the gold sections while still maintaining the commutativity properties of the gold section. This optimization has not been implemented for the models considered here, but may be considered for concrete future applications in order to take full advantage of the Tile Trotterization method.

\begin{figure*}
    \centering
    \includegraphics[width=0.7\linewidth]{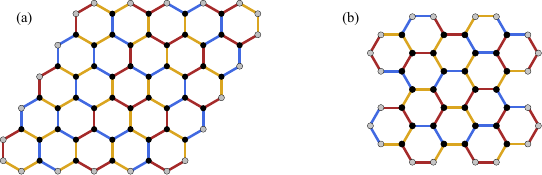}
    \caption{a) Example of a hexagonal lattice fragment covered by $S_2$ tiles in three lattice sections with $N_{ed} = 22$ edge sites (gray) and $N_{c} = 48$ center sites (black). (b) Another example of a hexagonal lattice fragment with a structure inspired by nanographene molecules studied in Ref.~\cite{Song2024HighlyFrustration} which show interesting $\pi$-spin properties. This structure is also covered by $S_2$ tiles in three lattice sections and has $N_{ed} = 20$ edge sites and $N_{c} = 28$ center sites.}
    \label{Fig:arb_hex_fragment}
\end{figure*}

\section{\label{APP:Tile_Trotter_applications} Tile Trotterization applications}

In the main text we considered Tile Trotterization of the extended Hubbard model on a periodic hexagonal lattice. Here, we consider the application of Tile Trotterization for two additional systems: the Hubbard model on a non-periodic hexagonal lattice (hexagonal lattice fragments) and the periodic hexagonal lattice Hubbard model.

Section \ref{sec:TILE_TROTTER_fragments} shows that Tile Trotterization is applicable also to non-periodic lattice models of arbitrary shapes. For this application, we provide gate counts and Trotter error norms for a specific class of hexagonal lattice fragments. Section~\ref{sec:PHUB} presents a simpler version of the main text application and discusses Tile Trotterization of the periodic hexagonal lattice Hubbard model. In our derivation of the gate counts, we consider this periodic model as a special case of the hexagonal lattice fragments in Section~\ref{sec:TILE_TROTTER_fragments}. The results obtained in Section~\ref{sec:PHUB} are used in the quantum phase estimation section of the main text (Section~\ref{sec:Phase_estimation}) to generate the numerical results used in Figs.~\ref{fig:QUBITIZATION_VS_TROTTER_plots} and \ref{fig:QPE_Trotter_HWP_comparison}(a).

\subsection{\label{sec:TILE_TROTTER_fragments} Tile Trotterization of the Hubbard model on hexagonal lattice fragments}

We consider all non-periodic hexagonal lattice fragments with the requirement that all lattice sites are part of at least one full hexagon. These fragments have two distinct types of sites: center sites where a lattice site has three nearest neighbors and edge sites where a lattice site has two nearest neighbors. We denote the number of center sites by $N_c$ and the number of edge sites by $N_{ed}$.

To apply Tile Trotterization, we cover the fragments by $S_2$ tiles and divide them into three sections of colors: blue (b), red (r) and gold (g). Two examples of fragments and the division into sections are shown in Fig.~\ref{Fig:arb_hex_fragment}. We use grey as the color for the edge sites and black as the color for the center sites. Continuing, we use $N_b$, $N_r$ and $N_g$ as the number of $S_2$ tiles used in sections b, r and g.

A single Trotter step of the Hubbard model, defined in Eq.~(\ref{eq:Hubbard_model}), on these fragments is implemented as
\begin{eqnarray}
      e^{-i(H_h+H_I) t} &\approx& e^{-iH_I\frac{t}{2}}e^{-iH_h^b \frac{t}{2}}e^{-iH_h^r \frac{t}{2}}e^{-iH_h^g t} \nonumber \\ &\times& e^{-iH_h^r \frac{t}{2}}e^{-iH_h^b \frac{t}{2}}e^{-iH_I\frac{t}{2}},
     \label{eq:fragment_Trotter_step}
\end{eqnarray}
where we have chosen to implement the sections in the order $H_1 = H_h^b$, $H_2 = H_h^r$ and $H_3 = H_h^g$. Performing $r$ repetitions of this Trotter step gives
\begin{eqnarray}
      \Big( e^{-i(H_h+H_I) t} \Big)^r &\approx& e^{-iH_I \frac{t}{2}} \Big(e^{-iH_h^b \frac{t}{2}}e^{-iH_h^r \frac{t}{2}}e^{-iH_h^g t} \nonumber \\ &\times& e^{-iH_h^r \frac{t}{2}}e^{-iH_h^b t}e^{-iH_It} \Big)^r e^{iH_I\frac{t}{2}},
     \label{eq:fragment_Trotter_step_r_rep}
\end{eqnarray}
so that the cost of one Trotter step for large $r$ is two applications of $e^{-iH_h^b t/2}$, two applications of $e^{-iH_h^r t/2}$, one application of $e^{-iH_h^g t}$ and one application of $e^{-iH_It}$.

Using the Jordan-Wigner transformation, the system is represented by $2N$ qubits and $e^{-iH_It}$ contains $N$ terms, and can be implemented with one layer of $N$ arbitrary Z-axis rotations of the same angle and two layers of $N$ CNOT gates.

The costing of implementing the time evolution of each hopping section is performed by counting the number of applications of $e^{-iH^{S_2}_\sigma t}$ in each section. This Trotter implementation requires $4N_b$, $4N_r$ and $2N_g$ applications of the time evolution operator of the $S_2$ tile Hamiltonian, accounting for both spin sectors. Only counting the non-Clifford gates, the time evolution of the $S_2$ tile Hamiltonian can be implemented using two arbitrary Z-axis rotations and 4 T gates as shown in Appendix~\ref{APP:tile_trotterization}. This leads to a total non-Clifford cost per Trotter step of $N_R$ arbitrary rotations and $N_T$ T gates,
\begin{eqnarray}
    & & N_R = N + 8N_b + 8N_r + 4N_g \label{eq:N_R_fragment} ,\\
    & & N_T = 16N_b + 16N_r + 8N_g \label{eq:N_T_fragment} .
\end{eqnarray}

The Tile Trotterization error norm is evaluated as shown in Eq.~(\ref{eq:Tile_Trotter_error_norm}), using $W_{\mathrm{SO2}}$ and $W_h$. First, we find an expression for $W_{\mathrm{SO2}}$ from Eq.~(\ref{eq:WSO2}), which shows that we only need to evaluate $\lVert [[H_I,H_h],H_I] \rVert$ and $\lVert [[H_I,H_h],H_h] \rVert$. These two commutator bounds are given by Lemma 1 and Lemma 2 of Ref.~\cite{Campbell2022EarlyModel}. The first commutator bound is given by
\begin{equation}
    \big\lVert [[H_I,H_h],H_I] \big\rVert \leq U^2 \lVert H_h \rVert = U^2 \tau \lVert R \rVert_1,
    \label{eq:IhI_fragment}
\end{equation}
where $\lVert \, \cdot \, \rVert_1$ is the Schatten one-norm and $R$ is the adjacency matrix of the fragments. The second commutator bound can be evaluated as 
\begin{equation}
    \big\lVert [[H_I,H_h],H_h] \big\rVert  \le \frac{U}{2} \sum_i \Big( \big\lVert [T_i,H_h] \big\rVert + 2 \lVert T_i \rVert^2 \Big), \label{eq:E6}
\end{equation}
where $T_i$ is an operator containing all hopping terms that interact with site $i$. Throughout this paper, we use the notation $H^i_k$ (instead of $T_i$ as in Ref.~\cite{Campbell2022EarlyModel}) to represent all hopping operators interacting with site $i$, where $k$ denotes the number of nearest neighbors of site $i$. The norm $\lVert H^i_k \rVert$ is only dependent on the character of the specific site $i$. The fragments have two distinct sites: center and edge sites, therefore 
\begin{equation}
    \sum_i 2 \lVert H^i_k \rVert^2= 2\lVert H^i_3 \rVert^2 N_c+2 \lVert H^i_2 \rVert^2N_{ed}.
\end{equation} 
The two norms $\lVert H^i_3 \rVert$ and $\lVert H^i_2 \rVert$ are evaluated as
\begin{eqnarray}
    & & \lVert H^i_3 \rVert = 2\sqrt{3}\tau \label{eq:commboundTC}, \\
    & & \lVert H^i_2 \rVert = 2\sqrt{2}\tau \label{eq:commboundTed},
\end{eqnarray}
using the properties of free fermionic operators shown in Appendix A in Ref.~\cite{Campbell2022EarlyModel}.

The norm of the commutator $\lVert [H^i_k,H_h] \rVert$ is not only dependent on the character of site $i$ but also on neighbor sites of $i$ and can be evaluated for each site $i$ as shown in Appendix A in Ref.~\cite{Campbell2022EarlyModel}. For these models, we find that $\max ( \lVert [H^i_k,H_h] \rVert) =2\sqrt{6} \tau^2$, using the periodic hexagonal lattice model with $L_x=L_y=L=4$ as $H_h$. Note that the norm is independent of lattice size for $L \geq 4$. The maximum value of the commutator norm is found for the center sites that only has center sites as neighbors. We use this to bound $\sum_i \lVert [H^i_k,H_h] \rVert$ as
\begin{equation}
    \sum_i \lVert [H^i_k,H_h] \rVert \leq 2\sqrt{6} \tau^2N, 
\end{equation}
where the equal sign only holds for periodic hexagonal lattices where all sites have $3$ nearest neighbors (with $N_c = N$ and $N_{ed}=0$).

This leads to the final expression for the second commutator bound
\begin{equation}
    \big\lVert [[H_I,H_h],H_h] \big\rVert  \le U \tau^2 \Big( 12N_c + 8N_{ed} +  \sqrt{6}N \Big).
\end{equation}

We obtain $W_h$ using Eq.~(\ref{eq:W}) and the order of the hopping sections in the Trotter step: $H_1 = H_h^b$, $H_2 = H_h^r$ and $H_3 = H_h^g$. Eq.~(\ref{eq:W}) is written out for this splitting in Eq.~(\ref{eq:W_h_h_h_h}) in Appendix~\ref{APP:W_h}. Eq.~(\ref{eq:W_h_written_out_for_RBG}) shows a formula for efficiently calculating $W_h$ for all Tile Trotterization applications where the lattice can be divided into three sections. 

\subsection{\label{sec:PHUB} Tile Trotterization of the Hubbard model on the periodic hexagonal lattice}

We use the periodic hexagonal lattice model and the division into sections shown in Figs.~\ref{Fig:hex_lattices} and \ref{fig:sections,L=6,8,10}, and described in Appendix~\ref{APP:hexagonallatticedivision}. This corresponds to decomposing the hopping Hamiltonian as $H_h = H_h^{\mathrm{b}} + H_h^{\mathrm{r}} + H_h^{\mathrm{g}}$, using $N_{\mathrm{b}}=N_{\mathrm{r}}=N_{\mathrm{g}}=N/4$ $S_2$ tiles in each section.

A single Trotter step of the Hubbard model on the periodic hexagonal lattice is implemented in the same way as for the arbitrary fragments described in Eqs.~(\ref{eq:fragment_Trotter_step})--(\ref{eq:fragment_Trotter_step_r_rep}), so that the cost of one Trotter step (when performing $r$ Trotter steps total) consists of two applications of $e^{-iH_h^b t/2}$, two applications of $e^{-iH_h^r t/2}$, one application of $e^{-iH_h^g t}$ and one application of $e^{-iH_It}$. Therefore, the cost per Trotter step can again be evaluated as in Eqs.~(\ref{eq:N_R_fragment})--(\ref{eq:N_T_fragment}), but for the periodic case we can replace the number of tiles in each section by $N/4$, simplifying the non-Clifford cost to
\begin{eqnarray}
    & & N_R = 6N, \label{eq:N_R_periodic} \\
    & & N_T = 10N. \label{eq:N_T_periodic}
\end{eqnarray}

Arbitrary rotations are expensive to perform on fault-tolerant quantum computers because such operations cannot be protected for arbitrary rotation angles. Hamming weight phasing (HWP) \cite{Gidney2017HalvingAddition, Kivlichan2020ImprovedTrotterization, Campbell2022EarlyModel} is a method to reduce the number of arbitrary rotations required, which can be applied in the case where many rotations of the same angle are performed in parallel. HWP allows for trading $m$ arbitrary rotations of the same angle with $m-1$ clean ancilla qubits and $m-1$ Toffoli gates. The Hubbard model time evolution contains $6$ layers of $N$ arbitrary rotations, where the angle of all gates within a given layer are the same. We choose to implement $m$ of these arbitrary rotations simultaneously, where $N$ is an integer multiple of $m$, which reduces the number of arbitrary rotations per layer to $N_R = \lfloor \log_2(m) + 1 \rfloor$ using $N_{\mathrm{Tof}} = m-1$ Toffoli gates and $\alpha = m-1$ clean ancilla qubits. To implement the total amount of 6$N$ arbitrary rotations, we need $\frac{6N}{m}$ layers of $m$ arbitrary rotations. This leads to a total gate count per Trotter step of
\begin{eqnarray}
    & & N_R = \frac{6N}{m}\lfloor \log_2(m) + 1 \rfloor, \label{eq:periodic_Hubbard_N_R} \\
    & & N_{\mathrm{Tof}} = \frac{6N}{m}(m-1), \\
    & & N_T = 10N.
\end{eqnarray}
Each Toffoli gate can be converted into 4 T gates to obtain a total T gate cost of 
\begin{eqnarray}
    N_T = 10N + 4N_{\mathrm{Tof}} = 10N+4 \times \frac{6N}{m}(m-1)  .
    \label{eq:Periodic_Hubbard_N_T}
\end{eqnarray}
Given that we choose $m$ as a fraction of $N$, the number of arbitrary rotations per Trotter step scales logarithmically in $N$ and the number of T gates scale linearly in $N$. The total qubit count needed for this Tile Trotterization implementation is $2N+(m-1)$. 

Note that choosing HWP with $m<N$ reduces the parallelization of arbitrary rotations in each Trotter step because only $m$ arbitrary rotations can be performed simultaneously. Additionally, choosing $m=N$ comes with a significant additional qubit cost so $m=\frac{N}{4}$ or $m=\frac{N}{2}$ might be more advantageous depending on the available resources.

The Tile Trotterization error norm is again evaluated from Eq.~(\ref{eq:Tile_Trotter_error_norm}), and $W_{\mathrm{SO2}}$ can be obtained from the commutator bounds given by Lemma~1 and Lemma~2 in Appendix~C of Ref.~\cite{Campbell2022EarlyModel}. This leads to the following bounds for the periodic hexagonal lattice
\begin{eqnarray}
    \big\lVert [[H_I,H_h],H_I] \big\rVert \leq U^2 \lVert H_h \rVert = U^2 \tau \lVert R \rVert_1,
    \label{eq:comm_bound_IhI_campbell}
\end{eqnarray}
where $\lVert \, \cdot \, \rVert_1$ is the Schatten one-norm and $R$ is the adjacency matrix of the periodic hexagonal lattice. We note that $\lVert R \rVert_1$ has at worst case linear scaling in $N$. Using the strategies provided by Ref.~\cite{Campbell2022EarlyModel}, we can bound the second commutator by
\begin{eqnarray}
    \big\lVert [[H_I,H_h],H_h] \big\rVert &\leq& \frac{U}{2} \sum_i \Big( \big\lVert [H_k^i,H_h] \big\rVert + 2 \lVert H_k^i \rVert^2 \Big) \nonumber \\
    &=& (12+\sqrt{6})U\tau^2N. \label{eq:comm_bound_Ihh_campbell}
\end{eqnarray}
Using that for all sites $i$ in the periodic hexagonal lattice
\begin{eqnarray}
    & & \lVert H^i_3 \rVert = 2\sqrt{3}\tau, \\
    & & \lVert [H^i_3,H_h] \rVert = 2\sqrt{6}\tau^2 .\label{eq:E20}
\end{eqnarray}
We used the adjacency matrix of the periodic hexagonal lattice model with $L_x=L_y=L=4$ to evaluate Eq.~(\ref{eq:E20}). We note that $[H_k^i,H_h]$ is a local operator and its norm remains constant for $L \geq 4$. 

In this paper, we introduce a new way of evaluating the commutator bound $\lVert [[H_I,H_h],H_h] \rVert$ that improves upon the commutator bound given by Lemma 2 in Ref.~\cite{Campbell2022EarlyModel}. In Appendix~\ref{APP:COMMUTATORBOUNDS}, we introduce Lemma~\ref{Lemma:Hubbard_Ihh} that can be applied to obtain the following bound for the periodic hexagonal Hubbard model 
\begin{equation}
    \big\lVert [[H_I,H_h],H_h] \big\rVert \leq 9.9 U\tau^2N, \label{eq:comm_bound_Ihh_new}
\end{equation}
which is a $31.5\%$ improvement compared to the bound given in Eq.~(\ref{eq:comm_bound_Ihh_campbell}). For the QPE resource estimation results presented in Fig.~\ref{fig:QUBITIZATION_VS_TROTTER_plots} and Fig.~\ref{fig:QPE_Trotter_HWP_comparison}(a), we use the $\lVert [[H_I,H_h],H_I] \rVert$ given by Eq.~(\ref{eq:comm_bound_IhI_campbell}) and the $\lVert [[H_I,H_h],H_h] \rVert$ bound given by Eq.~(\ref{eq:comm_bound_Ihh_new}).

The lattice is divided into three sections and we can therefore obtain $W_h$ using Eq.~(\ref{eq:W_h_written_out_for_RBG}) in Appendix~\ref{APP:W_h}. The hopping Hamiltonian Trotter error norm for the periodic hexagonal lattice division used in this paper scales linearly with $N$, as shown in Fig.~\ref{fig:W_h_scaling}. In the numerical examples for the hexagonal lattice periodic Hubbard model given in this paper with $U=4$ and $\tau=1$, we find that $W_h$ constitutes around $16\%$ of the total Trotter error norm $W_{\mathrm{tile}}$.

The results presented in this section are used to obtain the QPE resource estimates for the Hubbard model in Sec.~\ref{sec:Phase_estimation}.

\section{\label{APP:COMMUTATORBOUNDS} Commutator bounds}

The Tile Trotterization error norm of generalized Hubbard models is a sum of $W_{\mathrm{SO2}}$ and $W_h$. In this appendix, we show how to evaluate commutator bounds used to calculate $W_{\mathrm{SO2}}$. The $W_{\mathrm{SO2}}$ Trotter error is evaluated using the expression
\begin{equation}
      W_{\mathrm{SO2}} = \frac{1}{24} \lVert [[H_C,H_h],H_C] \rVert + \frac{1}{12} \lVert [[H_C,H_h],H_h] \rVert.
\end{equation}
The generalized Hubbard models used throughout this paper are defined as
\begin{equation}
    H = H_h + H_C,
\end{equation}
with the Coulomb term given by $H_C=H_I$ for the Hubbard model and $H_C=H_I+H_V$ for the extended Hubbard model. In the commutator bound derivations for the extended Hubbard model, we limit ourselves to lattices where all sites have the same number of nearest neighbors ($k$-regular graphs), which allows us to use the modified interaction term defined in Eq.~(\ref{eq:HV'}). For the commutator bounds presented here, we use the following definitions of the terms $H_h$, $H_I$ and $H_V$
\begin{eqnarray}
    & &H_h = -\tau \sum_{i,j,\sigma} R_{ij} a^{\dagger}_{i\sigma} a_{j\sigma}, \label{eq:H_h_appendix} \\
    & &H_I = \frac{U}{4} \sum_{i=1}^N Z_{i\uparrow} Z_{i\downarrow}, \label{eq:H_I_appendix} \\
    & &H_{V}  = \frac{V}{4} \sum_{\langle ij \rangle} \sum_{\sigma,\sigma'} Z_{i\sigma}Z_{j\sigma'}. \label{eq:H_V_appendix}
\end{eqnarray}
For $i\neq j$, we also define the hopping operator $B_{ij\sigma}$ (hopping term from spin orbital $i \sigma$ to spin orbital $j \sigma$) as
\begin{equation}
    B_{ij\sigma} = - \tau a^{\dagger}_{i\sigma} a_{j\sigma},
    \label{eq:BIJsigma}
\end{equation}
and define $B_{ij}$ as 
\begin{equation}
    B_{ij} = \sum_{\sigma \in \{ \uparrow,\downarrow \} } B_{ij\sigma}.
    \label{eq:BIJ}
\end{equation}
Continuing, we write the sum over spins as $\sum_{\sigma \in \{ \uparrow,\downarrow \} } = \sum_{\sigma}$. Using the $B_{ij}$ notation, we can write $H_h$ as
\begin{eqnarray}
    & &H_h  = \sum_{\langle ij \rangle} (B_{ij}+B_{ji}), \label{eq:H_h_defined_from_B_ij}
\end{eqnarray}
where $\sum_{\langle ij \rangle}$ runs over $\frac{k}{2}N$ nearest-neighbor terms for all lattices where each lattice site has $k$ nearest neighbors.

Before evaluating the commutator bound $\lVert [[H_I,H_h], H_h] \rVert$ in Section~\ref{sec:Lemma1}, $\lVert [[H_C,H_h], H_h] \rVert$ in Section~\ref{sec:Lemma2} and $\lVert [[H_C,H_h], H_C] \rVert$ in Section~\ref{sec:Lemma3}, we establish a set of commutator and anti-commutator rules used for the derivations (Section~\ref{sec:APP_COMM_RULES}). In Section~\ref{sec:Numerical_results}, we describe our procedure for evaluating the spectral norms. Finally, in Section~\ref{sec:1D_Hubbard_bounds}, we evaluate the commutator bounds of interest exactly for 1D Hubbard and extended Hubbard model periodic chain to give some intuition for the tightness of the bounds.

\begin{widetext}

\subsection{Commutator and anti-commutator rules} \label{sec:APP_COMM_RULES}

In the following derivations, we will need to consider the commutator and anti-commutators between Coloumb terms (of the form $Z_{i\sigma} Z_{j \sigma'}$ for $i\sigma \ne j \sigma'$) and hopping terms (of the form $B_{ij\sigma} = - \tau a^{\dagger}_{i\sigma} a_{j\sigma}$ for $i \ne j$).

To aid in this task, we establish a set of commutator and anti-commutator rules,
\begin{eqnarray}
    & & [Z_{m\sigma_1}Z_{n\sigma_2},B_{ij\sigma}] = 0 \;\; \textrm{if spin orbitals $m\sigma_1$, $n\sigma_2$, $i\sigma$ and $j\sigma$ are all different}, \label{eq:comm_rule_1}
    \\ & & [Z_{i\sigma} Z_{j\sigma}, B_{ij\sigma}] = 0, \label{eq:comm_rule_2}
    \\ & & \{ Z_{i\sigma}Z_{l\sigma'},B_{ij\sigma} \} = 0 \;\; \textrm{if spin orbitals $i\sigma$, $j\sigma$ and $l\sigma'$ are all different}, \label{eq:comm_rule_3}
    \\ & & \{ Z_{i\sigma}Z_{l\sigma'},B_{ji\sigma} \} = 0 \;\; \textrm{if spin orbitals $i\sigma$, $j\sigma$ and $l\sigma'$ are all different}. \label{eq:comm_rule_4}
    \label{eq:comm_rules}
\end{eqnarray}

In words, if the spin orbitals in the Coulomb term are both the same or both different to the spin orbitals in the hopping term, then the \emph{commutator} is zero. In contrast, if one and only one of the spin orbitals is shared between the Coulomb and hopping term, then the \emph{anti-commutator} is zero. This gives the intuition for these results, which we now proceed to derive.

The commutator in Eq.~(\ref{eq:comm_rule_1}) is trivially true because all operators in the expression act on different spin orbitals. Eqs.~(\ref{eq:comm_rule_2}) and (\ref{eq:comm_rule_3}) can be shown by using the following anti-commutation relation that applies for $\sigma \neq \sigma'$ or $i \neq j$,
\begin{equation}
    \{ Z_{i\sigma} Z_{j\sigma'}, a_{i\sigma}^\dagger \} = \{ (2n_{i\sigma}-\mathbb{1})(2n_{j\sigma'}-\mathbb{1}), a_{i\sigma}^\dagger \} = (2n_{j\sigma'}-\mathbb{1})\{(2n_{i\sigma}-\mathbb{1}), a_{i\sigma}^\dagger \} = 2(2n_{j\sigma'}-\mathbb{1})( \{n_{i\sigma}, a_{i\sigma}^\dagger \}-a_{i\sigma}^\dagger).
    \label{eq:commutator_rule_derivation_1}
\end{equation}
This expression can be evaluated by calculating the anti-commutator
\begin{equation}
    \{n_{i\sigma}, a_{i\sigma}^\dagger \} = \{a_{i\sigma}^\dagger a_{i\sigma}, a_{i\sigma}^\dagger \} = a_{i\sigma}^\dagger a_{i\sigma}a_{i\sigma}^\dagger+a_{i\sigma}^\dagger a_{i\sigma}^\dagger a_{i\sigma} = a_{i\sigma}^\dagger a_{i\sigma}a_{i\sigma}^\dagger = (\mathbb{1}-a_{i\sigma}a_{i\sigma}^\dagger) a_{i\sigma}^\dagger = a_{i\sigma}^\dagger
\end{equation}
using $\{ a_{i\sigma}^\dagger, a_{i\sigma} \} = \mathbb{1}$. Inserting this result back into Eq.~(\ref{eq:commutator_rule_derivation_1}), we obtain
\begin{equation}
     \{ Z_{i\sigma} Z_{j\sigma'}, a_{i\sigma}^\dagger \} = 0,
     \label{eq:anti_comm_rule_a_dagger}
\end{equation}
which can also be shown for $a_{i\sigma}$ instead of $a_{i\sigma}^\dagger$  to obtain
\begin{equation}
    \{ Z_{i\sigma} Z_{j\sigma'}, a_{i\sigma} \} = 0.
    \label{eq:anti_comm_rule_a}
\end{equation}
Now we can prove Eq.~(\ref{eq:comm_rule_2}) using the identity $[A,BC] = \{A,B\}C - B\{A,C\}$ and Eqs.~(\ref{eq:anti_comm_rule_a_dagger})--(\ref{eq:anti_comm_rule_a})
\begin{equation}
    [Z_{i\sigma} Z_{j\sigma}, B_{ij\sigma}] = -\tau [Z_{i\sigma} Z_{j\sigma}, a_{i\sigma}^\dagger a_{j\sigma}] = -\tau \Big( \, \{Z_{i\sigma} Z_{j\sigma},a_{i\sigma}^\dagger\}a_{j\sigma} - a_{i\sigma}^\dagger\{Z_{i\sigma} Z_{j\sigma},a_{j\sigma}\} \, \Big) = 0.
\end{equation}
Eq.~(\ref{eq:comm_rule_3}) can be shown using Eq.~(\ref{eq:anti_comm_rule_a_dagger}) and that, under the conditions given, $Z_{i\sigma}Z_{l\sigma'}$ commutes with $a_{j\sigma}$:
\begin{eqnarray}
    \{ Z_{i\sigma}Z_{l\sigma'},B_{ij\sigma} \} = -\tau \{ Z_{i\sigma}Z_{l\sigma'}, a_{i\sigma}^\dagger a_{j\sigma} \}  = -\tau \{ Z_{i\sigma}Z_{l\sigma'}, a_{i\sigma}^\dagger \} a_{j\sigma} = 0.
\end{eqnarray}
Finally, Eq.~\eqref{eq:comm_rule_3} is equivalent to Eq.~\eqref{eq:comm_rule_4} by taking the Hermitian conjugate
\begin{equation}
    \{Z_{i\sigma}Z_{l\sigma'}, B_{ij\sigma}\} = 0\ \Leftrightarrow \{Z_{i\sigma}Z_{l\sigma'}, B_{ji\sigma}\}=0.
\end{equation}

\subsection{Evaluating spectral norms numerically}
\label{sec:Numerical_results}

In the following sections we will derive lemmas that can be used to bound Trotter error for Hubbard models, calculated as spectral norms of nested commutators. These lemmas take a nested commutator and partition it into a sum of operators acting on a reduced number of spin orbitals, or qubits. The triangle inequality is then applied to bound the Trotter error as a sum of spectral norms of these reduced operators. The spectral norms of the reduced operators are calculated numerically.

Our numerical calculations of the spectral norm are obtained using the density matrix renormalization group (DMRG) algorithm, using the Block2 \cite{Zhai2023} code. For each result, we construct the desired operator $O$ using OpenFermion \cite{McClean_2020}, converting it to a qubit operator using the Jordan-Wigner transformation. We ran Block2 in Pauli mode where quantum numbers are not used. Since the operators under consideration are always Hermitian, the spectral norm is equal to the maximum of the absolute values of the eigenvalues of $O$. We can calculate this numerically by finding the lowest eigenvalue of both $O$ and $-O$, and taking the maximum of the absolute values of the two. For operators acting on a sufficiently small number of qubits, this can be achieved using exact eigenvalue solvers. In the following we consider somewhat larger partitions such that this is not possible, and instead use the DMRG method.

It is possible that DMRG may fail to converge, or else converge to a local minimum. In general, it is also necessary to ensure the maximum singular value is found over all symmetry sectors. We include all input and output files on Zenodo (see data availability statement) for Block2 simulations to demonstrate convergence, and the simulation parameters used. We additionally verified this methodology by comparing DMRG results to those from exact eigenvalue solvers for smaller partitions. In testing, we also performed DMRG on operators in both fermionic and qubit operator representations, ensuring consistent results between the two.

We next proceed to derive lemmas to bound Trotter errors in the Hubbard and extended Hubbard models.

\subsection{Hubbard model: $\mathbf{\big\lVert [[H_I,H_h],H_h] \big\rVert}$} \label{sec:Lemma1}
In this section, we show that the commutator bound: $[[H_I,H_h],H_h]$ from Ref.~\cite{Campbell2022EarlyModel}, given in Eq.~(\ref{eq:comm_bound_Ihh_campbell}), can be further tightened by partitioning the nested commutator into a sum of operators acting on small subsystems for which we can compute the spectral norm exactly. We consider the commutator bound for $[[H_I,H_h],H_h]$, which can be expressed as
\begin{equation}
    [[H_I,H_h],H_h] = \frac{U}{4} \sum_{i} [[Z_{i\uparrow}Z_{i\downarrow},H_h],H_h].
    \label{eq:Hubbard_bound_improvement1}
\end{equation}
We wish to provide a tight bound on the spectral norm of this nested commutator. Evaluating the bound exactly is computationally intractable for all but the smallest lattices, as the operator acts on all $2N$ qubits. Instead, we choose to partition the expression into smaller sets, for which the spectral norm can be evaluated exactly. The simplest approach would be taking the norm of Eq.~(\ref{eq:Hubbard_bound_improvement1}) and applying the triangle inequality,
\begin{equation}
    \Big\lVert [[H_I,H_h],H_h] \Big\rVert \leq \frac{U}{4}\sum_{i} \Big\lVert [[Z_{i\uparrow}Z_{i\downarrow},H_h],H_h] \Big\rVert.
\end{equation}
However, for some lattices we may also be able to evaluate the spectral norm for larger sets of operators, acting on more qubits. In this case, we may include a set of neighboring on-site interactions and define a partition of this set. We write $P=\{ X_1, X_2, ... \}$ such that each $X_i$ is a set which may contain more than one lattice site. The triangle inequality then yields
\begin{equation}
    \Big\lVert [[H_I,H_h],H_h] \Big\rVert \leq \frac{U}{4} \sum_{X \in P} \Big\lVert \sum_{i \in X} [[Z_{i\uparrow}Z_{i\downarrow},H_h],H_h] \Big\rVert.
\end{equation}
In the derivation of the commutator bound expression $\lVert [[H_I,H_h],H_h] \rVert$ (as well as $\lVert [[H_C,H_h],H_h] \rVert$ for the extended Hubbard model), it is useful to consider the following local hopping operator around site $i$
\begin{equation}
    H^i_k = \sum_{l \sim i} (B_{il} + B_{li}),
    \label{eq:def:Hik}
\end{equation}
where $\sum_{l \sim i}$ is a sum over $l$ neighbor to $i$. The operator $H^i_k$ contains all hopping terms between site $i$ and its nearest neighbors, with the subscript indicating the number of nearest neighbours $k$.

\begin{lemma} \label{Lemma:Hubbard_Ihh} For a Hubbard model Hamiltonian with $H_{H} = H_h + H_I$, defined in Eqs.~(\ref{eq:H_h_appendix})--(\ref{eq:H_I_appendix}),
\begin{eqnarray}
    \big\lVert \, [[H_I ,H_h],H_h] \, \big\rVert \leq \frac{U}{2} \sum_{X \in P} \Big\lVert \sum_{i \in X} Z_{i\uparrow} Z_{i\downarrow}([H^i_{k}, H_h]+2(H^i_{k})^2) \Big\rVert.
     \label{eq:Lemma1}
\end{eqnarray}
\end{lemma}

\begin{proof}
We evaluate the nested commutator as 
\begin{equation}
    [[H_I,H_h],H_h] =\frac{U}{4} \sum_{X \in P} [[ \sum_{i \in X} Z_{i\uparrow}Z_{i\downarrow}, H_h],H_h] .
    \label{eq:HiHhHh_commutator}
\end{equation}
We begin by evaluating the commutator $[ \sum_{i \in X} Z_{i\uparrow}Z_{i\downarrow}, H_h]$ using the commutator and anti-commutator relations in Eqs.~(\ref{eq:comm_rule_1})--(\ref{eq:comm_rule_4})
\begin{equation}
    \sum_{i \in X}[Z_{i\uparrow}Z_{i\downarrow},H_h] = \sum_{i \in X} \Big( 2Z_{i\uparrow}Z_{i\downarrow}H_k^i \Big) .
    \label{eq:commutator_ZiZiZjZjH_h}
\end{equation}
We nest the commutator with $H_h$ and use the identity $[AB,C]=A[B,C]+[A,C]B$ 
\begin{equation}    \sum_{i \in X}[[Z_{i\uparrow}Z_{i\downarrow},H_h],H_h] = \sum_{i \in X} \Big( 2Z_{i\uparrow}Z_{i\downarrow} [H_k^i,H_h] + 2 [Z_{i\uparrow}Z_{i\downarrow},H_h] H^i_k  \Big)  .
\end{equation}
Then, we reapply Eq.~(\ref{eq:commutator_ZiZiZjZjH_h}) and simplify the expression
\begin{equation}
    \sum_{i \in X}[[Z_{i\uparrow}Z_{i\downarrow},H_h],H_h] = 2\sum_{i \in X} Z_{i\uparrow}Z_{i\downarrow} \Big( [H_k^i,H_h] + 2 (H^i_k)^2\Big)  .
    \label{eq:28orb}
\end{equation}
Next, we multiply the above expression by $U/4$, sum over the partitions, take the spectral norm and apply the triangle inequality, proving Lemma \ref{Lemma:Hubbard_Ihh}.
\end{proof}

When making use of Lemma~\ref{Lemma:Hubbard_Ihh} in practice, it is useful to define the active space of the operator on the right-hand side of Eq.~(\ref{eq:28orb}). To do this, note that the surviving terms of $[H^i_{k},H_h]$ will be localized around site $i$ up to next-nearest neighbors.

\begin{numresult}
\textbf{Periodic hexagonal lattice:} 
We choose to partition the periodic hexagonal lattice sites into $N/4$ equivalent parts containing four lattice sites in each $X$ with an $S_3$ star graph connectivity. The resulting $S_3$ partition operator is a 38 spin-orbital operator and we compute its norm numerically using DMRG as described in Section~\ref{sec:Numerical_results}, resulting in the following spectral norm of the $S_3$ partition operator
\begin{equation}
 \Big\lVert \sum_{i \in X} Z_{i\uparrow} Z_{i\downarrow}([H^i_{3,\uparrow}, H_h]+2(H^i_{3})^2) \Big\rVert \leq 79.1 \tau^2.
\end{equation}
This commutator bound is identical for all $S_3$ partitions meaning that the entire commutator bound $\lVert [[H_I,H_h],H_h] \rVert$ can be evaluated by summing over $N/4$ $S_3$ partition commutator bounds, resulting in
\begin{equation}
 \Big\lVert [[H_I,H_h],H_h] \Big\rVert \leq \frac{U}{2} \cdot \frac{N}{4}  79.1 \tau^2 =9.9U\tau^2 N. 
 \label{eq:cor:hex_lattice_Hubbard}
\end{equation}
Note that this result is valid for all periodic hexagonal lattices with $L_x = L_y = L \geq 4$ as defined in Appendix~\ref{APP:hexagonallatticedivision}. The commutator bound in Eq.~(\ref{eq:cor:hex_lattice_Hubbard}) improves upon the commutator bound that can be obtained using Lemma 2 from Ref.~\cite{Campbell2022EarlyModel} by $31.5\%$.
\end{numresult}

\subsection{Extended Hubbard model: $\mathbf{\big\lVert [[H_C,H_h],H_h] \big\rVert}$} \label{sec:Lemma2}

We begin by rewriting the on-site interaction term as 
\begin{equation}
    H_I = \frac{U}{4}\sum^N_{i=1} Z_{i\uparrow}Z_{i\downarrow} = \frac{U}{4k}\sum_{\langle ij \rangle} (Z_{i\uparrow}Z_{i\downarrow} + Z_{j\uparrow} Z_{j\downarrow}),
\end{equation}
where $k$ is the number of nearest neighbors of each lattice site and $\sum_{\langle ij \rangle}$ runs over $kN/2$ nearest-neighbor terms. This allows us to write the full Coulomb term as a sum over lattice bonds, as
\begin{equation}
    H_C = \sum_{\langle ij \rangle} \Big(  \frac{U}{4k}(Z_{i\uparrow}Z_{i\downarrow} + Z_{j\uparrow} Z_{j\downarrow}) + \frac{V}{4} \sum_{\sigma,\sigma'} Z_{i\sigma}Z_{j\sigma'} \Big).
    \label{eq:H_C_new}
\end{equation}
We define a constant, $\alpha = V/U$, and write the Coulomb term as
\begin{equation}
    H_C = \sum_{\langle ij \rangle} \Big(  \frac{U}{4k}(Z_{i\uparrow}Z_{i\downarrow} + Z_{j\uparrow} Z_{j\downarrow}) + \frac{\alpha U}{4} \sum_{\sigma,\sigma'} Z_{i\sigma}Z_{j\sigma'} \Big) = \frac{U}{4}\sum_{\langle ij \rangle} \Big(  \frac{1}{k}(Z_{i\uparrow}Z_{i\downarrow} + Z_{j\uparrow} Z_{j\downarrow}) + \alpha \sum_{\sigma,\sigma'} Z_{i\sigma}Z_{j\sigma'} \Big)
    \label{eq:H_C_new_2}
\end{equation}

Once again, to provide a bound on the spectral norm of this operator, we partition the lattice into smaller sets, for which we can evaluate the spectral norm exactly. The simplest partition would be the following choice,
\begin{equation}
    \Big\lVert [[H_C,H_h],H_h] \Big\rVert \leq \frac{U}{4} \sum_{\langle ij \rangle} \Big\lVert [[\frac{1}{k}(Z_{i\uparrow}Z_{i\downarrow} + Z_{j\uparrow} Z_{j\downarrow}) + \alpha \sum_{\sigma,\sigma'} Z_{i\sigma}Z_{j\sigma'},H_h],H_h] \Big\rVert.
\end{equation}
To further tighten this bound, we may include multiple nearest-neighbour interaction terms inside the norm. We define the partitioning $P=\{ X_1, X_2, ...\}$ such that $X_i$ may contain more than one nearest-neighbor pair of sites. This allows us to write a bound which will be tighter in general,
\begin{equation}
    \Big\lVert [[H_C,H_h],H_h] \Big\rVert \leq  \frac{U}{4} \sum_{X \in P} \Big\lVert \sum_{(i,j) \in X}[[\frac{1}{k}(Z_{i\uparrow}Z_{i\downarrow} + Z_{j\uparrow} Z_{j\downarrow}) + \alpha \sum_{\sigma,\sigma'} Z_{i\sigma}Z_{j\sigma'},H_h],H_h] \Big\rVert.
\end{equation}
We also define two additional local hopping operators around lattice site $i$, similarly to the operator defined in Eq.~(\ref{eq:def:Hik}),
\begin{equation}
     H^i_{k,\sigma} = \sum_{l \sim i} (B_{il\sigma} + B_{li\sigma}) \; \; \; , \; \; \, H^{ij}_{k-1,\sigma} = \sum_{l \sim i, l \neq j} (B_{il\sigma} + B_{li\sigma}), 
    \label{eq:def:Hijkminus1}
\end{equation}
where $H^i_{k\sigma}$ contain all hopping terms of spin-$\sigma$ between site $i$ and its $k$ nearest neighbors and $ H^{ij}_{k-1,\sigma}$ contains hopping terms of spin-$\sigma$ between site $i$ and $k-1$ of its neighbors (all hopping terms except for the terms between neighboring sites $i$ and $j$). The local hopping operators $H^i_k$, $H^i_{k,\sigma}$ and $H^{ij}_{k-1,\sigma}$ are used along with the rewritten Coulomb term in Eq.~(\ref{eq:H_C_new}) in the following lemma on the spectral norm of the nested commutator $[[H_C, H_h],H_h]$, for extended Hubbard models.

\begin{lemma} \label{Lemma:Extended_Hubbard_Chh} For an extended Hubbard model Hamiltonian $H_{H} = H_h + H_I+H_V$, defined in Eqs.~(\ref{eq:H_h_appendix})--(\ref{eq:H_V_appendix}), with $\alpha = V/U$, on a lattice with $N$ lattice sites where all sites have $k$ nearest neighbors, then
\begin{eqnarray}
    \Big \lVert [[H_C,H_h],H_h] \Big\rVert &\leq& \frac{U \alpha}{2} \sum_{X \in P} \Big\lVert \sum_{(i,j) \in X} \Big (\frac{1}{k \alpha}Z_{i\uparrow}Z_{i\downarrow}([H_k^i,H_h]+2(H_k^i)^2) + \frac{1}{k \alpha}Z_{j\uparrow}Z_{j\downarrow}([H_k^j,H_h]+2(H_k^j)^2) \nonumber \\ &+& Z_{i\uparrow}Z_{j\downarrow}([H^{i}_{k,\uparrow} + H^{j}_{k,\downarrow},H_h]+2(H^{i}_{k,\uparrow} + H^{j}_{k,\downarrow})^2) + Z_{i\downarrow}Z_{j\uparrow}([H^{i}_{k,\downarrow} + H^{j}_{k,\uparrow},H_h]+2(H^{i}_{k,\downarrow} + H^{j}_{k,\uparrow})^2) \nonumber \\ &+& \sum_\sigma Z_{i\sigma}Z_{j\sigma}([H^{ij}_{k-1,\sigma} + H^{ji}_{k-1,\sigma},H_h]+2(H^{ij}_{k-1,\sigma} + H^{ji}_{k-1,\sigma})^2) \Big) \Big\rVert.
\label{eq:Lemma2}
\end{eqnarray}
\end{lemma}

\begin{proof}
We evaluate the nested commutator as
\begin{equation}
    [[H_C,H_h],H_h] =  \frac{U}{4} \sum_{X \in P} \Big[ \big[ \sum_{(i,j) \in X} \Big (\frac{1}{k}(Z_{i\uparrow}Z_{i\downarrow} + Z_{j\uparrow} Z_{j\downarrow}) + \alpha \sum_{\sigma,\sigma'} Z_{i\sigma}Z_{j\sigma'} \Big),H_h \big], H_h \Big].
\end{equation}
We first evaluate the commutator of the Coulomb terms and $H_h$ using the commutator and anti-commutator relations in Eqs.~(\ref{eq:comm_rule_1})--(\ref{eq:comm_rule_4}),
\begin{eqnarray}
    [ \sum_{(i,j) \in X} \Big (\frac{1}{k}(Z_{i\uparrow}Z_{i\downarrow} + Z_{j\uparrow} Z_{j\downarrow}) + \alpha \sum_{\sigma,\sigma'} Z_{i\sigma}Z_{j\sigma'} \Big),H_h ] &=&  2 \alpha
    \sum_{(i,j) \in X} \Big (\frac{1}{k \alpha}(Z_{i\uparrow}Z_{i\downarrow}H_k^i + Z_{j\uparrow} Z_{j\downarrow}H_k^j) +  Z_{i\uparrow}Z_{j\downarrow}(H^{i}_{k,\uparrow} + H^{j}_{k,\downarrow}) \nonumber \\ &+& Z_{i\downarrow}Z_{j\uparrow}(H^{i}_{k,\downarrow} + H^{j}_{k,\uparrow}) + \sum_\sigma Z_{i\sigma}Z_{j\sigma}(H^{ij}_{k-1,\sigma} + H^{ji}_{k-1,\sigma}) \Big),
\end{eqnarray}
where we used the $H^i_k$, $H^i_{k,\sigma}$ and $H^{ij}_{k-1,\sigma}$ defined in Eqs.~(\ref{eq:def:Hik}) and (\ref{eq:def:Hijkminus1}) to simplify the expressions. Note that for the two interaction terms between spin orbitals of same spin, the commutation relation in Eq.~(\ref{eq:comm_rule_2}) shows that the hopping terms between those spin orbitals commutes with the interaction term, leading to the difference between opposite spin interaction terms and same spin interaction terms in the commutator.

We nest the above commutator with $H_h$ and then sum over the partitions to obtain $[[H_C, H_h],H_h]$. We use the identity $[AB,C]=A[B,C]+[A,C]B$ and reapply the above equation, leading to
\begin{eqnarray}
    [[H_C,H_h],H_h] &=& \frac{U \alpha}{2} \sum_{X \in P} \sum_{(i,j) \in X} \Big (\frac{1}{k \alpha}Z_{i\uparrow}Z_{i\downarrow}([H_k^i,H_h]+2(H_k^i)^2) + \frac{1}{k \alpha}Z_{j\uparrow}Z_{j\downarrow}([H_k^j,H_h]+2(H_k^j)^2) \nonumber \\ &+& Z_{i\uparrow}Z_{j\downarrow}([H^{i}_{k,\uparrow} + H^{j}_{k,\downarrow},H_h]+2(H^{i}_{k,\uparrow} + H^{j}_{k,\downarrow})^2) + Z_{i\downarrow}Z_{j\uparrow}([H^{i}_{k,\downarrow} + H^{j}_{k,\uparrow},H_h]+2(H^{i}_{k,\downarrow} + H^{j}_{k,\uparrow})^2) \nonumber \\ &+& \sum_\sigma Z_{i\sigma}Z_{j\sigma}([H^{ij}_{k-1,\sigma} + H^{ji}_{k-1,\sigma},H_h]+2(H^{ij}_{k-1,\sigma} + H^{ji}_{k-1,\sigma})^2) \Big).
    \label{eq:HCHhHh}
\end{eqnarray}
We take the norm and apply the triangle inequality.
\end{proof}

We wish to evaluate the spectral norm of the $X$-subspace numerically. Therefore, it is useful to define the active space of the operator on the right-hand side of Eq.~(\ref{eq:HCHhHh}). The surviving terms of the commutator $[H^i_{k},H_h]$ contain operators acting on lattice sites localized around site $i$ and $j$ up to next-nearest neighbors. This allows us to choose a partition of the entire lattice of which the spectral norm of each individual and localized part can be evaluated exactly.

\begin{numresult}
\textbf{Periodic hexagonal lattice:}
We choose to partition the lattice into $S_3$ parts such that each $X$ contains three $(i,j)$ pairs forming an $S_3$ star graph. We have to sum over $N/2$ completely equivalent partitions to obtain the entire commutator bound. We evaluate the norm of the $S_3$ partition for parameters $V/U = \alpha = 1/2$ and $k=3$ (each site in the periodic hexagonal lattice has 3 neighbors). The resulting operator acts on a $38$ spin-orbital or qubit subspace and we compute its spectral norm numerically using DMRG as described in Section~\ref{sec:Numerical_results}, resulting in the following bound
\begin{eqnarray}
        & & \lVert [[H_C,H_h],H_h] \rVert \Big\rvert_{\frac{U}{V}=2} \leq 12.5 U \tau^2 N.
        \label{eq:Chh_Hex_lattice_numerical_norm}
\end{eqnarray}    
Note that this result is valid for all periodic hexagonal lattices with $L_x = L_y = L \geq 4$ as defined in Appendix~\ref{APP:hexagonallatticedivision}.

\end{numresult}

\subsection{Extended Hubbard model: $\mathbf{\big\lVert [[H_C,H_h],H_C] \big\rVert}$} \label{sec:Lemma3}

We next consider $[[H_C, H_h],H_C]$, which can be expressed as
\begin{equation}
    [[H_C, H_h],H_C] = \sum_{\langle ij \rangle} \sum_{\sigma} [[H_C, B_{ij\sigma} + B_{ji\sigma}],H_C].
    \label{eq:HcH_initial}
\end{equation}
Our goal is to provide a tight bound for the spectral norm of this operator. We again choose to partition the expression into smaller sets for which the spectral norm can be evaluated exactly. A simple approach would be to take
\begin{equation}
    \Big\lVert \sum_{\langle ij \rangle} \sum_{\sigma} [[H_C, B_{ij\sigma} + B_{ji\sigma}],H_C] \Big\rVert \leq \sum_{\langle ij \rangle} \Big\lVert \sum_{\sigma} [[H_C, B_{ij\sigma} + B_{ji\sigma}],H_C] \Big\rVert.
    \label{eq:basic_partitioning}
\end{equation}
For some lattices we may also be able to evaluate the spectral norm for larger sets of operators, acting on more qubits. In this case, we may take the set of all bonds, $\langle ij \rangle = \{ (i,j) \; | \; \textrm{$i$ and $j$ are neighbors} \} $, and again define a partition of this set. We label this partition as $P$. Then, we can write $P = \{ X_1, X_2, \ldots \}$, such that each $X_i$ may contain more than one bond on the lattice, for example we might have $X_1 = \{ (1, 2), (2, 3) \}$ which contains two lattice bonds. Note that this partitioning of the bonds on the lattice is simply a form of tiling of the lattice. Then,
\begin{equation}
    \Big\lVert \sum_{\langle ij \rangle} \sum_{\sigma} [[H_C, B_{ij\sigma} + B_{ji\sigma}],H_C] \Big\rVert \leq \sum_{X \in P} \Big\lVert \sum_{\sigma} \sum_{(i,j) \, \in \, X} [[H_C, B_{ij\sigma} + B_{ji\sigma}],H_C] \Big\rVert.
    \label{eq:general_partitioning}
\end{equation}
If each element $X \in P$ contains multiple bonds then Eq.~(\ref{eq:general_partitioning}) will in general provide a tighter bound than Eq.~(\ref{eq:basic_partitioning}). We then need only ensure that the operators, $O$, inside the spectral norm $\lVert O \rVert$, act on sufficiently few qubits, or are sufficiently sparse, such that this norm can be calculated numerically. In general, this will depend on the lattice under consideration.

We write an expression for all the Coulomb interaction terms (from $H_I$ and $H_V$) that interact with spin orbitals on lattice sites $i$ and $j$
\begin{equation}
    H_{C,ij} = \frac{U}{4} \Big( Z_{i\uparrow}Z_{i\downarrow} + Z_{j\uparrow}Z_{j\downarrow} \Big) + \frac{V}{4} \Big ( Z_{i\uparrow}Z_{j\uparrow}+Z_{i\uparrow}Z_{j\downarrow}+Z_{i\downarrow}Z_{j\uparrow}+Z_{i\downarrow}Z_{j\downarrow} + (Z_{i\uparrow}+Z_{i\downarrow})\Sigma^i+(Z_{j\uparrow}+Z_{j\downarrow})\Sigma^j \Big) , \label{eq:Coulomb_terms_ij}
\end{equation}
where we define $\Sigma^i$ as the sum over the $Z$-operators on the nearest neighbors of $i$ \textit{other than} site $j$, and similarly we define $\Sigma^j$ as the sum over $Z$-operators on the nearest neighbors of $j$ \textit{other than} $i$. These operators are defined as
\begin{equation}
\Sigma^i \equiv \sum_{\sigma \in \{ \uparrow,\downarrow \} } \sum_{l \sim i,l \neq j} Z_{l\sigma},  \; \; \;\Sigma^j \equiv \sum_{\sigma \in \{ \uparrow,\downarrow \} } \sum_{l \sim j,l \neq i} Z_{l\sigma}.
\end{equation}
The operators $\Sigma^i$ and $\Sigma^j$ contain $2(k-1)$ operators of type $Z_{l\sigma}$. To further simplify the commutator bound expressions, we choose to give the bounds for the case where the factor between the interaction parameters $U$ and $V$ is constant. We define $\frac{V}{U} =\alpha$ and use this to rewrite $H_{C,ij}$ as
\begin{equation}
    H_{C,ij} = \frac{U \alpha}{4} \Big( \frac{1}{\alpha}Z_{i\uparrow}Z_{i\downarrow} + \frac{1}{\alpha}Z_{j\uparrow}Z_{j\downarrow} + Z_{i\uparrow}Z_{j\uparrow}+Z_{i\uparrow}Z_{j\downarrow}+Z_{i\downarrow}Z_{j\uparrow}+Z_{i\downarrow}Z_{j\downarrow} + (Z_{i\uparrow}+Z_{i\downarrow})\Sigma^i+(Z_{j\uparrow}+Z_{j\downarrow})\Sigma^j \Big),
\end{equation}
We also establish a commutator relation which is useful for this commutator bound. Using the commutator and anti-commutator relations given by Eqs.~(\ref{eq:comm_rule_2})--(\ref{eq:comm_rule_4}), we evaluate the commutator between $H_{C,ij}$ and the hopping operators $B_{ij\sigma}$ and $B_{ji\sigma}$
\begin{equation}
    [H_{C,ij},B_{ij\sigma}+B_{ji\sigma}] = \frac{U \alpha}{2} \Big( \frac{1}{\alpha}Z_{i\uparrow}Z_{i\downarrow} + \frac{1}{\alpha}Z_{j\uparrow}Z_{j\downarrow} +Z_{i\uparrow}Z_{j\downarrow}+Z_{i\downarrow}Z_{j\uparrow}+ Z_{i\sigma} \Sigma^i+Z_{j\sigma}\Sigma^j \Big)(B_{ij\sigma}+B_{ji\sigma}).
    \label{eq:commutator_HCij_Bijsigma_Bjisigma}
\end{equation}
Finally, we define $C_{ij} \equiv  \frac{1}{\alpha}Z_{i\uparrow}Z_{i\downarrow} + \frac{1}{\alpha}Z_{j\uparrow}Z_{j\downarrow} +Z_{i\uparrow}Z_{j\downarrow}+Z_{i\downarrow}Z_{j\uparrow}$, which is used to rewrite the commutator in Eq.~(\ref{eq:commutator_HCij_Bijsigma_Bjisigma}) as
\begin{equation}
    [H_{C,ij},B_{ij\sigma}+B_{ji\sigma}] =\frac{U \alpha}{2} \Big( C_{ij}+Z_{i\sigma}\Sigma^i+Z_{j\sigma}\Sigma^j \Big)(B_{ij\sigma}+B_{ji\sigma}).
    \label{eq:commutator_HCij_Bijsigma_Bjisigma2}
\end{equation}
Using these relations, we provide a lemma for the commutator bound $[[H_C,H_h],H_C]$ for extended Hubbard models.

\begin{lemma} \label{Lemma:extended_Hubbard_ChC} For an extended Hubbard model Hamiltonian $H_{EH} = H_h + H_I + H_V$, defined in Eqs.~(\ref{eq:H_h_appendix})--(\ref{eq:H_V_appendix}), with $\alpha = \frac{V}{U}$, on a lattice with $N$ lattice sites where all sites have $k$ nearest neighbors,
\begin{eqnarray}
    \big\lVert \, [[H_C ,H_h],H_C] \, \big\rVert \leq \frac{U
    ^2 \alpha^2}{4} \sum_{X \in P} \Big\lVert \sum_\sigma \sum_{(i,j) \in X} \Big(C_{ij} + Z_{i\sigma}\Sigma^i + Z_{j\sigma} \Sigma^j \Big)^2 \Big( B_{ij\sigma}+B_{ji\sigma} \Big) \Big\rVert,
     \label{eq:Lemma3}
\end{eqnarray}
\end{lemma}

\begin{proof}
We evaluate the commutator as 
\begin{equation}
    [[H_C,H_h],H_C] = \sum_{X \in P} [[H_C,\sum_{(i,j) \, \in \, X}(B_{ij} + B_{ji})],H_C].
\end{equation}
We begin by evaluating the commutator: $[H_C,\sum_{(i,j) \, \in \, X} (B_{ij\sigma}+B_{ji\sigma})]$. We know from the commutation relation in Eq.~(\ref{eq:comm_rule_1}) that $B_{ij\sigma}$ and $B_{ji\sigma}$ commute with all Coulomb terms that do not interact with sites $i$ and $j$, meaning we can evaluate the commutator as $\sum_{(i,j) \, \in \, X}[H_{C,ij},B_{ij\sigma}+B_{ji\sigma}]$. Using the result of Eq.~(\ref{eq:commutator_HCij_Bijsigma_Bjisigma2}), we obtain
\begin{equation}
    \sum_{(i,j) \, \in \, X}[H_{C,ij},B_{ij\sigma}+B_{ji\sigma}] = \sum_{(i,j) \, \in \, X}  \frac{U \alpha}{2} \Big( C_{ij}+ Z_{i\sigma}\Sigma^i + Z_{j\sigma} \Sigma^j \Big)(B_{ij\sigma}+B_{ji\sigma})
    \label{eq:comm_HCBijsigma}
\end{equation}
The commutator in (\ref{eq:comm_HCBijsigma}) is nested with $H_C$. We use that the Coulomb terms commute to evaluate the nested commutator
\begin{eqnarray}
        [[H_C,\sum_{(i,j) \, \in \, X} (B_{ij\sigma}+B_{ji\sigma})],H_C] &=& \sum_{(i,j) \, \in \, X}  \frac{U \alpha}{2}\Big(C_{ij} + Z_{i\sigma} \Sigma^i + Z_{j\sigma} \Sigma^j \Big)[B_{ij\sigma}+B_{ji\sigma},H_C], \nonumber \\ &=&
        -  \sum_{(i,j) \, \in \, X}  \frac{U^2 \alpha^2}{4} \Big( C_{ij}+Z_{i\sigma}\Sigma^i + Z_{j\sigma} \Sigma^j \Big)^2(B_{ij\sigma}+B_{ji\sigma}) .
        \label{eq:HC_BIJ_HC}
\end{eqnarray}
Then, we obtain $[[H_C,\sum_{(i,j) \, \in \, X} (B_{ij}+B_{ji})],H_C]$ by summing over $\sigma$ in Eq.~(\ref{eq:HC_BIJ_HC}),
\begin{eqnarray}
        & & [[H_C,\sum_{(i,j) \, \in \, X} (B_{ij}+B_{ji})],H_C] = - \frac{U^2 \alpha^2}{4} \sum_\sigma \sum_{(i,j) \, \in \, X} \Big(C_{ij} + Z_{i\sigma}\Sigma^i + Z_{j\sigma} \Sigma^j \Big)^2 \Big( B_{ij\sigma}+B_{ji\sigma} \Big) .
        \label{eq:comm_HCBij}
\end{eqnarray}
Next, we sum over the partitions to obtain
\begin{eqnarray}
        & & [[H_C, H_h],H_C] = - \frac{U^2 \alpha^2}{4}\sum_{X \in P} \sum_\sigma \sum_{(i,j) \, \in \, X} \Big(C_{ij} + Z_{i\sigma}\Sigma^i +  Z_{j\sigma} \Sigma^j \Big)^2 \Big( B_{ij\sigma}+B_{ji\sigma} \Big) ,
\end{eqnarray}
and take the norm and apply the triangle inequality
\begin{eqnarray}
        & & \Big\lVert [[H_C, H_h],H_C] \Big\rVert \le \frac{U^2 \alpha^2}{4} \sum_{X \in P} \Big\lVert \sum_\sigma \sum_{(i,j) \, \in \, X} \Big(C_{ij} + Z_{i\sigma}\Sigma^i + Z_{j\sigma} \Sigma^j \Big)^2 \Big( B_{ij\sigma}+B_{ji\sigma} \Big) \Big\rVert,
        \label{eq:ChC_final}
\end{eqnarray}
which concludes the proof of Lemma \ref{Lemma:extended_Hubbard_ChC}.
\end{proof}

In the above we have considered a strict partition of the set $\langle ij \rangle = \{ (i,j) \; | \; \textrm{$i$ and $j$ are neighbors} \} $ into $P = \{ X_1, X_2, \ldots \}$. However, we could also define a set $P = \{ X_1, X_2, \ldots \}$ such that each bond in $\langle ij \rangle$ appears $n$ times in total, for some integer $n$. In this case, we only need to divide the right-hand side of Eq.~(\ref{eq:ChC_final}) by $n$ to account for this multiple counting. This is equivalent to simply multiplying and dividing by $n$ on the right-hand side of Eq.~(\ref{eq:HcH_initial}). In terms of terminology, this is no longer strictly a partition of the set $\langle ij \rangle$, but this does not affect its practical use (equivalently, this could be viewed a partition of the multi-set where each bond on the lattice appears $n$ times). We will use this in the numerical result below. For simplicity, we continue to use the term ``partition''.

\begin{numresult}
\textbf{Periodic hexagonal lattice:}
We choose to partition the lattice into $C_6$ parts such that each $X$ contains six $(i,j)$ pairs forming a $C_6$ circle graph. Each bond is contained in two $C_6$ partitions, meaning we have to sum over $N/2$ completely equivalent $C_6$ partitions and divide by two to avoid double counting the bonds when evaluating the entire commutator bound. We evaluate the norm of the $C_6$ partition for $V/U = \alpha = 1/2$. The resulting $C_6$ partition commutator is a $24$-qubit operator and we compute its spectral norm numerically using DMRG as described in Section~\ref{sec:Numerical_results}, resulting in the following bound 
\begin{eqnarray}
        & & \lVert [[H_C,H_h],H_C] \rVert \Big\rvert_{\frac{U}{V}=2} \leq 4.80U^2\tau  N.
         \label{eq:ChC_Hex_lattice_numerical_norm}
\end{eqnarray}
Note that this result is valid for all periodic hexagonal lattices with $L_x = L_y = L \geq 4$ as defined in Appendix~\ref{APP:hexagonallatticedivision}.

\end{numresult}

\subsection{Example: One-dimensional lattices} \label{sec:1D_Hubbard_bounds}

To demonstrate the tightness of the bounds in Lemmas~\ref{Lemma:Hubbard_Ihh}, \ref{Lemma:Extended_Hubbard_Chh} and \ref{Lemma:extended_Hubbard_ChC} for an example, we consider the Hubbard and extended Hubbard model on periodic one-dimensional lattices. Here, it is straightforward to calculate the spectral norms exactly for a number of small lattice sizes, which can then be compared to bounds from the above lemmas.

\begin{figure}
    \centering
    \includegraphics[width=0.5\linewidth]{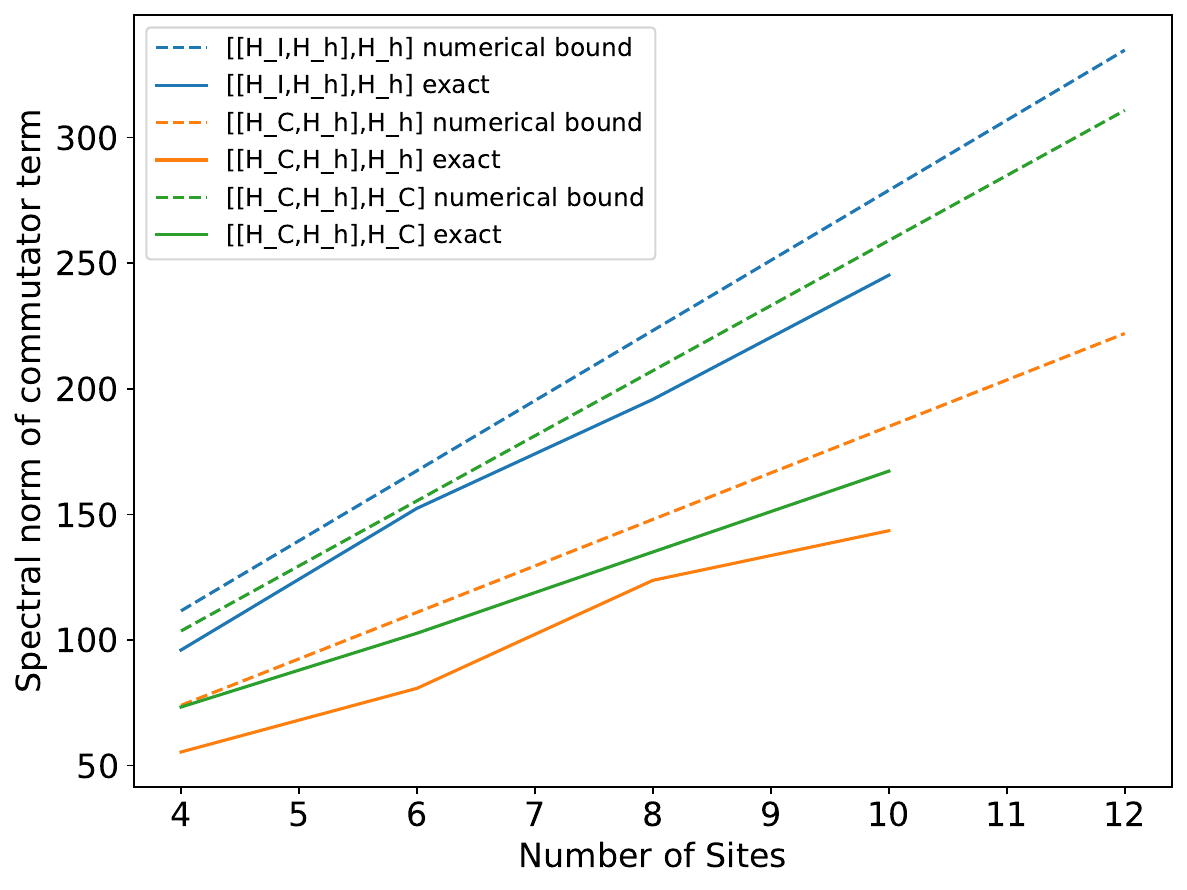}
    \caption{Numerical results for the spectral norm of commutators considered in Appendix~\ref{APP:COMMUTATORBOUNDS}, presented for a one-dimensional periodic lattice with varying number of sites. Exact norms are calculated for lattices up to $10$ sites, and plotted as solid lines. These are then compared to bounds derived in Lemmas~\ref{Lemma:Hubbard_Ihh}, \ref{Lemma:Extended_Hubbard_Chh} and \ref{Lemma:extended_Hubbard_ChC} (plotted as dashed lines). We take a partition with two neighboring lattice sites for $\lVert [[H_I,H_h],H_h] \rVert$ and a partition containing two neighboring bonds for the extended Hubbard model bounds.}
    \label{fig:Hubbard_chain_comm_bounds}
\end{figure}

The numerical results are plotted in Fig.~\ref{fig:Hubbard_chain_comm_bounds}. Exact commutator norms are calculated up to 10 lattice sites, which can be compared to the bounds calculated using Eqs.~(\ref{eq:cor:hex_lattice_Hubbard}), (\ref{eq:Chh_Hex_lattice_numerical_norm}) and (\ref{eq:ChC_Hex_lattice_numerical_norm}). For the Hubbard model commutator bound, we take a partition containing two neighboring lattice sites, and for the extended Hubbard model bounds we take a partition with two lattice bonds. The corresponding norms in these equations are calculated exactly by first constructing a sparse matrix representation of the operators. The results demonstrate that the bounds are always tight within a factor of at most 2, and often by a much smaller factor. These bounds could be made tighter by taking larger partitions in the above expressions, at the cost of performing spectral norm calculations for operators acting on more qubits.

\section{\label{APP:Trotter_tables} Tile Trotterization Error Norms and Resource Requirements}

In Table~\ref{Table:Trotter_error_norms} we provide the Tile Trotterization error norms ($W_{\mathrm{tile}}$), the total number of qubits ($N_Q$), and the number of arbitrary rotations ($N_R$) and T gates ($N_T$) required per Trotter step for the hexagonal lattice Hubbard model simulations considered in Figs.~\ref{fig:QUBITIZATION_VS_TROTTER_plots} and \ref{fig:QPE_Trotter_HWP_comparison}. We use Hubbard model parameters $U=4$, $V=2$ and $\tau=1$. The nearest-neighbor interaction parameter, $V$, is only used in the extended Hubbard model. All systems use the periodic hexagonal lattice model described in Appendix~\ref{APP:hexagonallatticedivision} with parameters $L_x = L_y = L$ and $4 \leq L \leq 18$, for even $L$. The hopping Hamiltonian Trotter error constitutes $16 \%$ of $W_{\mathrm{tile}}$ for the Hubbard model and $10 \%$ of $W_{\mathrm{tile}}$ for the extended Hubbard model.

\begin{table*}
\caption{Tile Trotterization error norms ($W_{\mathrm{tile}}$), qubit counts ($N_Q$), arbitrary rotation costs ($N_R$) and T gate costs ($N_T$) for performing a single Trotter step for periodic hexagonal lattice Hubbard models with lattice parameters $L_x=L_y=L$, where $4\leq L \leq 18$ for even $L$, and Hubbard model parameters $U=4$, $V=2$ and $\tau=1$. The Hubbard model parameters only affects the Trotter error norms. We show data for the resources required to implement Tile Trotterization for the periodic Hubbard model and the periodic extended Hubbard model without HWP ($\alpha=0$) and with HWP using $\alpha=N/4-1$, $\alpha=N/2-1$ and $\alpha=N-1$ ancilla qubits. The data from this table can be used to reproduce the Tile Trotterization costings shown in Figs.~\ref{fig:QUBITIZATION_VS_TROTTER_plots} and \ref{fig:QPE_Trotter_HWP_comparison}.}
\begin{ruledtabular}
\begin{tabular}{l|c|r|r|r|r|r|r|r|r}
& $N=2L^2$  & 32 & 72 & 128 & 200 & 288 & 392 & 512 & 648 \\
\hline
Hubbard model & $W_{\mathrm{tile}}$ & 166 & 374& 665& 1041& 1497& 2039& 2663 & 3371\\
Extended Hubbard model & $W_{\mathrm{tile}}$ & 262& 592& 1052 & 1644& 2367& 3222& 4208 & 5326 \\
\hline
Hubbard model, $\alpha=0$ & $N_Q$ & 64 & 144 & 256 & 400 & 576 & 784 & 1024 & 1296 \\
& $N_R$ & 192& 432 & 768 & 1200 & 1728 & 2352 & 3072 & 3888\\
& $N_T$ & 320 & 720 & 1280 & 2000 & 2880 & 3920 & 5120 & 6480\\
\hline
Hubbard model, $\alpha=\frac{N}{4}-1$ & $N_Q$ & 71& 161& 287& 449& 647& 881& 1151 & 1457\\
& $N_R$ & 96 & 120& 144& 144& 168& 168 & 192 & 192\\
& $N_T$ & 992& 2352& 4256& 6704& 9696& 13232& 17312 & 21936\\
\hline
Hubbard model, $\alpha=\frac{N}{2}-1$ & $N_Q$ & 79 & 179 & 319 & 499 & 719 & 979 & 1279 & 1619 \\
& $N_R$ & 60& 72 & 84 & 84 & 96 & 96 & 108 & 108\\
& $N_T$ & 1040 & 2400 & 4304 & 6752 & 9744 & 13280 & 17360 & 21984\\
\hline
Hubbard model, $\alpha=N-1$ & $N_Q$ & 95 & 215 & 383 & 599 & 863 & 1175 & 1535 & 1943 \\
& $N_R$ & 36& 42 & 48 & 48 & 54 & 54 & 60 & 60\\
& $N_T$ & 1064 & 2424 & 4328 & 6776 & 9768 & 13304 & 17384 & 22008\\
\hline

\hline
Extended Hubbard model, $\alpha=0$ & $N_Q$ & 64 & 144 & 256 & 400 & 576 & 784 & 1024 & 1296 \\
& $N_R$ & 384& 864 & 1536 & 2400 & 3456 & 4704 & 6144 & 7776\\
& $N_T$ & 320 & 720 & 1280 & 2000 & 2880 & 3920 & 5120 & 6480 \\
\hline
Extended Hubbard model, $\alpha=\frac{N}{4}-1$ & $N_Q$ & 71& 161& 287& 449& 647& 881& 1151 & 1457\\
& $N_R$ & 192& 240& 288& 288& 336& 336& 384 & 384\\
& $N_T$ & 1664& 3984& 7232& 11408& 16512& 22544& 29504 & 37392 \\
\hline
Extended Hubbard model, $\alpha=\frac{N}{2}-1$ & $N_Q$ & 79 & 179 & 319 & 499 & 719 & 979 & 1279 & 1619\\
& $N_R$ & 120& 144& 168& 168& 192& 192& 216 & 216\\
& $N_T$ & 1760& 4080& 7328& 11504& 16608& 22640& 29600 & 37488 \\
\hline
Extended Hubbard model, $\alpha=N-1$ & $N_Q$ & 95 & 215 & 383 & 599 & 863 & 1175 & 1535 & 1943 \\
& $N_R$ & 72& 84& 96& 96& 108& 108& 120 & 120\\
& $N_T$ & 1808& 4128& 7376& 11552& 16656& 22688& 29648 & 37536
\label{Table:Trotter_error_norms}
\end{tabular}
\end{ruledtabular}
\end{table*}

\section{\label{APP:qubitization} Qubitization circuits}

In this section we describe how qubitization circuits may be implemented for the Hubbard model on a periodic hexagonal lattice. Our qubitization approach builds upon previous work by Babbush \emph{et al.} in Ref.~\cite{Babbush2018EncodingComplexity}. In Section~\ref{sec:Phase_estimation} we compared the cost of QPE performed with both Tile Trotterization and qubitization-based approaches. Here, we provide a brief introduction to the qubitization approach, and present the implementations and costing of SELECT and PREPARE operators used for our QPE resource estimates.

In Trotter-based QPE, we perform phase estimation with a Trotterized approximation of the time evolution operator, $U = e^{-iHt}$. QPE allows us to estimate the eigenphases of this unitary, and therefore the energies of $H$. However, it is equally possible to perform phase estimation with other unitaries that encode the Hamiltonian. In qubitization, the unitary of interest is the walk operator, $\mathcal{W}$, which has eigenvalues
\begin{eqnarray}
    e^{\pm i \, \mathrm{arccos}(E_n/\lambda)},
\end{eqnarray}
where $E_n$ are the eigenvalues of $H$ and $\lambda$ is the L1 norm of $H$. From these, we can obtain estimates of energies, $E_n$.

The quantum walk operator is built from SELECT and PREPARE operators. Consider a Hamiltonian
\begin{equation}
H = \sum_{l=0}^{L-1} w_l P_l,\;\;\; \lambda =\sum_{l=0}^{L-1}|w_l|,
\end{equation}
where $w_l > 0$ are coefficients and $P_l$ are (tensor products of) Pauli operators. Then, SELECT defines a block-encoding of $H / \lambda$. In particular,
\begin{eqnarray}
    \mathrm{SELECT} = \sum_{l=0}^{L-1} | l \rangle \langle l | \otimes P_l,
\end{eqnarray}
where $|l\rangle$ are flag qubit states. Each state $|l\rangle$ flags a corresponding term in the Hamiltonian, $P_l$. The PREPARE operator acts on the $|0\rangle$ state of the flag qubits and prepares a state that encodes the coefficients of $H$. More formally, it prepares the signal state that flags the block encoding of $H / \lambda$. It can be defined by
\begin{eqnarray}
    \mathrm{PREPARE} |0\rangle = \sum_{l=0}^{L-1} \sqrt{\frac{w_l}{\lambda}} | l \rangle.
\end{eqnarray}

The walk operator can be expressed in terms of the SELECT and PREPARE operations, and a reflection operator. In particular, it has been shown that the walk operator controlled on a single ancilla qubit can be expressed by the following operations:
\begin{center}
\begin{quantikz}
                         & \qw            & \ctrl{1}    & \qw & \qw \\
\lstick{$|l\rangle$}     & \qw\qwbundle{} & \gate[2]{\mathcal{W}} & \qw & \qw \\
\lstick{$|\psi\rangle$}  & \qw\qwbundle{} &             & \qw & \qw
\end{quantikz}
\;\;\;\; = \;\;
\begin{quantikz}
                         & \qw            & \ctrl{1}                  & \qw                                & \gate{Z}   & \qw                     & \qw \\
\lstick{$|l\rangle$}     & \qw\qwbundle{} & \gate[2]{\mathrm{SELECT}} & \gate{\mathrm{PREPARE}^{\dagger}}  & \octrl{-1} & \gate{\mathrm{PREPARE}} & \qw \\
\lstick{$|\psi\rangle$}  & \qw\qwbundle{} &                           & \qw                                & \qw        & \qw                     & \qw
\end{quantikz}
\end{center}
Below we will describe the implementation and costing of each of these operations.

We will consider the following form of the Hubbard Hamiltonian, where the Jordan-Wigner mapping and a chemical potential shift are applied,
\begin{equation}
H^{\mathrm{JW}}_{H} = - \frac{\tau}{2} \sum_{\langle p,q \rangle, \sigma} \Big( X_{p\sigma} \overrightarrow{Z} X_{q\sigma} + Y_{p\sigma} \overrightarrow{Z} Y_{q\sigma} \Big) + \frac{U}{4} \sum_{p=1}^N Z_{p \uparrow} Z_{p \downarrow},
\end{equation}
where $X$, $Y$ and $Z$ are Pauli operators. We label the Hamiltonian by JW and relabel the site indices as $p$ and $q$ in order to distinguish from the fermionic Hamiltonians defined in Section~\ref{sec:Hubbard_models}. The operators $\overrightarrow{Z}$ indicate a string of $Z$ operators acting on all qubits between $p\sigma$ and $q\sigma$ in the JW ordering.

The gate cost of qubitization-based QPE scales with the L1 norm of the Hamiltonian. For the Hubbard model Hamiltonian, the L1 norm, $\lambda$, is given by
\begin{equation}
    \lambda = 2\tau \times \textrm{\# of bonds} + \frac{U}{4} \times \textrm{\# of sites}.
\end{equation}
The number of sites is $N$ and for the periodic hexagonal lattice the number of bonds is $3N/2$, and therefore,
\begin{equation}
    \lambda = \Big( 3\tau + \frac{U}{4} \Big) N.
\end{equation}

\subsection{\label{APP:SELECT} SELECT}

We first consider the implementation of SELECT for the Hubbard model on a periodic hexagonal lattice. We index terms in the Hamiltonian using the flag registers $\ket{U}\ket{p_x}\ket{p_y}\ket{p_c}\ket{\alpha}\ket{q_x}\ket{q_y}\ket{q_c}$. We then define the action of SELECT as
\begin{eqnarray}
\mathrm{SELECT} |U,p,\alpha,q\rangle |\psi\rangle = \nonumber |U,p,\alpha,q\rangle \begin{cases}
Z_{p0} Z_{q1} |\psi\rangle & U \wedge (p = q) \wedge (\alpha=0) \\
-X_{p\alpha} \overrightarrow{Z} X_{q\alpha} |\psi\rangle & \neg U \wedge (p < q) \\
-Y_{q\alpha} \overrightarrow{Z} Y_{p\alpha} |\psi\rangle & \neg U \wedge (p > q) \\
\mathrm{UNDEFINED} & \textrm{otherwise}.
\end{cases}
\end{eqnarray}
The labels $p$ and $q$ are lattice site indices that also include the ``color'' label, $c$, coming from the white and grey labels of the lattice points in each $(l_x,l_y)$ pair, as shown in Fig.~\ref{Fig:hex_lattices}(a). We define the ordering $p = p_x + p_y L_y + p_c L_x L_y$ and $q = q_x + q_y L_y + q_c L_x L_y$. Note that all $c=0$ terms come before all $c=1$ terms within this ordering. Also note that for use in qubitized QPE, the action of the UNDEFINED block must be such that the total action of SELECT is Hermitian.

The quantum circuit to achieve this definition of SELECT, controlled on an ancilla qubit, is presented in Fig.~\ref{fig:SELECT_CIRCUIT}, and consists of three unary iterators. The Toffoli cost of the first two unary iterators are $4 L_x L_y - 1$ each, while the Toffoli cost of the final iterator, which has two additional controls, is $(2 L_x L_y - 1) + 2 = 2 L_x L_y + 1$, and so the total Toffoli cost combined is $10 L_x L_y - 1$. This leads to a T gate cost of $C_S = 40 L_x L_y - 4$. Expressing this in terms of the number of lattice points, $N$, we obtain $C_S = 20 N - 4$. This is asymptotically the most expensive subroutine in the walk operator.
\begin{figure*}
\begin{quantikz}
\lstick{control}        & \qw & \gate{S^{\dagger}} & \ctrl{9}                            & \ctrl{9}                                         & \ctrl{5}                 & \qw \\
\lstick{$U$}            & \qw & \qw & \qw                                                & \qw                                              & \control{}               & \qw \\
\lstick{$p_x$}          & \qw\qwbundle{\textrm{$\log$ } L_x} & \qw & \gate{\mathrm{In}_{p_x}} & \qw                                              & \qw                      & \qw \\
\lstick{$p_y$}          & \qw\qwbundle{\textrm{$\log$ } L_y} & \qw & \gate{\mathrm{In}_{p_y}} & \qw                                              & \qw                      & \qw \\
\lstick{$p_c$}          & \qw & \qw & \gate{\mathrm{In}_{p_c}}                           & \qw                                              & \qw                      & \qw \\
\lstick{$\alpha$}       & \qw & \qw & \gate{\mathrm{In}_{\alpha}}                        & \gate{\mathrm{In}_{\alpha}}                      & \octrl{4}              & \qw \\
\lstick{$q_x$}          & \qw\qwbundle{\textrm{$\log$ } L_x} & \qw & \qw                      & \gate{\mathrm{In}_{q_x}}                         & \gate{\mathrm{In}_{q_x}} & \qw \\
\lstick{$q_y$}          & \qw\qwbundle{\textrm{$\log$ } L_y} & \qw & \qw                      & \gate{\mathrm{In}_{q_y}}                         & \gate{\mathrm{In}_{q_y}} & \qw \\
\lstick{$q_c$}          & \qw & \qw & \qw                                                & \gate{\mathrm{In}_{q_c}}                         & \gate{\mathrm{In}_{q_c}} & \qw \\
\lstick{$|\psi\rangle$} & \qw & \qw & \gate{\overrightarrow{Z} Y_{p_x p_y p_c \alpha}}   & \gate{\overrightarrow{Z} X_{q_x q_y q_c \alpha}} & \gate{Z_{q_x q_y q_c 1}} & \qw
\end{quantikz}
\caption{Controlled SELECT circuit for the Hubbard model on an $L_x \times L_y$ periodic hexagonal lattice.}
\label{fig:SELECT_CIRCUIT}
\end{figure*}

\subsection{\label{APP:PREPARE} PREPARE}

The action of PREPARE is defined to act on the flag qubits prepared in state $\ket{0}$ as
\begin{equation}
\begin{aligned}
    \mathrm{PREPARE} \, | 0 \rangle = \sqrt{\frac{U}{4\lambda}} \sum_{p_x=0}^{L_x-1} \sum_{p_y=0}^{L_y-1} \sum_{p_c=0}^1 &|1\rangle_U |p_x, p_y, p_c, 0 \rangle | p_x, p_y, p_c \rangle  \\
    + \; \sqrt{\frac{t}{2\lambda}} \sum_{p_x=0}^{L_x-1} \sum_{p_y=0}^{L_y-1} \sum_{\sigma=0}^1 &|0\rangle_U \biggr(
    |p_x, p_y, 0, \sigma \rangle | p_x, p_y, 1 \rangle + |p_x, p_y, 1, \sigma \rangle | p_x, p_y, 0 \rangle \\
    + \; &|p_x, p_y, 0, \sigma \rangle | p_x-1, p_y, 1 \rangle + |p_x-1, p_y, 1, \sigma \rangle | p_x, p_y, 0 \rangle \\
    + \; &|p_x, p_y, 0, \sigma \rangle | p_x, p_y-1, 1 \rangle + |p_x, p_y-1, 1, \sigma \rangle | p_x, p_y, 0 \rangle \biggr) .
\end{aligned}
\end{equation}

\begin{figure*}
\begin{quantikz}
\lstick{$|0\rangle_U$}        & \qw                           & \qw & \gate{R_Y}                         & \octrl{4} & \qw      & \qw      & \qw      & \octrl{7} & \octrl{8} & \octrl{9}  & \qw        & \qw        & \qw & U \\
\lstick{$|0\rangle_{p_x}$}    & \qw\qwbundle{\textrm{$\log$ } L_x} & \qw & \gate{\textrm{UNIFORM}_{L_x}}          & \qw       & \ctrl{4} & \qw      & \qw      & \qw       & \qw       & \qw        & \swap{4}   & \qw        & \qw & p_x \\
\lstick{$|0\rangle_{p_y}$}    & \qw\qwbundle{\textrm{$\log$ } L_y} & \qw & \gate{\textrm{UNIFORM}_{L_y}}          & \qw       & \qw      & \ctrl{4} & \qw      & \qw       & \qw       & \qw        & \qw        & \swap{4}   & \qw & p_y \\
\lstick{$|0\rangle_{p_c}$}    & \qw                           & \qw & \gate{H}                           & \qw       & \qw      & \qw      & \ctrl{4} & \qw       & \qw       & \qw        & \ctrl{}    & \ctrl{}    & \qw & p_c \\
\lstick{$|0\rangle_{\alpha}$} & \qw                           & \qw & \qw                                & \gate{H}  & \qw      & \qw      & \qw      & \qw       & \qw       & \qw        & \qw        & \qw        & \qw & \alpha \\
\lstick{$|0\rangle_{q_x}$}    & \qw\qwbundle{\textrm{$\log$} L_x} & \qw & \qw                                & \qw       & \targ{}  & \qw      & \qw      & \qw       & \gate{-1} & \qw        & \targX{}   & \qw        & \qw & q_x \\
\lstick{$|0\rangle_{q_y}$}    & \qw\qwbundle{\textrm{$\log$} L_y} & \qw & \qw                                & \qw       & \qw      & \targ{}  & \qw      & \qw       & \qw       & \gate{-1}  & \qw        & \targX{}   & \qw & q_y \\
\lstick{$|0\rangle_{q_c}$}    & \qw                           & \qw & \qw                                & \qw       & \qw      & \qw      & \targ{}  & \targ{}  & \qw       & \qw        & \qw        & \qw        & \qw & q_c \\
\lstick{$|0\rangle$}          & \qw                           & \qw & \gate[2]{\textrm{PREP}_{00+01+10}} & \qw       & \qw      & \qw      & \qw      & \qw       & \ctrl{}   & \qw        & \qw        & \qw        & \qw & \mathrm{temp} \\
\lstick{$|0\rangle$}          & \qw                           & \qw &                                    & \qw       & \qw      & \qw      & \qw      & \qw       & \qw       & \ctrl{}    & \qw        & \qw        & \qw & \mathrm{temp} \\
\end{quantikz}
\caption{PREPARE circuit for the Hubbard model on an $L_x \times L_y$ hexagonal lattice. Note that the controlled ``$-1$'' operations on the $| q_x \rangle$ and $| q_y \rangle$ registers can be performed $\textrm{mod } 2^{\lceil \log_2 L_x \rceil}$ and $\textrm{mod } 2^{\lceil \log_2 L_y \rceil}$, respectively. They do not need to be performed $\textrm{mod } L_x$ and $\textrm{mod } L_y$, as one might expect. This is discussed further in the text of Appendix~\ref{APP:qubitization}.}
\label{fig:PREPARE_CIRCUIT}
\end{figure*}

\begin{table*}
\begin{tabular}{l|l|l|l}
Circuit element & Toffoli count & T count (incl. Toffolis) & Ancilla qubits \\
\hline
{\bf Controlled SELECT} & $10 L_x L_y - 1$ & $40 L_x L_y - 4$ & $\lceil \mathrm{log}_2 L_x \rceil + \lceil \mathrm{log}_2 L_y \rceil + 3$ \\
\hline
{\bf PREPARE:} & & & \\
UNIFORM $\frac{1}{\sqrt{L_x}}\sum_{l=0}^{L_x-1}\ket{l}$ & $3\lceil \log_2 L_x\rceil - 3\eta_{L_x} - 3$ & $4 \times$ Toffoli count $ + 2 \Theta$ & $\lceil \mathrm{log}_2 L_x \rceil - \eta_{L_x} + 2$ \\
UNIFORM $\frac{1}{\sqrt{L_y}}\sum_{l=0}^{L_y-1}\ket{l}$ & $3\lceil \log_2 L_y\rceil - 3\eta_{L_y} - 3$ & $4 \times$ Toffoli count $ + 2 \Theta$ & $\lceil \mathrm{log}_2 L_y \rceil - \eta_{L_y} + 2$ \\
concat. success qubits & 1 & 4 & 1 \\
controlled Hadamard & $1$ & $4$ & $2$ \\
controlled -1 on $q_x$ register & $\lceil \mathrm{log}_2 L_x \rceil$ & $4 \lceil \mathrm{log}_2 L_x \rceil$ & $\lceil \mathrm{log}_2 L_x \rceil$ \\
controlled -1 on $q_y$ register & $\lceil \mathrm{log}_2 L_y \rceil$ & $4 \lceil \mathrm{log}_2 L_y \rceil$ & $\lceil \mathrm{log}_2 L_y \rceil$ \\
Two controlled swaps & $\lceil \mathrm{log}_2 L_x \rceil + \lceil \mathrm{log}_2 L_y \rceil$ & $7 \lceil \mathrm{log}_2 L_x \rceil + 7 \lceil \mathrm{log}_2 L_y \rceil$ & 0 \\
$R_Y(\theta)$ & - & $\Gamma$ & 1 \\
$\mathrm{PREP}_{00+01+10}$ & - & $3 \Gamma $ & 1 \\
\hline
{\bf Reflection} & $2(2\lceil \log_2 L_x\rceil + 2\lceil \log_2 L_y\rceil + 10) - 3$ & $4 \times$ Toffoli count & $1$ \\
\end{tabular}
\caption{The cost of various circuit elements to perform the controlled quantized walk operator for the Hubbard model on the $L_x \times L_y$ periodic hexagonal Hubbard model. The T gate counts include the cost of converting Toffoli gates into T gates. The ``ancilla qubits'' column is the total number of ancilla qubits for each circuit element, without reusing qubits, and not including flag qubits. The value $\eta_{L_x}$ ($\eta_{L_y}$) is the largest power of $2$ that is a factor of $L_x$ ($L_y$), $\Theta$ is the number of T gates per rotation in the UNIFORM state preparation circuits, and $\Gamma$ is the number of T gates used to implement each of the other rotation gates in PREPARE. The ``concat. success qubits'' element refers to concatenating the success qubits from the two UNIFORM operations, which SELECT must be controlled on. The ``reflection'' step uses the result from \cite{Khattar2024RiseCircuits} to perform a $Z$ gate with $n$ controls in $2n - 3$ Toffolis, with a single ancilla. In practice, ancilla qubits can mostly be shared between subroutines. The only ancillas that are not reused are: the rotation qubit and success flag qubit from each UNIFORM state preparation, the concatenated success qubit, and a qubit prepared in the $|T\rangle$ state in the controlled Hadamard operation. Otherwise, subroutines in PREPARE can use the $\lceil \mathrm{log}_2 L_x \rceil + \lceil \mathrm{log}_2 L_y \rceil + 3$ ancillas required by SELECT. Therefore, the total number of ancilla qubits required is $\lceil \mathrm{log}_2 L_x \rceil + \lceil \mathrm{log}_2 L_y \rceil + 9$, in addition to $2 \lceil \mathrm{log}_2 L_x \rceil + 2 \lceil \mathrm{log}_2 L_y \rceil + 6$ flag qubits. Compared to the Hubbard model qubitization circuits of Ref.~\cite{Babbush2018EncodingComplexity}, our implementation removes a term in SELECT that is logarithmic in the lattice size, and also a flag qubit.}
\label{tab:walk_circuit_cost}
\end{table*}

\begin{figure*}
\begin{quantikz}
\lstick{$|0\rangle$} & \qw                  & \gate{R_y(\theta_2)}       & \gate{R_y(\theta_3)}      & \qw \\
\lstick{$|0\rangle$} & \gate{R_y(\theta_1)} & \octrl{-1}                 & \ctrl{-1}                 & \qw \\
\end{quantikz}
\hspace{5mm} = \hspace{2mm}
\begin{quantikz}
\lstick{$|0\rangle$} & \gate{R_y(\frac{\theta_2+\theta_3}{2})} & \targ{1}   & \gate{R_y(\frac{\theta_2-\theta_3}{2})} & \targ{1} & \qw \\
\lstick{$|0\rangle$} & \gate{R_y(\theta_1)}                    & \ctrl{-1}  & \qw                                     & \ctrl{-1}  & \qw \\
\end{quantikz}
\caption{Demonstration of how $\mathrm{PREP}_{00+01+10}$ can be implemented. This requires $3$ rotation gates, which can be implemented to precision $\epsilon$ with an expected T gate count of $1.15 \, \mathrm{\log}(\frac{1}{\epsilon}) + 9.2$ each. In the text, we denote the number of T gates used to implement each of these rotations by $\Gamma$, which we set to $40$. The required rotation angles are $\theta_1 = 2 \, \mathrm{tan}^{-1}(1/\sqrt{2})$, $\theta_2 = \pi / 4$ and $\theta_3 = 0$, with $R_y(\theta) = e^{-i \theta Y / 2}$.}
\label{fig:prep_rotation_circuit}
\end{figure*}

The quantum circuit to achieve this definition of PREPARE is shown in Fig.~\ref{fig:PREPARE_CIRCUIT}. This consists of a number of circuit elements that we will briefly explain. The final Toffoli and T gate costs and ancilla costs for each circuit element are summarized in Table~\ref{tab:walk_circuit_cost}.

The UNIFORM operations perform uniform state preparation, and can be implemented using amplitude amplification by the approach described in Appendix~A of Ref.~\cite{Lee2021EvenHypercontraction}. However, we make a slight modification to the approach described there. In particular, Ref.~\cite{Lee2021EvenHypercontraction} implements the ancilla rotation by using a phase-gradient ancilla register, which requires $b_n$ ancilla qubits and $b_n - 3$ Toffoli gates. However, as noted in Appendix~A of Ref.~\cite{Sanders2020CompilationOptimization}, when the rotation angle $\beta$ is known classically, as is the case here, this can also be implemented efficiently using rotation synthesis. Here, we use this alternative approach, requiring just a single ancilla for the rotation synthesis. For this gadget, the rotation synthesis can be lower precision than elsewhere; we define $\Theta$ to be the number of T gates used for each rotation in each UNIFORM state preparation, which is separate from the number of T gates used in synthesis of other rotations in PREPARE, which we denote $\Gamma$. As described in \cite{Lee2021EvenHypercontraction}, the Toffoli cost consists of two inequality tests and a reflection. Of the ancillas required, the rotation qubit and success flag qubit and cannot be reused by other routines. We also note that the walk operator's reflection is controlled on the rotation and success qubits, which are included in the final Toffoli count.

The circuit to perform a controlled Hadamard gate is shown in Fig.~17 of the same paper by Lee \emph{et al.} \cite{Lee2021EvenHypercontraction}, which requires two ancilla qubits and one Toffoli gate. One of the ancilla qubits is prepared in a $|T\rangle$ state, which we choose to preserve for future applications of the walk operator; therefore, this ancilla qubit cannot be reused, which is accounted for in the total qubit count.

It is also necessary to prepare the state
\begin{equation}
    |\Psi_{00+01+10}\rangle = \frac{1}{\sqrt{3}} ( |00\rangle + |01\rangle + |10\rangle ),
\end{equation}
which we achieve using a circuit of the form in Fig.~\ref{fig:prep_rotation_circuit}, using $3$ rotation gates. We denote the number of T gates per rotation in this subroutine as $\Gamma$. For our costing in Section~\ref{sec:Phase_estimation}, we set $\Gamma = 40$.

We next consider how to perform the controlled ``$-1$'' operation. One might expect that we actually have to perform ``$-1$ mod $L_x$'' or ``$-1$ mod $L_y$'' operations to properly enforce periodic boundary conditions, similarly to the PREPARE circuit of Ref.~\cite{Babbush2018EncodingComplexity}. However, we can avoid this complication. To see this, consider the case of an $L \times L$ periodic lattice, and let us consider the ``$-1$'' operation on the $| q_x \rangle$ register. This register will consist of $n = \lceil \log_2 L \rceil$ qubits. Before this operation, the register will hold a uniform superposition of states $| l \rangle$, for $0 \le l \le L - 1$. For $l \ge 1$ the ``$-1$'' operation will act in the expected manner. For $l = 0$ it will instead give the state $| 2^n - 1 \rangle$. By periodic boundary conditions, we need lattice sites with $x=0$ to be connected to sites with $x = L - 1$, and so this may seem incorrect. However, provided that SELECT is implemented appropriately, this ultimately leads to the correct circuit. In particular, the SELECT circuit consists of unary iterators which are designed to act on $| l \rangle$ for $l \le L$. The action of a controlled unary iterator on $| l \rangle | \Psi \rangle$ is
\begin{equation}
    | c \rangle | l \rangle | \Psi \rangle \rightarrow | c \rangle | l \rangle ( P_l )^c | \Psi \rangle,
\end{equation}
where $| c \rangle$ is the control qubit. In Ref.~\cite{Babbush2018EncodingComplexity}, the authors introduce optimizations to the basic SELECT circuit. This is achieved by removing ``runs'' of the OFF controls on the right side of the circuit. These result in the well-established ``sawtooth'' circuits, which can be implemented in $L - 1$ Toffoli gates. However, removing these OFF controls also allows us to simplify our circuits as described above. In particular, in this case it can be seen that
\begin{equation}
    | c \rangle | 2^n - 1 \rangle | \Psi \rangle \rightarrow | c \rangle | 2^n - 1 \rangle ( P_{L-1} )^c | \Psi \rangle.
\end{equation}
In words, if the value $l = 2^n - 1$ is provided to the unary iterator, it is guaranteed to select the $l = L-1$ term, which is the desired behavior to enforce periodic boundary conditions. This is a convenient benefit of the optimized iterators, which avoids us needing to perform modular ``$-1$ mod $L$'' addition, and instead allows us to simply perform ``$-1$ mod $2^n$'' addition.

The adder circuit can be further simplified, due to the fact that it always subtracts $1$. For an $n$-bit binary number, subtracting $1$ is equivalent to adding $1_1 1_2 \ldots 1_n$. Therefore each bit of the value added is $1$, and we can prepare and unprepare a single ancilla qubit by using two Toffoli gates to incorporate the two controls for the adders. Then we use the circuit from Ref.~\cite{Sanders2020CompilationOptimization}~(Fig.~18) to perform addition with respect to this ancilla, the Toffoli cost of which is $\lceil \log_2 L \rceil - 2$. Therefore, the total Toffoli costs of the two controlled adders are $\lceil \log_2 L_x \rceil$ and $\lceil \mathrm{\log}_2 L_y \rceil$.

\end{widetext}

\end{document}